\documentclass[%
 reprint,
 amsmath,amssymb,
 aps, showkeys,
]{revtex4-1}

\usepackage{graphicx}% Include figure files
\usepackage{dcolumn}% Align table columns on decimal point
\usepackage{bm}% bold math
\usepackage{hyperref}
\usepackage{lipsum}
\usepackage{subfigure}
\usepackage{amsmath}% add hypertext capabilities
\graphicspath{ {./images/} }

\hypersetup{
    colorlinks = true,
    linkcolor = blue,
    citecolor = blue,
    urlcolor = blue,
}

\begin{document}

\preprint{APS/123-QED}

\title{On the interplay between plasma triangularity and micro-tearing turbulence}% Force line breaks with \\

\author{A. Balestri$^{1,*}$, J. Ball$^1$, S. Coda$^1$}

\affiliation{$^1$École Polytechnique Fédérale de Lausanne (EPFL), Swiss Plasma Center (SPC), 1015 Lausanne, Switzerland}%Lines break automatically or can be forced with \\


\email{alessandro.balestri@epfl.ch}

\begin{abstract}
In this work, we study the interplay between triangularity and micro-tearing turbulence using linear and nonlinear flux tube GENE simulations. We consider scenarios with negative and positive triangularity plasma shaping taken from existing tokamaks (TCV, DIII-D, MAST-U and SMART) and EU-DEMO. The study of all these tokamaks reveals a coherent picture. Negative triangularity geometry is more susceptible to micro-tearing modes (MTM), which, when present, make transport much worse than in positive triangularity. At sufficiently large $\beta$ (the ratio of plasma pressure over magnetic pressure), magnetic shear and ratio of electron to ion temperature gradient, all the scenarios with negative triangularity are dominated by MTM turbulence. In contrast, the corresponding scenarios with positive triangularity remain dominated by electrostatic turbulence and MTMs are subdominant or stable. We observe that conventional tokamaks usually operate in a parameter space far away from the onset of this MTM-dominated regime in negative triangularity, thus preserving the beneficial effect of negative triangularity on turbulent transport. In contrast, spherical tokamaks operate close to this regime and may ultimately exhibit worse transport at negative triangularity than positive triangularity. We find that lowering the magnetic shear in spherical tokamaks can preserve the beneficial effect of negative triangularity on electrostatic turbulence and prevent strong MTM transport. Finally, linear and nonlinear simulations reveal the reason for stronger MTMs: the magnetic drifts are faster in the negative triangularity geometry. 

\end{abstract}

%Use showkeys class option if keyword
\maketitle

%\tableofcontents

\section{Introduction}\label{1}

Since the first experiments on the TCV tokamak \cite{Weisen_1997}, many advances have been made in the study of Negative Triangularity (NT) plasma shaping. It is now well established that NT closes the access to H-mode \cite{Merle_2017,Nelson_2024}, creating the most robust ELM-free scenario, while reducing turbulent transport sufficiently to achieve H-mode-like performance \cite{9,7,Austin_2019,Paz-Soldan_2021,Coda_2022,Happel_2023,Nelson_PRL,Balestri_2024,Aucone_2024,Mariani_2024,Thome_2024,Paz-Soldan_2024}. These properties make NT a viable alternative to Fusion Power Plant (FPP) concepts based on the more common Positive Triangularity (PT) H-mode scenario. However, the conceptualization of an NT-FPP remains in its early stages \cite{MANTA_2024,Wilson_2025} and some important questions remain unanswered. In particular, the observed reduction in turbulent transport has been primarily observed and studied in the electrostatic limit, or in regimes dominated by electrostatic turbulence, e.g Trapped Electron Mode (TEM) turbulence \cite{Marinoni_2009,5,Balestri_2024,Balestri_2024_2,Di_Giannatale_2024,Hoffmann_2025,merlo_jenko_2023} and Ion Temperature Gradient (ITG) turbulence \cite{merlo_jenko_2023,Balestri_2024_2,Merlo_ITG,Hoffmann_2025,Di_Giannatale_2024}. However,  in comparison to most current experiments, reactor-relevant plasmas are expected to have large values of $\beta$, making electromagnetic turbulence more significant. The most important electromagnetic instabilities are  Micro-Tearing Modes (MTM) and Kinetic Ballooning Modes (KBM). To date, only \cite{Pueschel_2025} has analyzed the interplay between MTM and triangularity. The limited amount of work on this topic and its importance in view of an NT power plant motivates this study.

We highlight the fact that reliable saturated MTM-dominated nonlinear simulations have been obtained throughout this work, thus adding to the relatively small number reported in literature \cite{Guttenfelder_2011,2011_Doerk,Doerk_2012,Maeyama_2017,Hamed_2023}. 

The reminder of the paper is structured as follows. Section \ref{2} provides a brief overview of the theory behind MTM turbulence and summarizes key findings from recent years. Section \ref{3} introduces the GENE code and the numerical methodology that has been used in the rest of the paper. Section \ref{4} presents the results of linear simulations, where key parameters affecting MTM turbulence are scanned to compare behaviour in NT and PT. Section \ref{5} shows nonlinear simulations results to verify and expand the picture that emerged from linear simulations. Analogously to \cite{Balestri_2024_2}, section \ref{6} presents linear and nonlinear simulations to develop a simple physical picture for why MTMs are destabilized in NT. Finally, section \ref{7} provides conclusions and a summary.

\section{Overview of micro-tearing turbulence theory}\label{2}

A micro-tearing mode is an electromagnetic micro-instability that has been known in the plasma physics community for more than four decades \cite{Hazeltine_1975}. However, despite its long history, it remains understudied and has received more attention only in recent years thanks to its strong relation with H-mode. Since many readers may not be familiar with MTM turbulence, we will provide a brief overview.

MTMs rely on fluctuations of the radial magnetic field to create small-scale magnetic islands that increase the stochasticity of the magnetic field itself, enabling the motion of electrons along these perturbed field lines to result in radial transport. Because of their nature, MTMs can be considered as the gyroradius-scale counterpart of large-scale MHD tearing modes. However, contrary to their MHD counterparts, MTMs do not drain free energy from the relaxation of magnetic field lines after a reconnection event. Rather, their source of free energy comes from the electron temperature gradient $\mathbf{\nabla} T_e$, which couples with perturbations in the magnetic field, resulting in perturbations in the drag force $R_\parallel\propto\mathbf{B}\cdot\mathbf{\nabla}T_e$. This fundamental destabilizing mechanism was proposed for the first time by Hazeltine et al. \cite{Hazeltine_1975} in slab geometry. It was subsequently extended by \cite{1980_Gladd} with a more sophisticated linear theory that highlights the importance of magnetic shear in destabilizing MTMs and then adapted by \cite{Drake_1977} to cylindrical geometry. These three analytical works found that collisions play a fundamental role and predicted that MTMs can be unstable only at finite collisionality. Indeed,  the parallel drag force caused by collisions is made inhomogeneous by the presence of fluctuations in $\nabla_\parallel T_e$, which creates a parallel current that is able to reinforce the original radial perturbation of the magnetic field. Catto and Rosenbluth \cite{catto1981trapped} then showed that collisions are not strictly necessary to destabilize MTMs, which can be unstable also in a collisionless regime. Finally, \cite{1990_Garbet} showed analytically that MTMs can be unstable nonlinearly and lead to strong heat transport. In more recent years, MTMs have been investigated by means of gyrokinetic simulations. Ref. \cite{Applegate_2007} showed with linear gyrokinetic simulations that MTMs can be dominant in MAST discharges and are destabilized by $\beta$, collisionality and electron temperature gradient. In \cite{Guttenfelder_2011,2011_Doerk}, some of the first nonlinear gyrokinetic simulations with dominant MTMs were performed, showing that MTMs break magnetic field lines, which leads to a stochastization of the magnetic field and an associated increase in the radial transport of electrons. Ref. \cite{Doerk_2012} showed with nonlinear simulations that MTMs can dominate in standard aspect ratio tokamaks like ASDEX-U. This work also confirms that the most important parameters for the destabilization of MTMs are $\beta$ and the electron temperature gradient, while a very weak dependence on collisionality is found. Finally, \cite{2019_Hamed,Hamed_2023} show the importance of the electrostatic potential and magnetic drifts in linearly and nonlinearly destabilizing MTMs.

\begin{figure}
    \centering
    \begin{subfigure}
    {\includegraphics[width=1\linewidth]{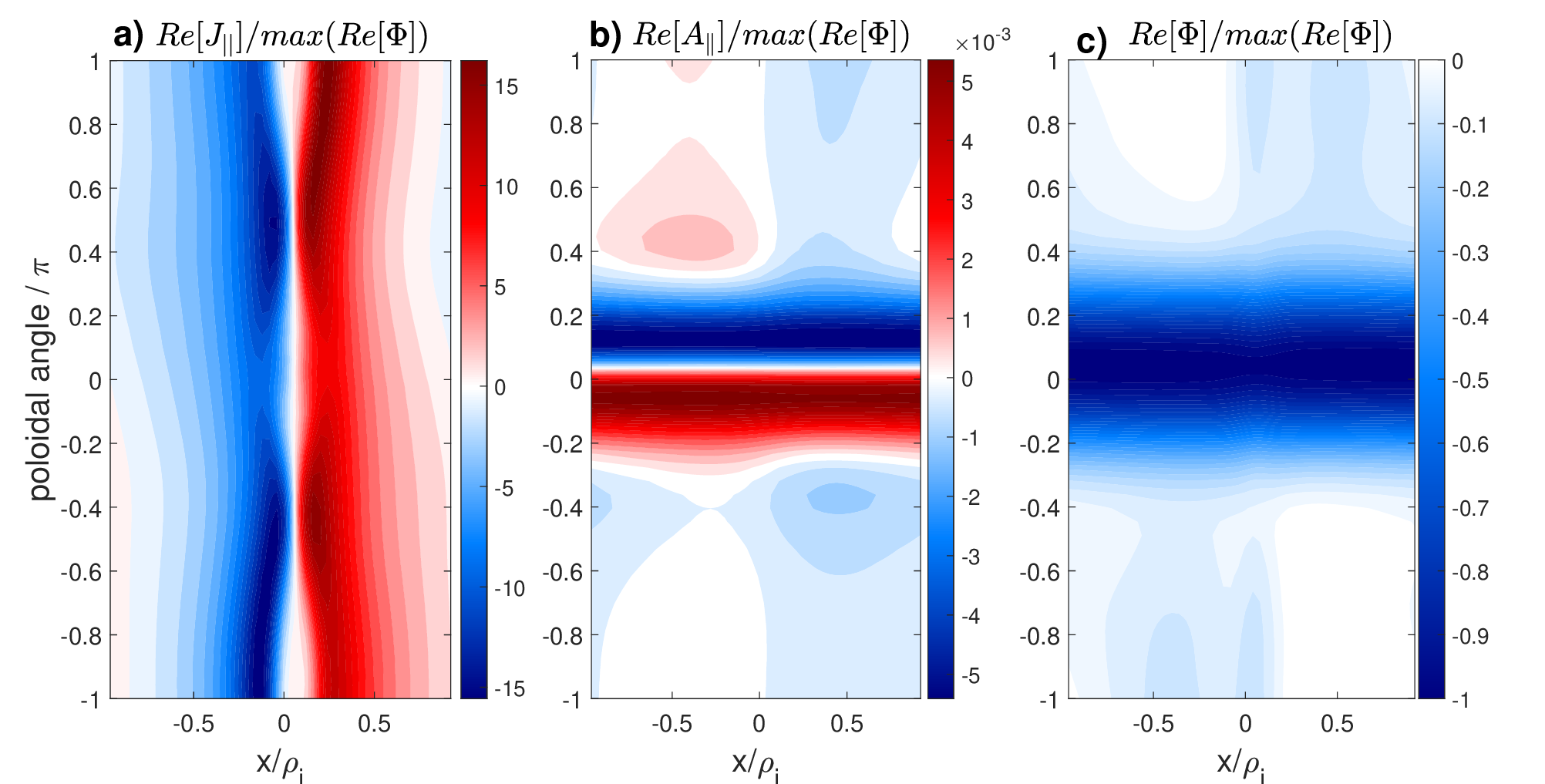}}
    \end{subfigure}
     \begin{subfigure}
    {\includegraphics[width=\linewidth]{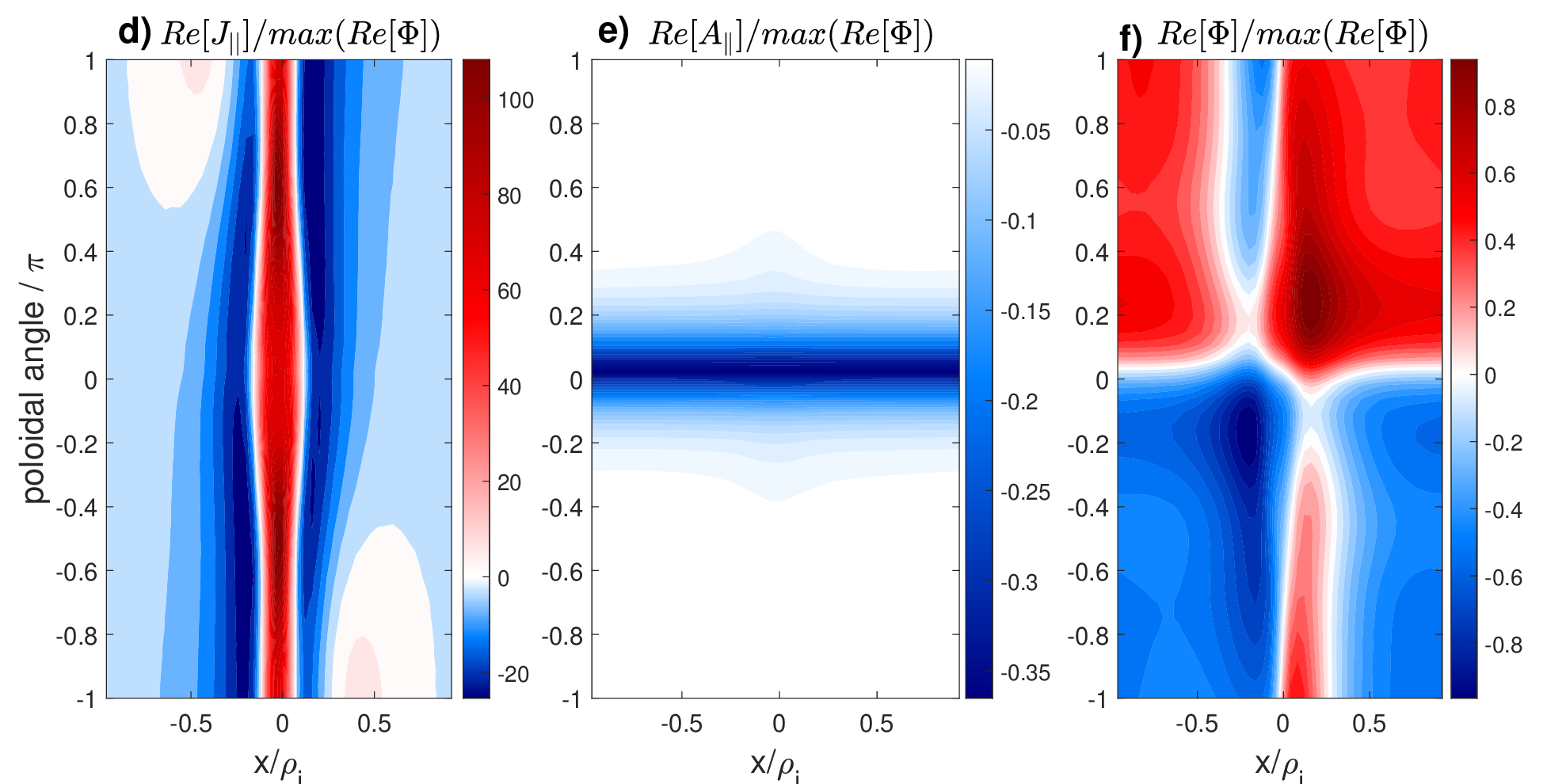}}
    \end{subfigure}
    \caption{Structures of parallel current $J_\parallel$ (left), parallel component of the vector potential $A_\parallel$ (center) and electrostatic potential $\Phi$ (right) for an ITG mode (top) and an MTM (bottom) as functions of the radial coordinate $x/\rho_i$ and the straight-field line poloidal angle. The outboard midplane corresponds to a poloidal angle of 0, while the inboard midplane to $\pm \pi$. In the first row the structures for an ITG mode are displayed, in the second row those for an MTM. $J_\parallel$, $A_\parallel$ and $\Phi$ are normalized with respect to the maximum of $\phi$}
    \label{Jpar}
\end{figure}

In summary, the literature on MTM turbulence indicates that a finite value of collisionality is essential to destabilize MTMs, even though the precise value of collisionality does not strongly influence the growth rate of these modes. The key parameters for triggering MTMs are $\beta$, magnetic shear and electron temperature gradient. Additionally, an important role in the destabilization is played by electrostatic effects and magnetic drifts. Finally, when MTMs interact nonlinearly, the small-scale magnetic islands formed by these modes can interact with each other and lead to magnetic stochastization. While the larger size of ion gyroradii enables them to average over such stochastic regions, the electrons that are carried by the stochastic field lines greatly increase the heat flux. Thus, MTM transport is typically dominated by the electron electromagnetic component. Finally, MTMs appear more commonly in STs, but can also be found in standard aspect ratio tokamaks.

To better understand the results showed in the rest of the paper, we give here a simplified and qualitative picture for the destabilization of MTMs. As previously mentioned, the fundamental mechanism underlying the destabilization of MTMs is essentially the same as in the formation of MHD tearing modes. The formation of a thin current layer within the plasma can shear the magnetic field (according to Ampere's law) and lead to the diffusion of magnetic field lines across the current layer itself, eventually producing a reconnection event. In MHD, this current layer has to preexist. By contrast, MTMs can self-generate this current thanks to the presence of a temperature gradient. A perturbation of the radial magnetic field can combine with the electron temperature gradient to create a perturbation in the electron thermal drag force ($\tilde{R}_\parallel\propto\mathbf{\tilde{B}}\cdot\mathbf{\nabla}T_e$). This creates an electric field ($\tilde{E}_\parallel$) and hence a current ($\tilde{J}_\parallel$). This thin current sheet then generates a time-varying perturbation in the radial magnetic field that can reinforce the original perturbation and drive an instability. Once the magnetic field lines break and form a magnetic island, electrons following field lines will move in the radial direction, thus increasing radial transport. This is why $\beta$ is a key parameter in the destabilization of MTMs. Indeed, $\beta$ increases the amplitude of magnetic field perturbations so that they create stronger perturbations in the drag force. Finally, if nonlinear dynamics is involved, different magnetic islands can interact with each other to make the magnetic field stochastic.

Figure \ref{Jpar} shows a comparison between the parallel current $J_\parallel$, vector potential $A_\parallel$ and electrostatic potential $\Phi$ in the $(x,z)$ plane for an electrostatic mode (ITG) and an MTM from linear gyrokinetic flux tube simulations. More details on the coordinates can be found in the next section. As mentioned, MTMs exhibit a thin current layer, which shears the magnetic field and allows the MTM to grow. We can clearly see this structure in figure \ref{Jpar}(d), where a current layer with a width $\ll\;\rho_i$ is localized in the center of the simulation domain and extends across the whole poloidal domain. This behaviour contrasts with electrostatic turbulence (figure \ref{Jpar}(a)), where $j_\parallel$ features two areas with equal magnitude and opposite sign that cancel each other. These structures in $J_\parallel$ are reflected in the structure of $A_\parallel$. For MTM (figure \ref{Jpar}(b)) we see that $A_\parallel$ is even with respect to $z=0$, while it is odd for ITG.

\section{Numerical model and methodology}\label{3}

This work consists of numerical simulations performed with the flux tube gradient-driven version of the GENE code. GENE is a physically comprehensive Eulerian gyrokinetic code that solves the Vlasov-Maxwell equations discretized on a 5-dimensional (5D) grid. The code can be run in a linear mode, if the nonlinear terms of the GK equation are neglected, or in a nonlinear mode if the terms are retained. The former is much less computationally expensive and is especially useful to develop an understanding of the nature of the turbulence that dominates in the system. The latter is much more expensive but is needed to have a physically accurate model of micro-turbulence. Indeed, one must consider the nonlinear interaction for turbulence to saturate and to assess the transport level. The real space is parametrized by a 3D set of non-orthogonal field-aligned coordinates
\begin{equation}
    \begin{cases}
        x=x(\psi)\\
        y=C_y\left(q(\psi)\theta-\varphi\right),\\
        z=\theta
    \end{cases}
    \label{coordinates}
\end{equation}
which correspond respectively to the radial, binormal, and parallel (to B) coordinates. Here $\psi$ is the radial coordinate, $\varphi$ the toroidal angle, $\theta$ the poloidal angle, $q$ the safety factor and $C_y$ is a constant. The remaining 2 coordinates of the 5D discretized grid, $v_\parallel$ and $\mu$, parametrize the velocity space and correspond to the parallel velocity and the magnetic moment, respectively. In the flux tube, the kinetic profiles are Taylor expanded around the flux surface of interest so that their values and gradients are kept constant across the simulation domain. This choice is justified if the radial turbulence scale is much smaller than the radial variation of equilibrium quantities (machine scale). This is the case for most medium-size tokamaks, and is especially true for reactor-scale devices. Therefore, the real-space simulation domain corresponds to a small rectangle extended in $x$ and $y$ at the outboard midplane, which follows the magnetic field lines along $z$. Periodic boundary conditions are used for the perpendicular coordinates $x$ and $y$, while pseudo-periodic boundary conditions that account for magnetic shear are applied in $z$.

GENE has various options to reconstruct the magnetic equilibrium at a specific flux surface. For all the simulations presented in this work, we used the local equilibrium Miller model \cite{miller,Candy_2009}, because, as an analytical model, it allows us to easily change the geometry by changing a few parameters. A Miller equilibrium is completely defined by a total of 14 scalar quantities. The parametrization of the shape, given by
\begin{equation}
    \begin{cases}
        R(\theta)=R_0[1+A^{-1}\cos{\left(\theta+\arcsin{(\delta)}\sin{\theta}\right)}]\\
        Z(\theta)=Z_0+\kappa \frac{R_0}{A}\sin{\left(\theta+\zeta\sin{(2\theta)}\right)}
    \end{cases}\text{,}
    \label{eq_Miller}
\end{equation}
requires 6 parameters: the major radius $R_0$ and the elevation $Z_0$ of the geometric center of the flux surface, the local aspect ratio $A=R_0/r$, the elongation $\kappa$, the triangularity $\delta$ and the squareness $\zeta$. Here $R$ and $Z$ are the radial and vertical cylindrical coordinates as functions of the poloidal angle $\theta$. The calculation of the poloidal field needs 6 additional parameters: the elongation shear $s_\kappa$, the triangularity shear $s_\delta$, the squareness shear $s_\zeta$, the safety factor $q_0$ and the Shafranov shift given by $\partial_rR_0$ and $\partial_rZ_0$. Finally, to specify $p'$ and $FF'$, which are needed to solve the local Grad-Shafranov equation, we need 2 additional parameters: the magnetic shear $\hat{s}$ and the radial derivative of the plasma pressure $\alpha$.

To perform our simulations, we used the Miller specification to approximate a flux surface from actual experimental equilibria. We flipped the triangularity $\delta$ (from NT to PT or vice versa) together with its shear $s_\delta$. The values of density and temperature, and their logarithmic gradients, were kept fixed. This procedure is commonly used in GK simulations and enables us to isolate the effect of geometry on the transport. Therefore, rather than predicting the gradients arising from a certain configuration, we predict the heat fluxes needed to sustain the imposed profiles and gradients. 

Given that several instabilities can arise in our simulations, table \ref{modes_criteria} shows the criteria we used to distinguish certain modes. TEM and ITG are electrostatic, thus the parity of the eigenfunction of $A_\parallel$ over ballooning space is odd, while for MTM it has to be even to allow field-line breaking. Thus, a useful quantity we can use to distinguish between these modes is $\mathcal{P}(A_\parallel)=1-\frac{\int_{-\infty}^\infty d\theta\;J\;A_\parallel}{\int_{-\infty}^\infty d\theta\;J}$, where $\theta$ is the ballooning angle and $J$ the Jacobian. Therefore, $\mathcal{P}(A_\parallel)=1$ for ITG and TEM, whereas  $\mathcal{P}(A_\parallel)<1$ for MTM. In addition, TEM and MTM generally propagate in the electron diamagnetic direction (negative frequency in GENE), while ITG propagates in the ion diamagnetic direction (positive frequency in GENE). MTMs predominantly transport heat flux and specifically in the electromagnetic channel. On the contrary, particle and heat transport are fairly similar in ITG and TEM, and the electrostatic component of the fluxes dominate over the electromagnetic one. Similar criteria can be found in \cite{Parisi_2024}.

\begin{table}
    \centering
    \begin{tabular}{l|cccc}
    \toprule
        Mode & $\mathcal{P}(A_\parallel)$ & $\omega$ & $Q_e^{em}/Q_e^{es}$ & $\Gamma_{tot}/Q_{tot}$ \\ \hline
        TEM  & $1$ & $<0$ & $\ll1$ & $\sim1$ \\
        ITG  & $1$ & $>0$ & $\ll1$ & $\sim1$ \\
        MTM  & $<1$ & $<0$  & $\gg1$ & $\ll1$ \\\toprule
    \end{tabular}
    \caption{\label{modes_criteria} The properties used to distinguish ITG, TEM and MTM. The first column defines the type of mode, the second is the parity of $A_\parallel$ in ballooning space, the third column is the sign of the real frequency of the mode, the fourth column is the ratio of the electromagnetic component of the electron heat flux over the electrostatic component and the last column is the ratio of the total particle flux over the total heat flux.}
\end{table}

\begin{table}[h]
    \centering
    \begin{tabular}{l|ccccc}
    \toprule
        &TCV & DIII-D & DEMO & SMART & MAST-U \\ \hline
        $\rho_{tor}$& 0.8 & 0.75 & 0.75 & 0.85 & 0.8\\
        $R/L_{Te}$& 17.37 & 11.39 & 10.33 & 15.3 & 11.74\\ 
        $R/L_{Ti}$& 13.94 & 9.27 & 10.33 & 9.21 & 11.52\\ 
        $R/L_{ne}$& 10.91 & 3.72 & 0.77 & 2.01 & 0.88\\
        $R/L_{nC}$& - & 5.24 & - & - & -\\
        $T_e$ [keV]& 0.35 & 0.46 & 1.5 & 0.09 & 0.19\\ 
        $T_i/T_e$& 1.4 & 1.7 & 1.0 & 0.99 & 1.78\\ 
        $n_e\,[10^{19}m^{-3}]$& 2.25 & 2.88 & 7.98 & 2.83 & 2.33 \\ 
        $n_C/n_e$& - & 0.04 & - & - & - \\
        $\beta [\%]$ &0.16 & 0.13 & 0.14 & 0.60 & 0.76\\
        $\nu_{C}$ & 0.0054 & 0.0079 & 0.011 & 0.05 & 0.01\\
        \hline
        $A$& 5 & 4 & 3.5 & 2.2 & 1.8\\ 
        $q$& 1.89 & 1.99 & 1.54 & 3.33 & 5.89\\ 
        $\hat{s}$& 2.46 & 1.75 & 2.01 & 3.74 & 4.91\\
        $\kappa$ & 1.22 & 1.28 & 1.35 & 1.76 & 1.58\\
        $|\delta|$ & 0.15 & 0.16 & 0.13 & 0.16 & 0.25\\
        $\zeta$ & 1.8$\cdot10^{-3}$ & -0.04 & -0.07 & -0.05 & -0.01\\
        $s_\kappa$ & 0.11 & 0.04 & 0.35 & 0.66 & 0.26\\
        $|s_\delta|$& 0.35 & 0.34 & 0.34 & 0.97 & 0.78\\
        $s_\zeta$ & 6.9$\cdot10^{-4}$ & -0.13 & -0.21 & -0.28 & -0.12\\\toprule
        % $\alpha_{MHD}$& 0.37 & 0.55 & 0.08 & 1.93 &2.51\\\toprule
    \end{tabular}
    \caption{\label{input} Key simulation parameters for various experimental scenarios including the radial location of the simulation expressed in $\rho_{tor}$, the logarithmic gradients of electron temperature $R/L_{Te}$, ion temperature $R/L_{Ti}$, electron density $R/L_{ne}$, carbon density $R/L_{nC}$, as well as the ion-electron temperature ratio $T_e/T_i$ , electron density $n_e$, the carbon to electron density $n_C/n_e$, the local electron $\beta$, normalized collisional frequency $\nu_{C}$ and geometric parameters needed to specify the Miller equilibrium.}
\end{table}

As a final remark, we note that this work consists primarly of linear simulations and only a few nonlinear simulations. The reason for the limited amount of nonlinear simulations is the challenging requirements of simulating MTM turbulence. To properly resolve MTM turbulence, we need very large simulation boxes in the radial domain to fully capture the eddies, which are extremely elongated. In addition, high radial resolutions are required because the current layers are very narrow in $x$. Moreover, MTMs usually transport most of the fluxes at low binormal wavenumber, thus requiring a large box also in $y$. These requirements make simulations extremely expensive and achieving convergence is challenging. For these reasons, we carried out a large number of linear simulations to develop a first understanding of how MTMs depend on a large set of parameters that we expect to be relevant. Then we performed a small set of nonlinear simulations to confirm that nonlinear dynamics does not change the trends observed with linear simulations.

\begin{figure*}
    \centering
    \begin{subfigure}
    {\includegraphics[width=0.95\textwidth]{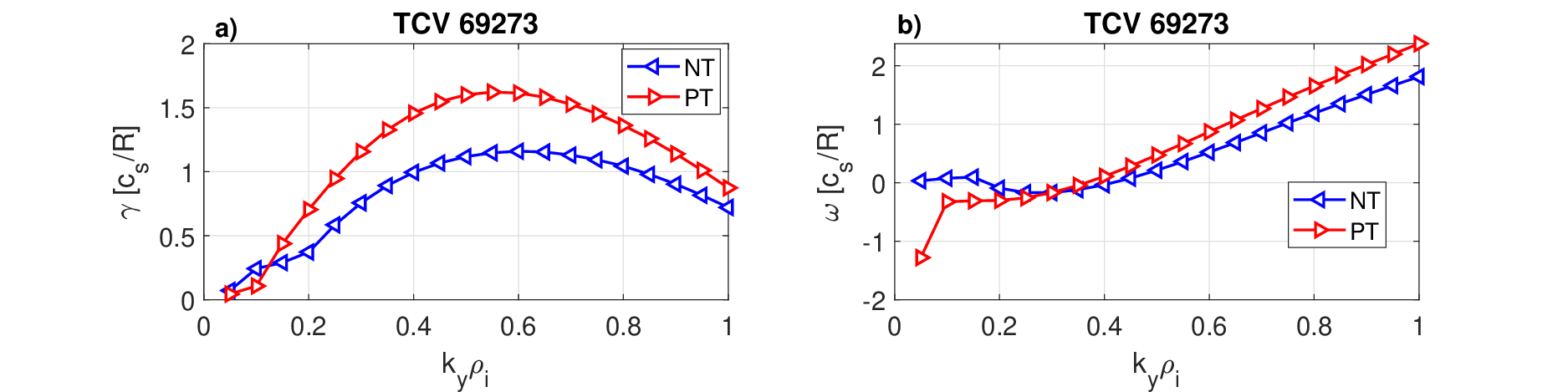}}
    \end{subfigure}
     \begin{subfigure}
    {\includegraphics[width=0.95\textwidth]{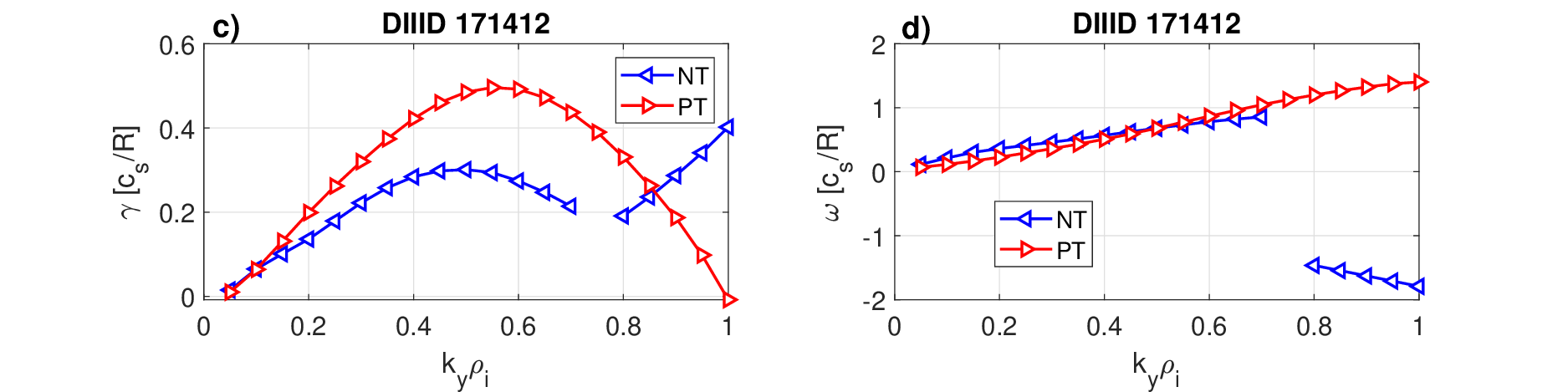}}
    \end{subfigure}
    \begin{subfigure}
    {\includegraphics[width=0.95\textwidth]{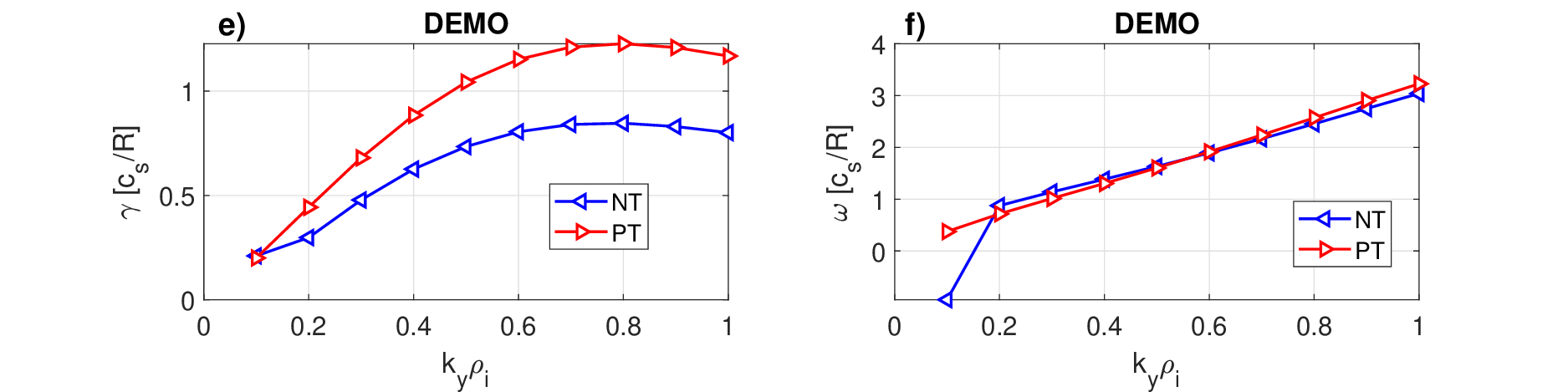}}
    \end{subfigure}
     \begin{subfigure}
    {\includegraphics[width=0.95\textwidth]{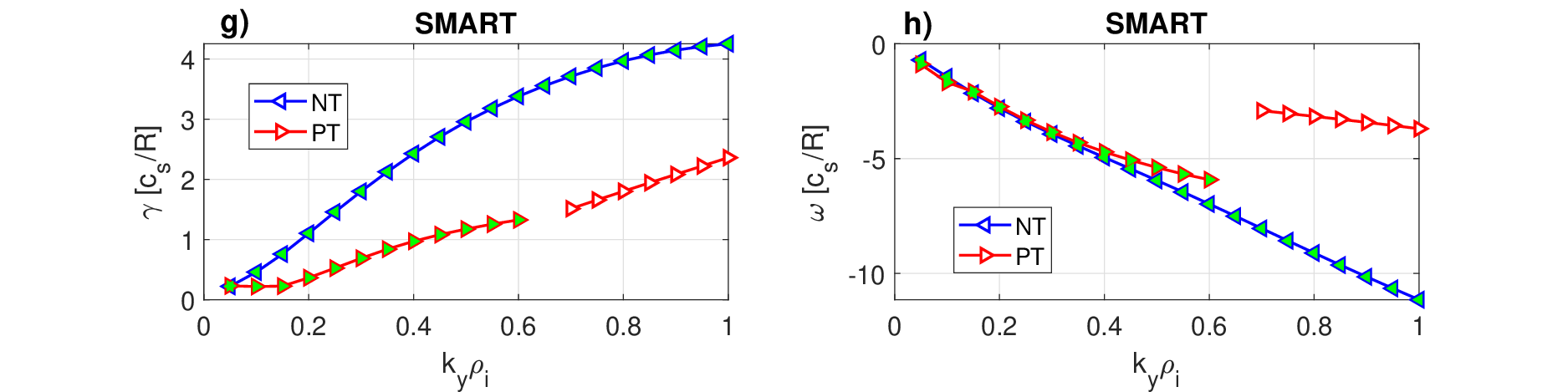}}
    \end{subfigure}
    \begin{subfigure}
    {\includegraphics[width=0.95\textwidth]{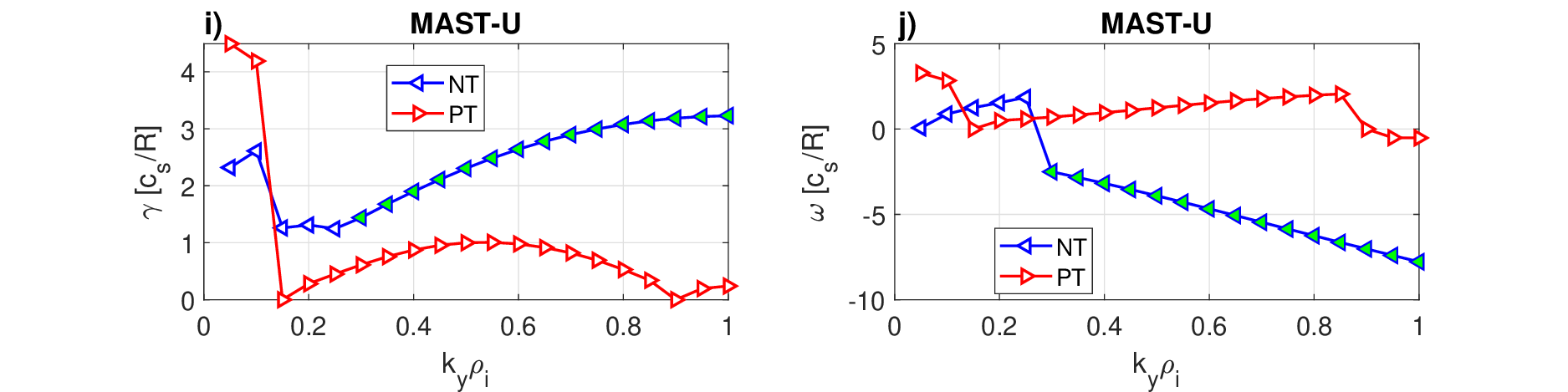}}
    \end{subfigure}
    \caption{Linear growth rates (left column) and real frequencies (right column) of the most unstable mode as functions of binormal wavenumber $k_y\rho_i$ for NT (blue) and PT (red) scenarios of different tokamaks. MTMs are highlighted by green markers. The growth rates and frequencies are normalized with respect to ion sound speed $c_s$ and major radius $R$.}
    \label{scanky_tot}
\end{figure*}

\section{Multi-machine linear simulations}\label{4}

In this section, we present linear flux tube simulations of PT and NT scenarios in several conventional tokamaks and spherical tokamaks (i.e. TCV, DIII-D, SMART, MAST-U and DEMO) to assess the existence window of MTMs. The reference parameters are listed in table \ref{input}.

The scenarios from TCV and DIII-D are actual NT experiments (shots \#69273 and \#171421) respectively. For MAST-U we considered a real PT H-mode discharge (\#47090). Since SMART \cite{SMART1, SMART2} is not yet operational, we used the same scenario described in \cite{Balestri_2024}, which was predicted by TRANSP for a PT scenario \cite{Cruz-Zabala_2024}. Finally, for DEMO, we used a preliminary NT scenario predicted with ASTRA \cite{ASTRA} coupled to TGLF \cite{TGLF}. As mentioned, the opposite triangularity scenarios have been built by flipping the sign of triangularity and its shear, while keeping all the other parameters in table \ref{input} fixed. All the simulations have been performed in the outer core region, i.e. $0.7<\rho_{tor}<0.9$, where the magnitude of triangularity is substantial and we expect it to have the largest impact on transport. Here $\rho_{tor}=\sqrt{\Phi/\Phi_{LCFS}}$, where $\Phi$ is the toroidal magnetic flux.

Using the base set of parameters shown in table \ref{input}, we first performed linear binormal wavenumber $k_y$ scans to assess the dominant type of instabilities at the ion scale. The scenarios are arranged in figure \ref{scanky_tot} by decreasing aspect ratio: from TCV to MAST-U. Figures \ref{scanky_tot}(a,b) indicate that TCV operates in a mixed ITG-TEM regime, characterized by a smooth transition from negative frequencies (TEM) to positive frequencies (ITG). In the range that we expect to be the most important for transport $k_y\rho_i=[0.1,1.0]$, NT proves to be more stable, with growth rates reduced by approximately 30\% relative to PT. A similar trend appears in figures \ref{scanky_tot}(c-f) for DIII-D and DEMO. In the two devices, both NT and PT are ITG-dominated at low $k_y\rho_i$ and transition sharply to a TEM regime around $k_y\rho_i=1.0$. In both cases, NT remains more stable, exhibiting a 30–50\% reduction in growth rates compared to PT. The results for the two STs, shown in the last two rows of figure \ref{scanky_tot}, present a contrasting picture. In SMART and MAST-U, NT is more unstable than PT across the entire ion scale, with growth rates exceeding those of PT by more than a factor of two. Based on the criteria outlined in table \ref{modes_criteria}, the dominant modes in NT MAST-U are ITG at very low $k_y\rho_i$ and MTMs for the rest of ion scale, while the PT scenario is dominated by ITG and TEM. In SMART, the NT geometry exhibits only MTMs in the considered range of $k_y\rho_i$, while PT is dominated by MTMs until $k_y\rho_i=0.6$ and transitions to TEM for larger wave numbers. 

\begin{figure}
    \centering
    \includegraphics[width=\linewidth]{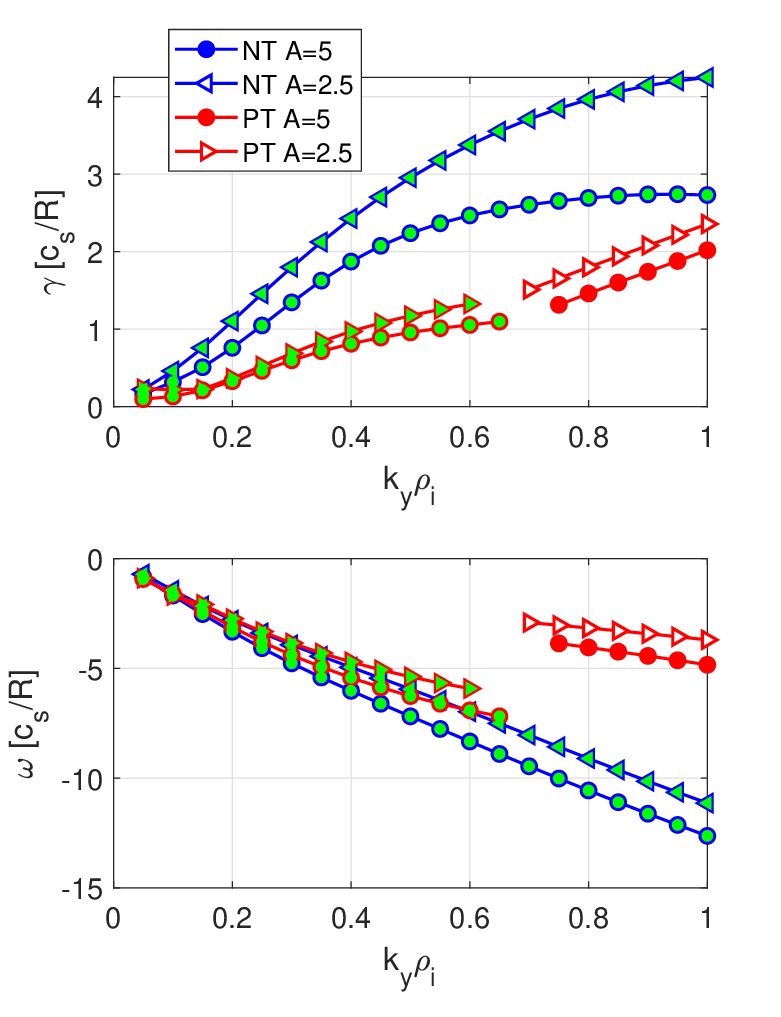}
    \caption{Linear growth rates (a) and real frequencies (b) of the most unstable mode as a function of binormal wavenumber $k_y\rho_i$ for SMART NT (blue) and PT (red) scenarios with nominal tight aspect ratio (empty triangles) and larger aspect ratio (full circles). MTMs are highlighted by green markers.}
    \label{scanky_scanA}
\end{figure}

\begin{figure*}
    \centering
    \begin{subfigure}
    {\includegraphics[width=0.48\textwidth]{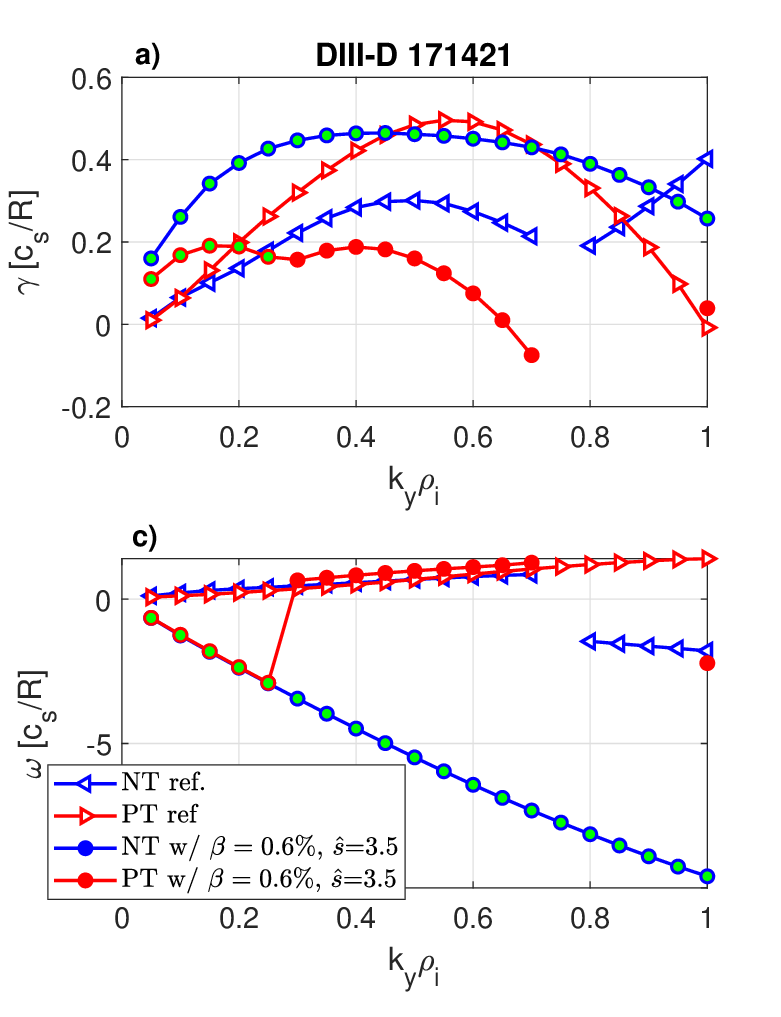}}
    \end{subfigure}
     \begin{subfigure}
    {\includegraphics[width=0.48\textwidth]{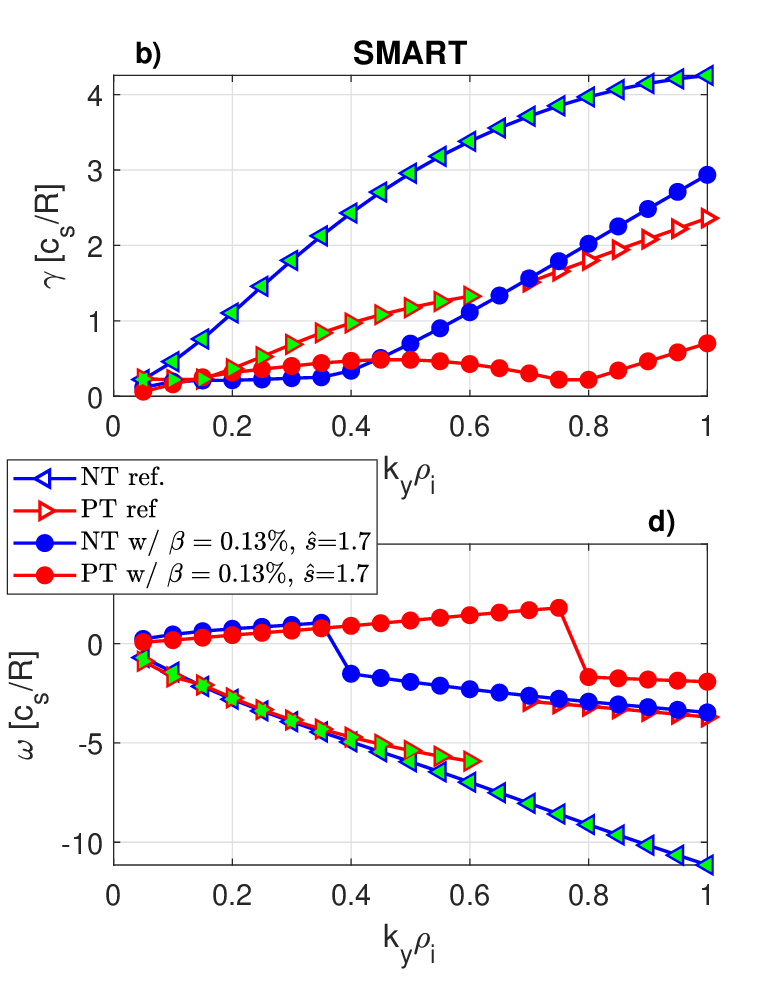}}
    \end{subfigure}
    \caption{Linear growth rates (a,b) and real frequencies (c,d) of the most unstable mode as functions of binormal wavenumber $k_y\rho_i$ for NT (blue) and PT (red) scenarios of DIII-D (left) and SMART (right). Triangles represent simulations with nominal and self-consistent parameters for the two devices. Circles represent simulations with swapped $\beta$ and $\hat{s}$ between the two machines. MTMs are highlighted by green markers.}
    \label{scanky_beta_shat_mod}
\end{figure*}

Therefore, from figure \ref{scanky_tot} we observe that, at nominal parameters, all the conventional aspect ratio tokamaks (TCV, DIII-D, DEMO) are dominated by electrostatic turbulence (ITG or TEM) and NT is more stable than PT in the $k_y\rho_i$ range where nonlinear fluxes usually peak. Conversely, the two STs are dominated by MTMs in NT geometry, while PT ones are only marginally affected by them. In this regime NT is less stable than PT, contrary to what is observed in the electrostatic regimes. To explore whether aspect ratio $A$ alone destabilizes MTMs in NT geometries, we artificially modified the aspect ratio of the SMART scenario by adjusting the minor radius. These linear simulations (figure \ref{scanky_scanA}) indicate that MTMs remain dominant in SMART even at a conventional aspect ratio of $A=5$ (remembering that $A$ is the \textit{local} aspect ratio of the flux surface being simulated). In addition, NT remains more unstable than PT, suggesting that MTMs are in general stronger in NT geometry, independently of $A$. Thus, we cannot conclude that aspect ratio alone is responsible for the destabilization of MTM and other parameters must be involved in the onset of these modes. Looking at table \ref{input}, we note that the considered conventional $A$ tokamaks differ from the STs primarily in $\beta$ and $\hat{s}$. The higher magnetic shear in the considered STs appears to stem from the fact that the scenarios we considered for SMART and MAST-U are Double-Null (DN), whereas TCV, DIII-D and DEMO are Single-Null (SN). The presence of two X-points in the STs scenarios seems to increase magnetic shear. On the other hand, the much larger values of $\beta$ in SMART and MAST-U are indirectly but intrinsically related to the lower values of $A$. The reduced $A$ of STs loosens MHD limits on the plasma pressure, which allows for increased $\beta$. Therefore, to further investigate the role played by magnetic shear and $\beta$, we performed additional linear simulations for SMART and DIII-D swapping their nominal values of $\beta$ and $\hat{s}$ at the chosen radial location (keeping other input parameters fixed). 

Figure \ref{scanky_beta_shat_mod} confirms that $\beta$ and $\hat{s}$ play crucial roles in destabilizing MTM. Figures \ref{scanky_beta_shat_mod}(a,c) show that, using the values of $\beta$ and $\hat{s}$ from SMART, MTMs are triggered in both NT and PT DIII-D discharges. Still, MTMs are unstable for a wider range of the ion scale in NT and are much more unstable than in PT. Therefore, increasing $\beta$ and $\hat{s}$ caused the DIII-D NT scenario, which was more stable than PT, to be dominated by electromagnetic modes and to be much more unstable than the PT counterpart. Figures \ref{scanky_beta_shat_mod}(b,d) show a consistent picture also for SMART. When $\beta$ and $\hat{s}$ are reduced, MTMs are stabilized and electrostatic turbulence dominates across the whole ion scale. In addition, NT becomes more stable than PT in the range $k_y\rho_i=[0.05,0.4]$, where nonlinear fluxes usually peak. Thus, we conclude that MTM destabilization is not directly related to aspect ratio, but depends strongly on $\beta$ and $\hat{s}$. Moreover, whenever MTMs dominate in the NT scenario, the growth rates are higher than in the corresponding PT scenario.

\begin{figure*}
    \centering
    \begin{subfigure}
    {\includegraphics[width=0.32\textwidth]{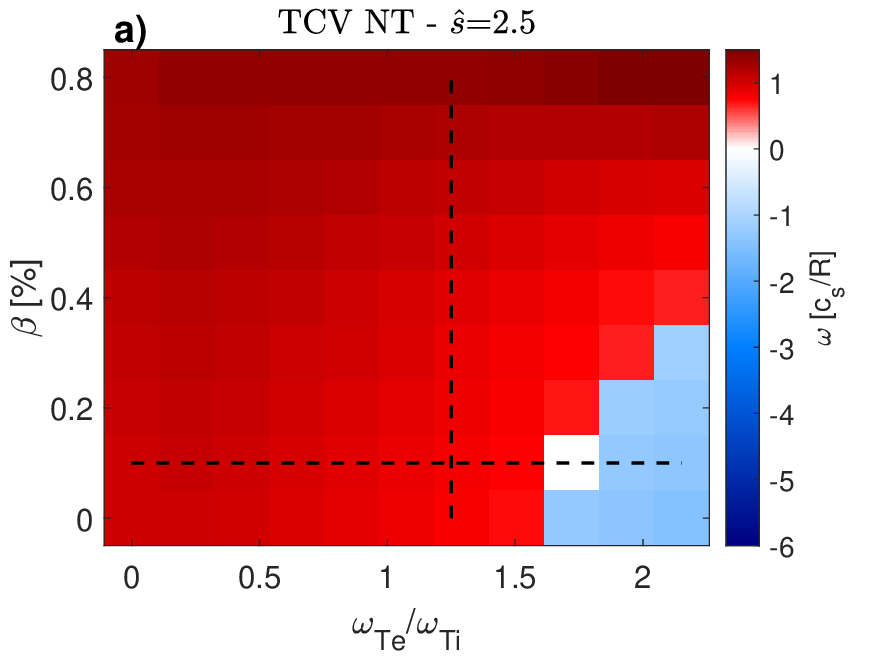}}
    \end{subfigure}
     \begin{subfigure}
    {\includegraphics[width=0.32\textwidth]{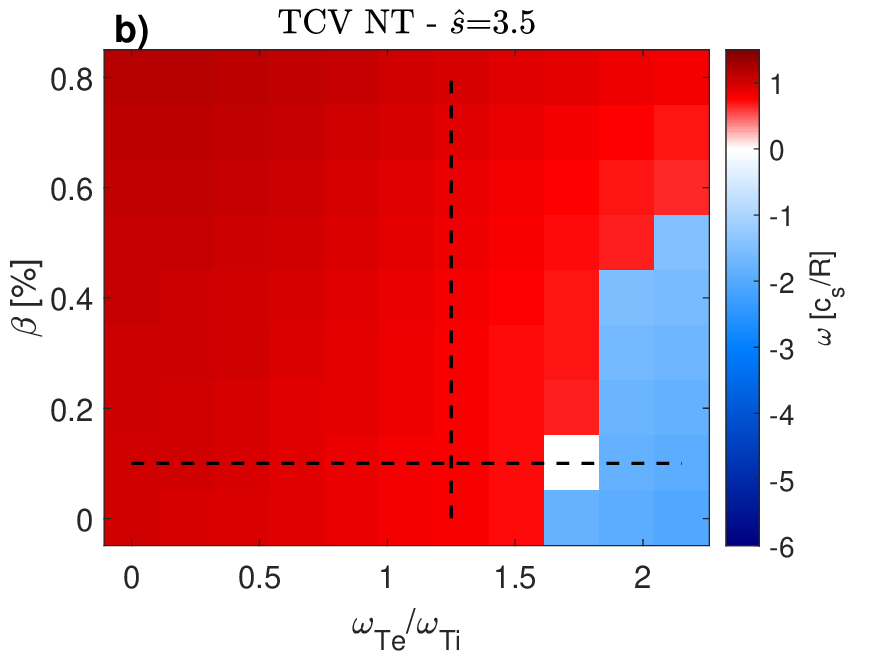}}
    \end{subfigure}
    \begin{subfigure}
    {\includegraphics[width=0.32\textwidth]{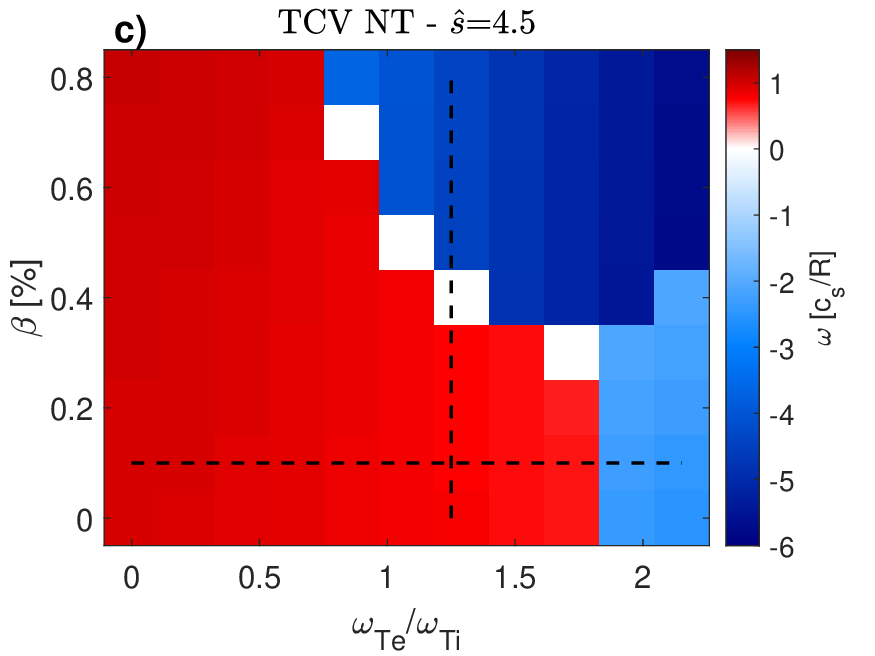}}
    \end{subfigure}
    \begin{subfigure}
    {\includegraphics[width=0.32\textwidth]{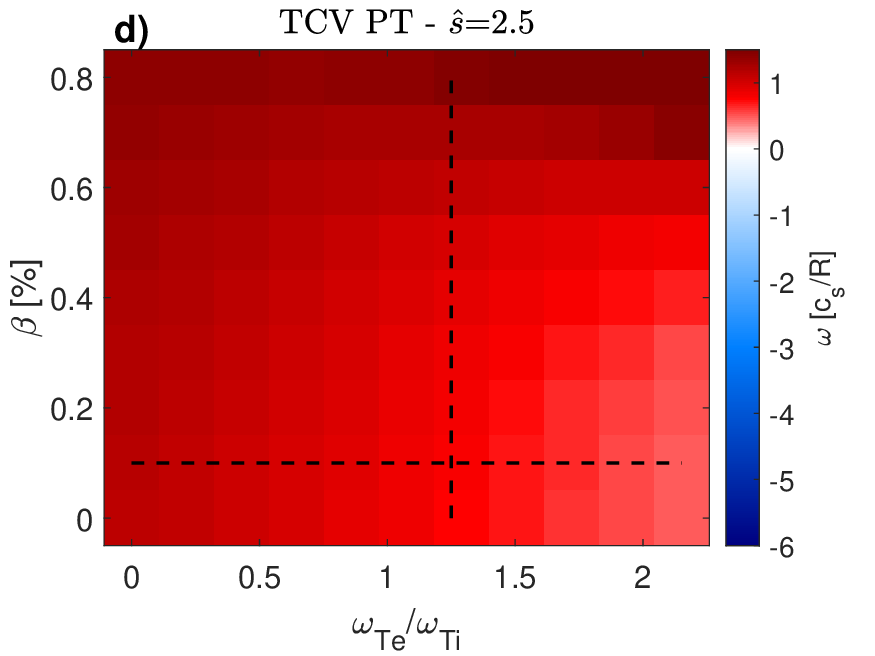}}
    \end{subfigure}
     \begin{subfigure}
    {\includegraphics[width=0.32\textwidth]{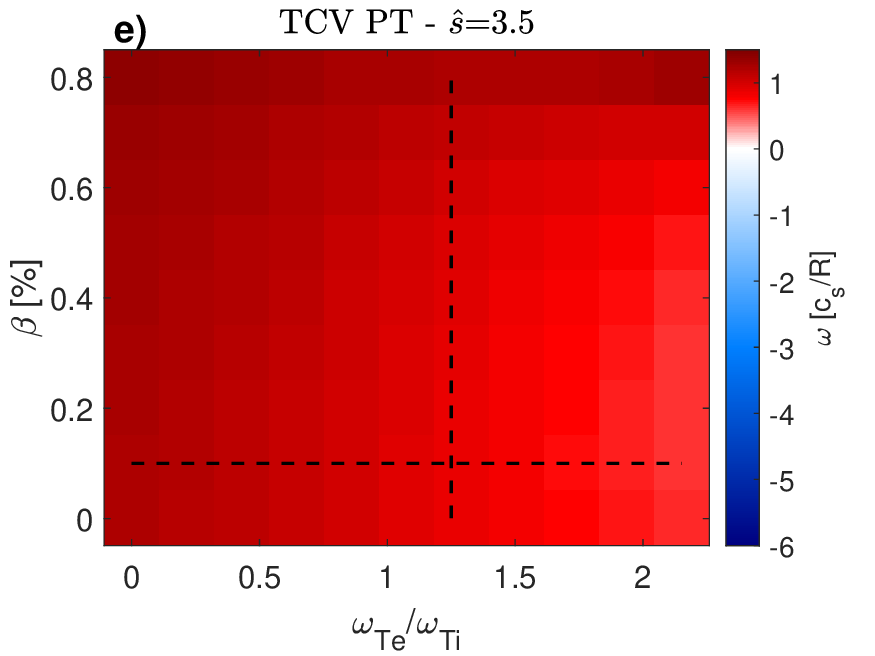}}
    \end{subfigure}
    \begin{subfigure}
    {\includegraphics[width=0.32\textwidth]{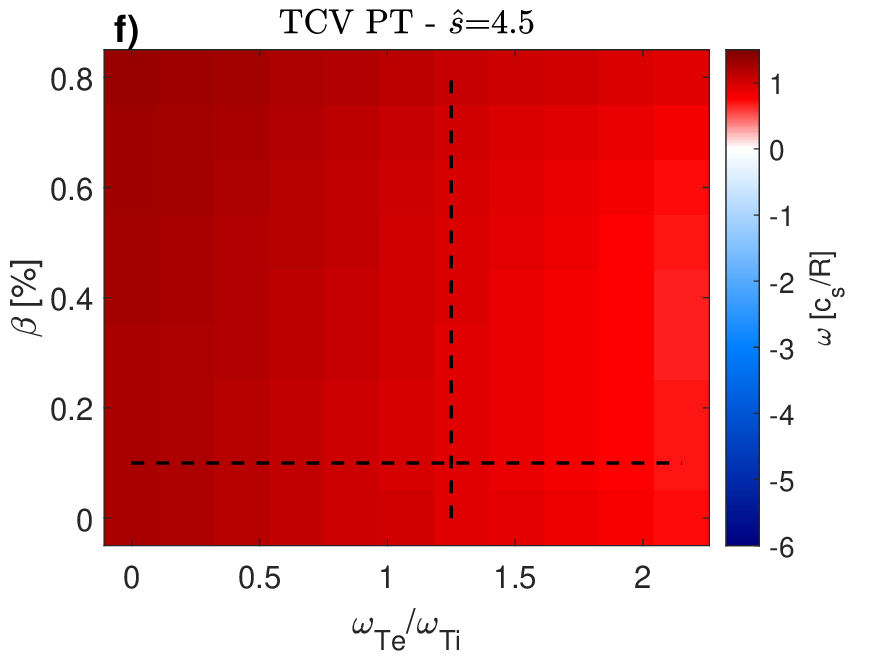}}
    \end{subfigure}
    \caption{Real frequency colormaps for TCV NT (top) and PT (bottom) scenarios as functions of $\beta$ and the ratio of electron temperature gradient over ion temperature gradient $\omega_{Te}/\omega_{Ti}$. Different columns correspond to different values of magnetic shear $\hat{s}$, which increases from left to right. The black dashed lines represent the nominal values of $\omega_{Te}/\omega_{Ti}$ and $\beta$, while the nominal shear is $\hat{s}=2.5$.}
    \label{TCV_omega}
\end{figure*}

\begin{figure*}
    \centering
    \begin{subfigure}
    {\includegraphics[width=0.32\textwidth]{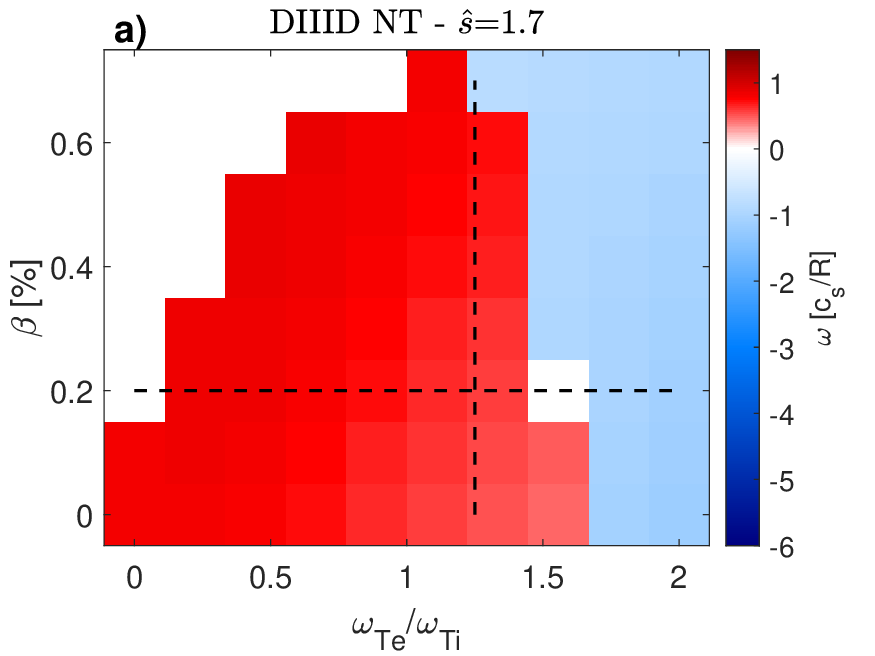}}
    \end{subfigure}
     \begin{subfigure}
    {\includegraphics[width=0.32\textwidth]{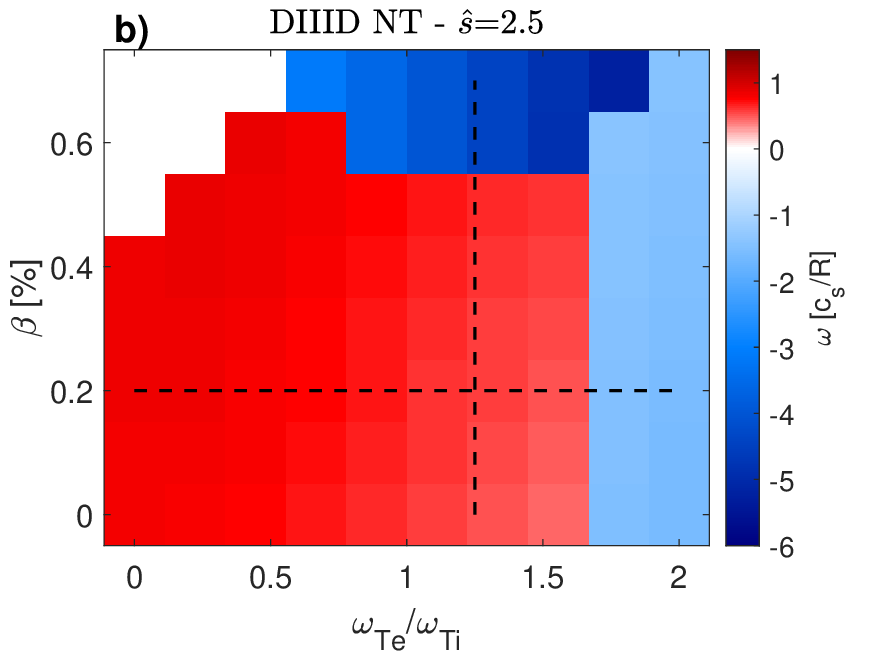}}
    \end{subfigure}
    \begin{subfigure}
    {\includegraphics[width=0.32\textwidth]{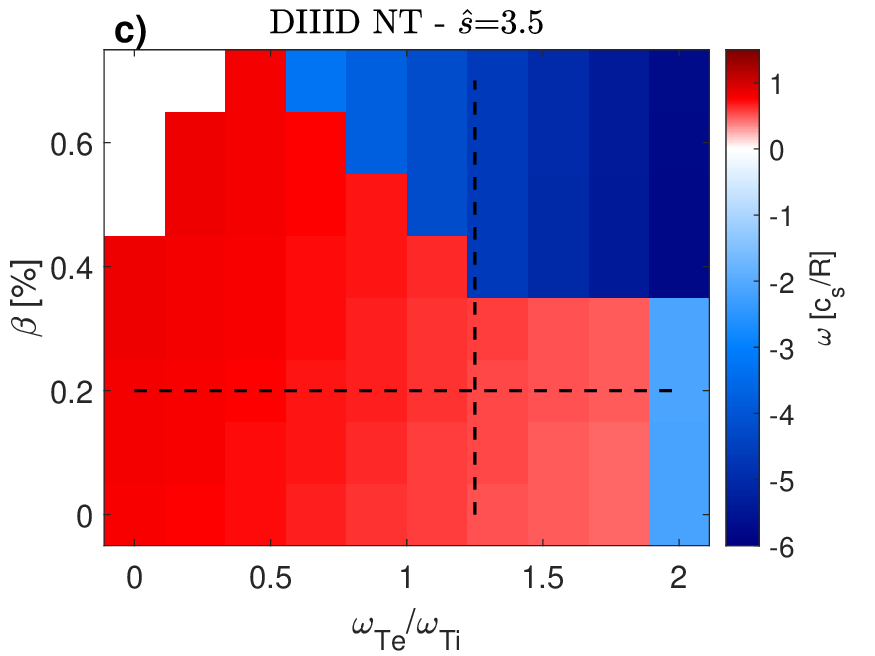}}
    \end{subfigure}
    \begin{subfigure}
    {\includegraphics[width=0.32\textwidth]{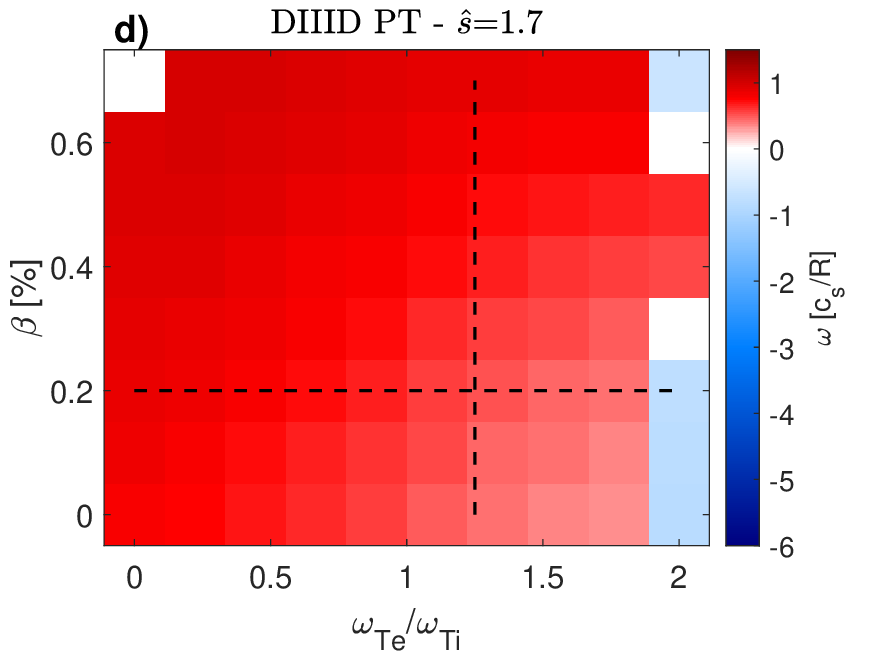}}
    \end{subfigure}
     \begin{subfigure}
    {\includegraphics[width=0.32\textwidth]{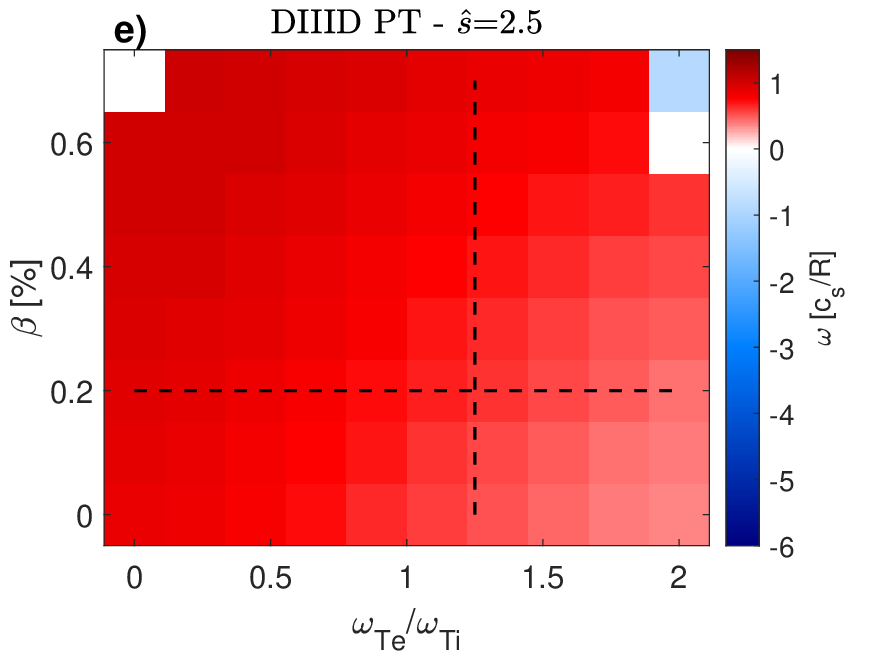}}
    \end{subfigure}
    \begin{subfigure}
    {\includegraphics[width=0.32\textwidth]{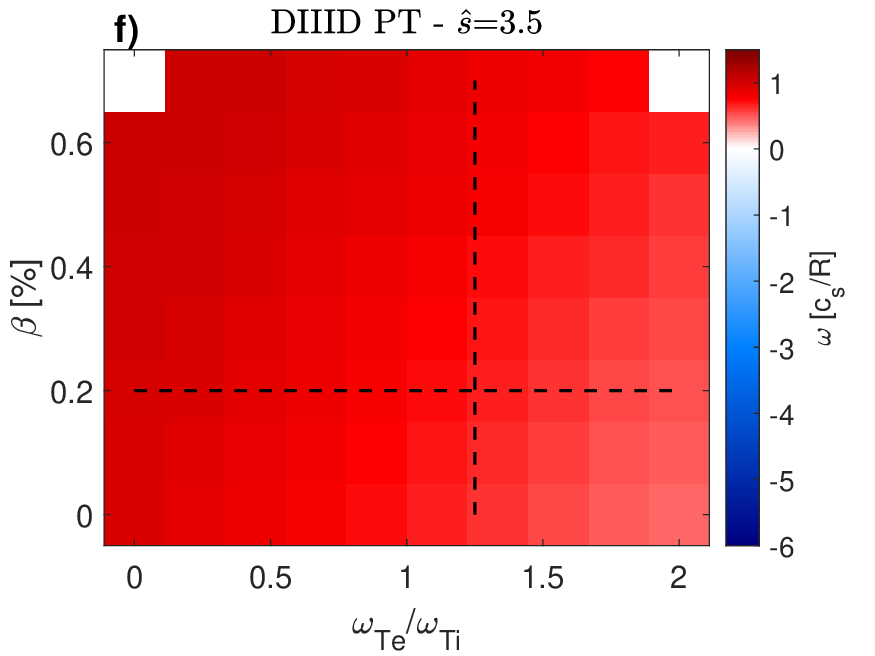}}
    \end{subfigure}
    \caption{Same as figure \ref{TCV_omega}, except for the DIII-D scenarios. The nominal shear is $\hat{s}=1.7$.}
    \label{DIIID_omega}
\end{figure*}

\begin{figure*}
    \centering
    \begin{subfigure}
    {\includegraphics[width=0.32\textwidth]{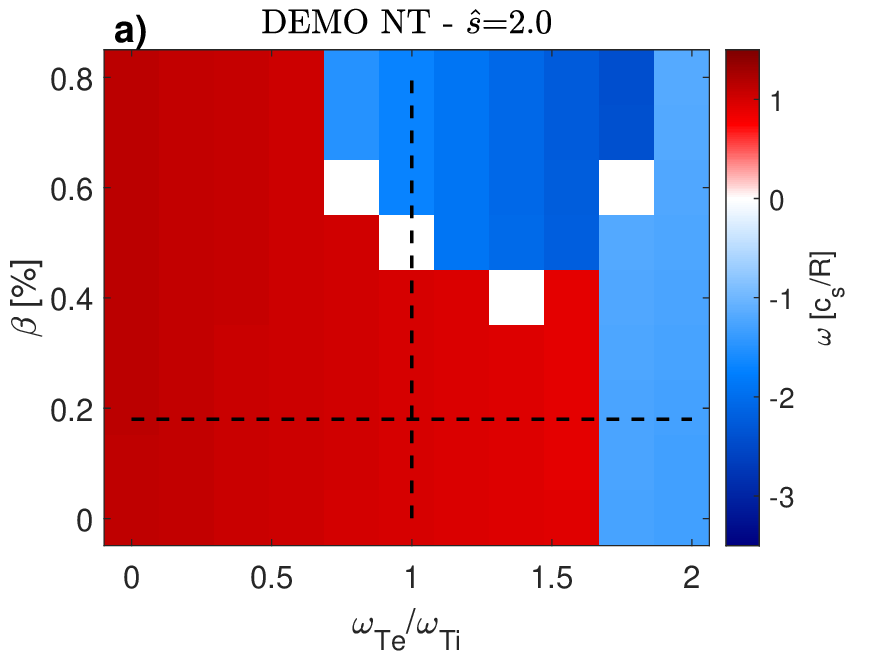}}
    \end{subfigure}
     \begin{subfigure}
    {\includegraphics[width=0.32\textwidth]{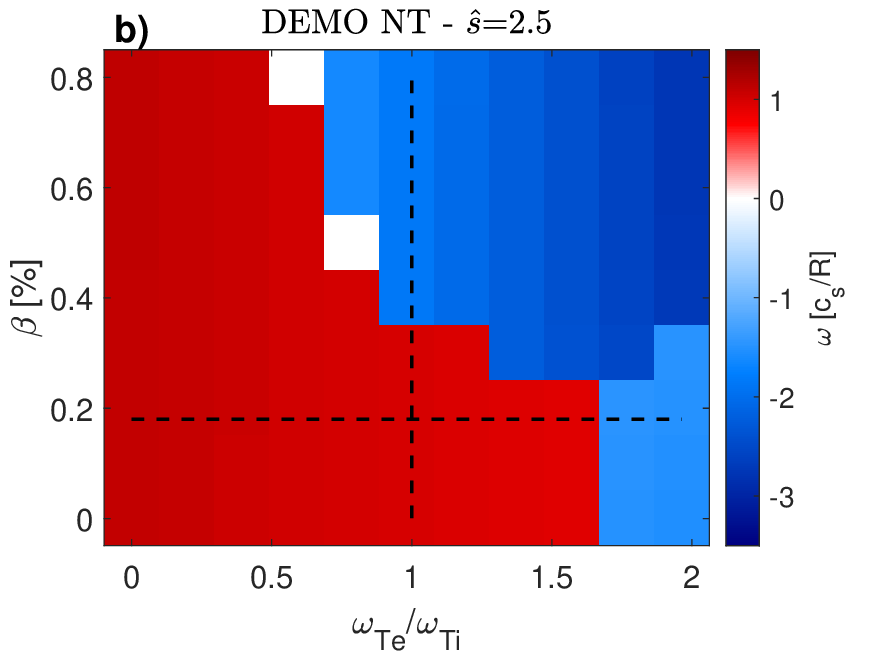}}
    \end{subfigure}
    \begin{subfigure}
    {\includegraphics[width=0.32\textwidth]{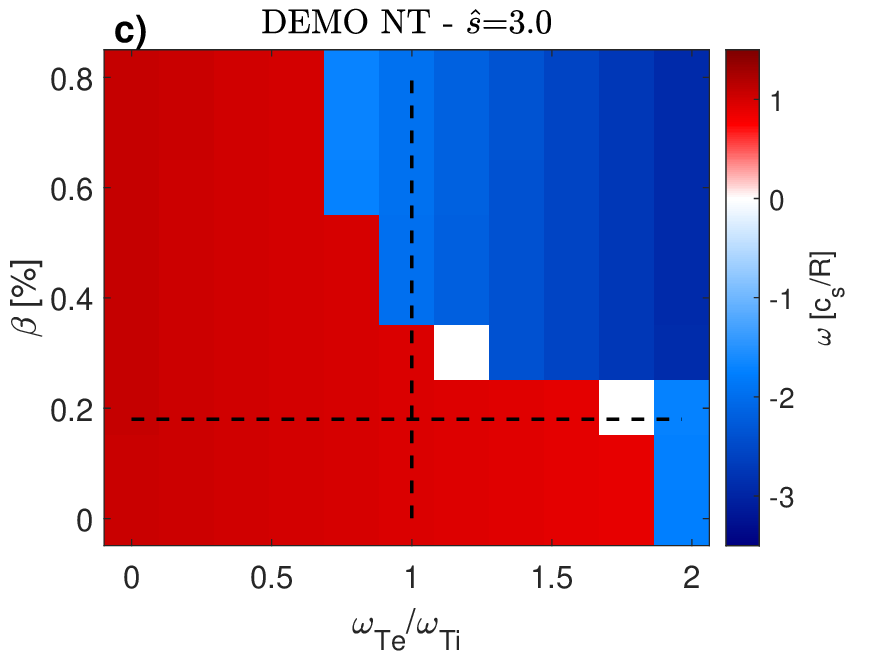}}
    \end{subfigure}
    \begin{subfigure}
    {\includegraphics[width=0.32\textwidth]{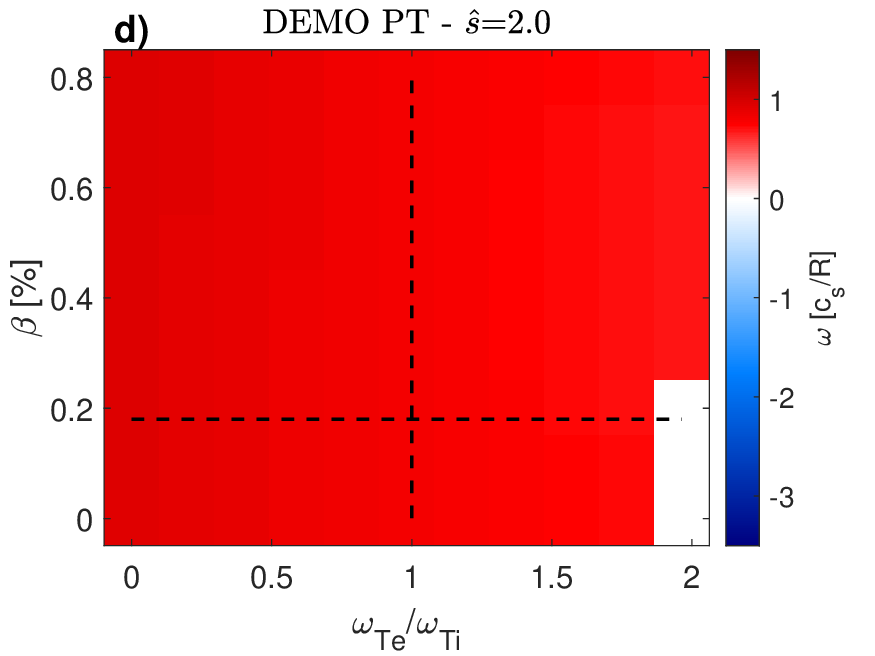}}
    \end{subfigure}
     \begin{subfigure}
    {\includegraphics[width=0.32\textwidth]{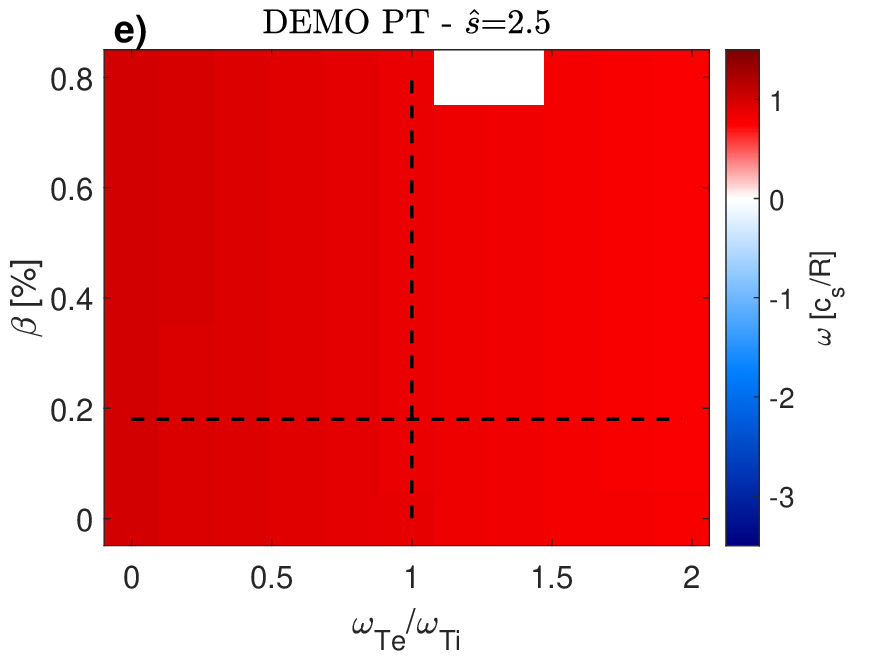}}
    \end{subfigure}
    \begin{subfigure}
    {\includegraphics[width=0.32\textwidth]{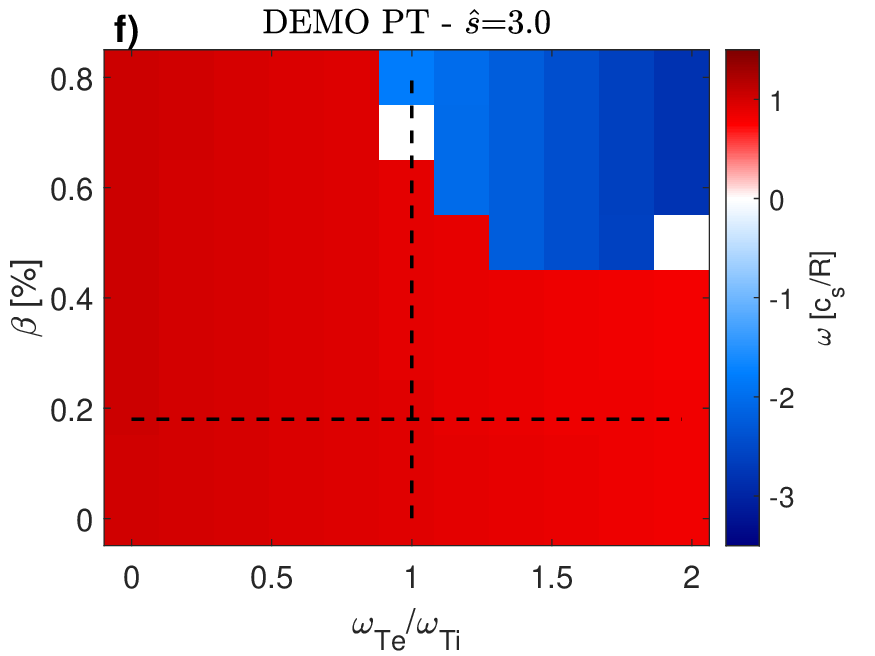}}
    \end{subfigure}
    \caption{Same as figure \ref{TCV_omega}, except for the DEMO scenarios. The nominal shear is $\hat{s}=2.0$.}
    \label{DEMO_omegaNT}
\end{figure*}

\begin{figure*}
    \centering
    \begin{subfigure}
    {\includegraphics[width=0.32\textwidth]{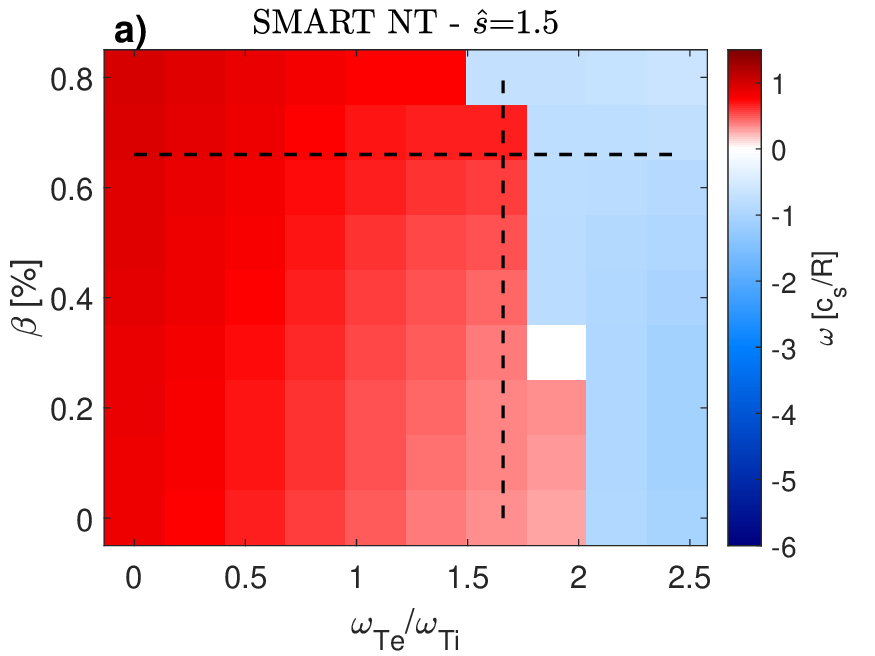}}
    \end{subfigure}
     \begin{subfigure}
    {\includegraphics[width=0.32\textwidth]{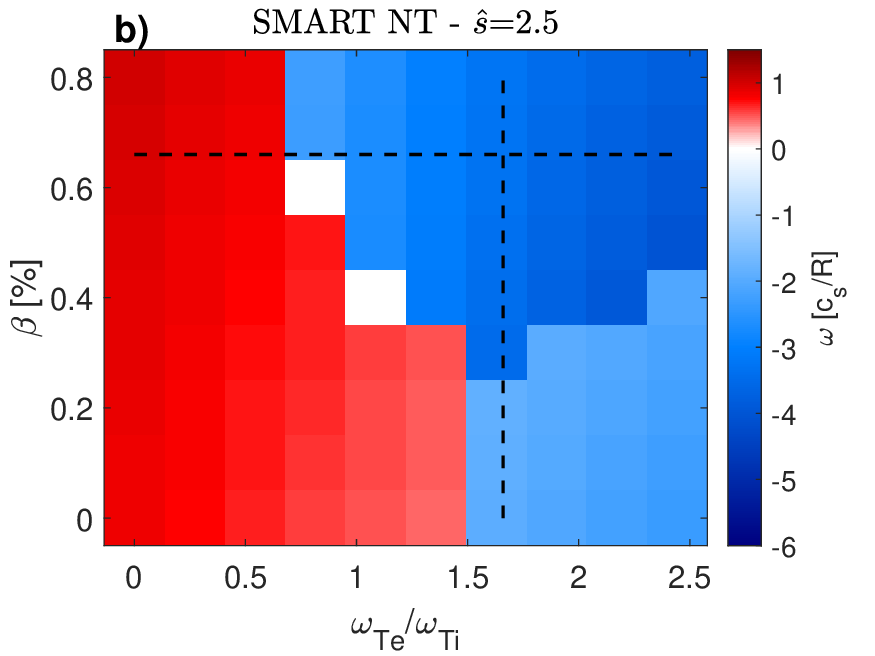}}
    \end{subfigure}
    \begin{subfigure}
    {\includegraphics[width=0.32\textwidth]{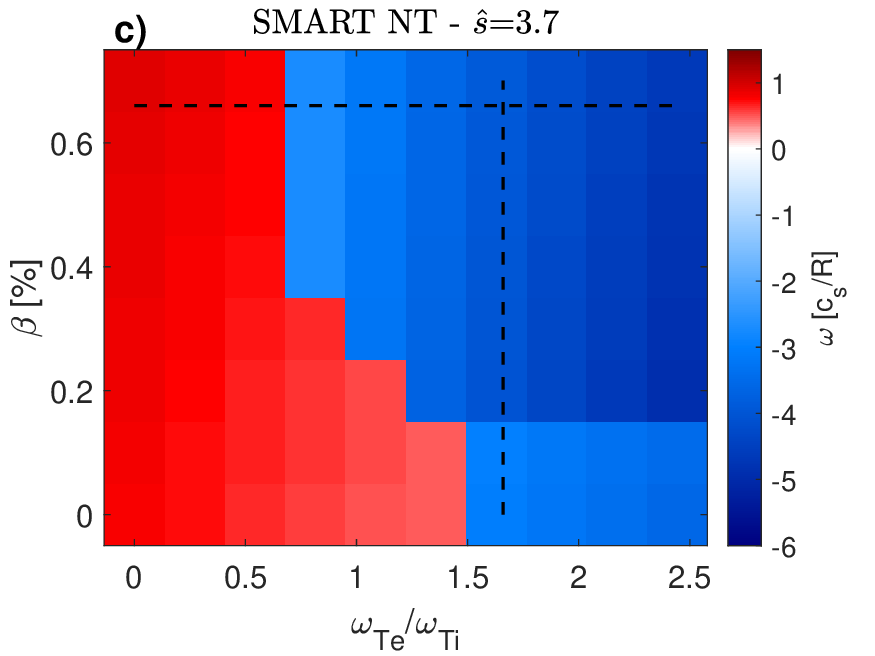}}
    \end{subfigure}
    \begin{subfigure}
    {\includegraphics[width=0.32\textwidth]{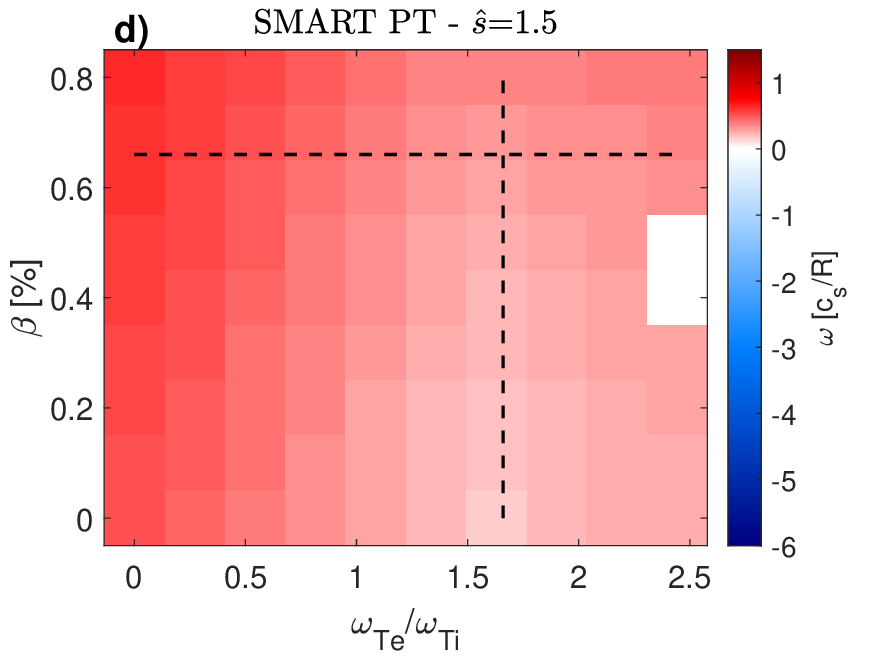}}
    \end{subfigure}
     \begin{subfigure}
    {\includegraphics[width=0.32\textwidth]{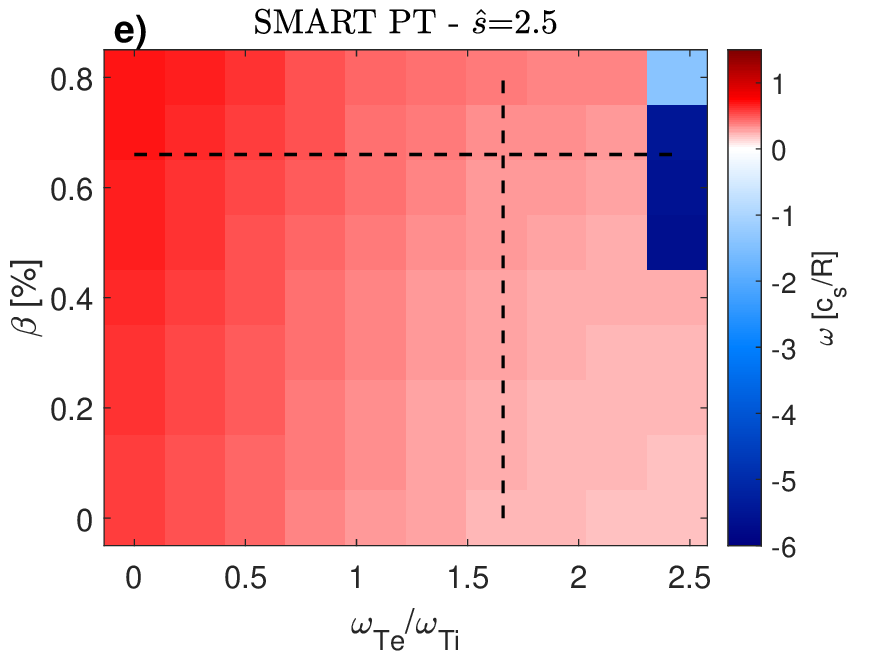}}
    \end{subfigure}
    \begin{subfigure}
    {\includegraphics[width=0.32\textwidth]{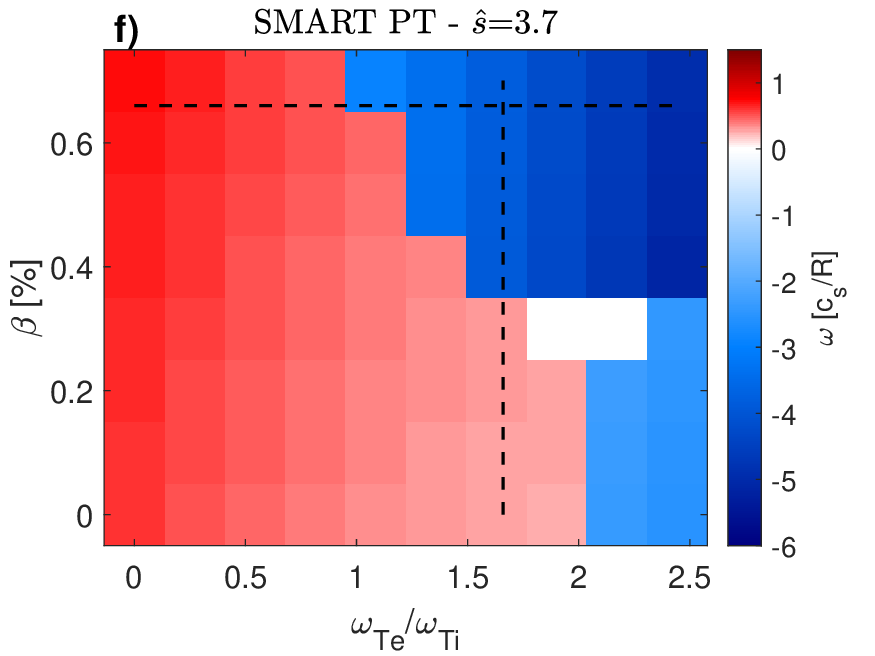}}
    \end{subfigure}
    \caption{Same as figure \ref{TCV_omega}, except for the SMART scenarios. The nominal shear is $\hat{s}=3.7$.}
    \label{SMART_omega}
\end{figure*}

\begin{figure*}
    \centering
    \begin{subfigure}
    {\includegraphics[width=0.32\textwidth]{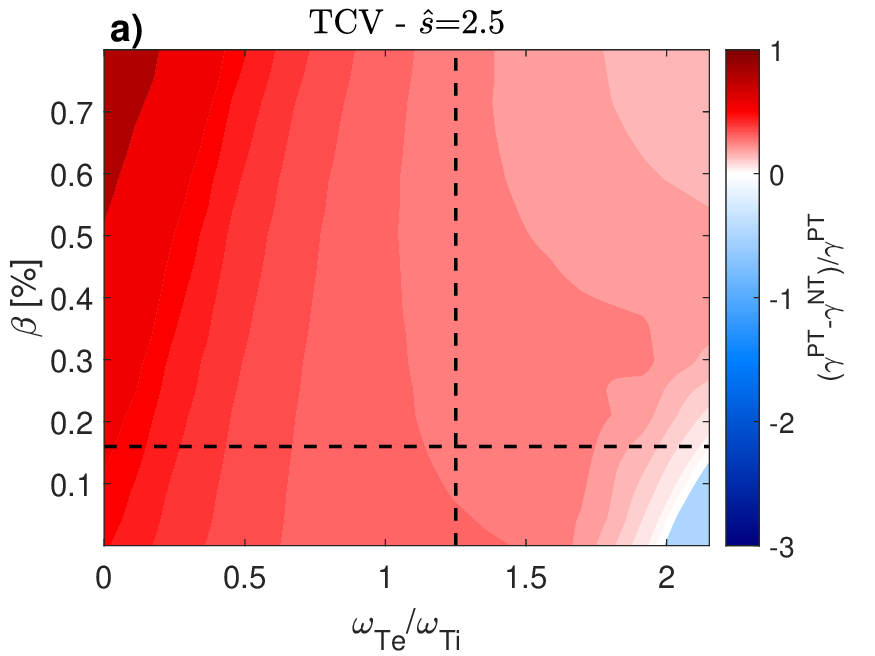}}
    \end{subfigure}
     \begin{subfigure}
    {\includegraphics[width=0.32\textwidth]{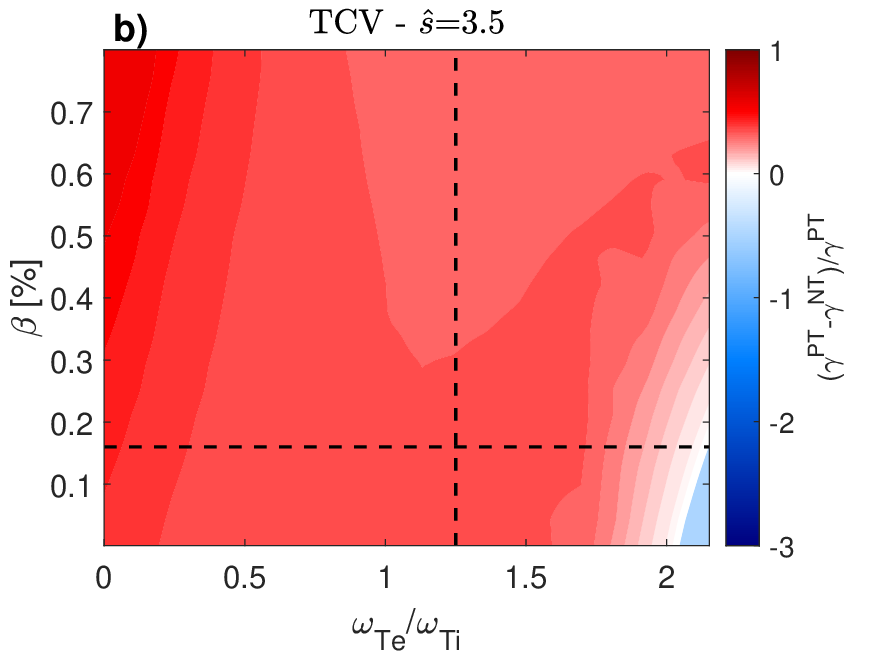}}
    \end{subfigure}
    \begin{subfigure}
    {\includegraphics[width=0.32\textwidth]{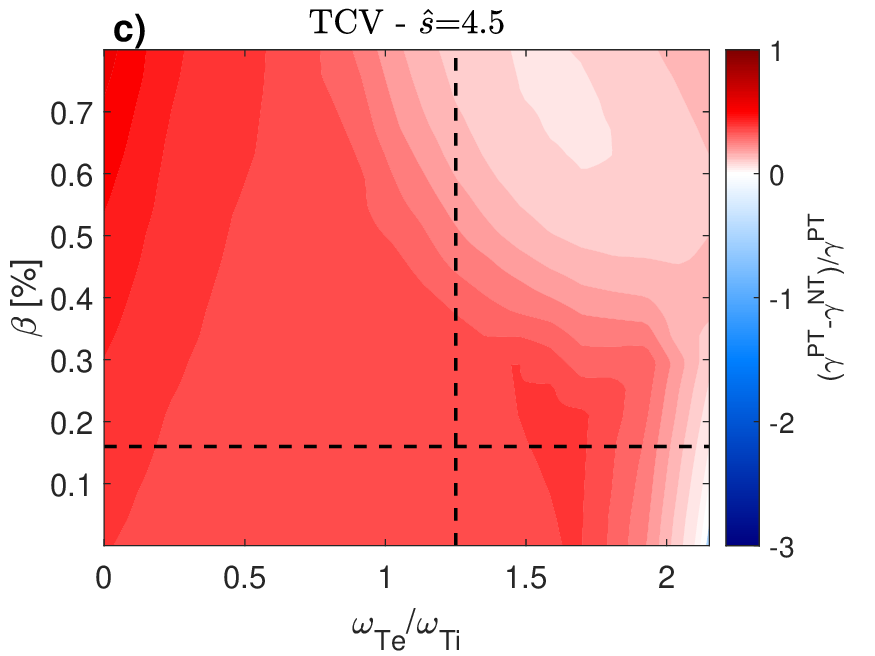}}
    \end{subfigure}
    \begin{subfigure}
    {\includegraphics[width=0.32\textwidth]{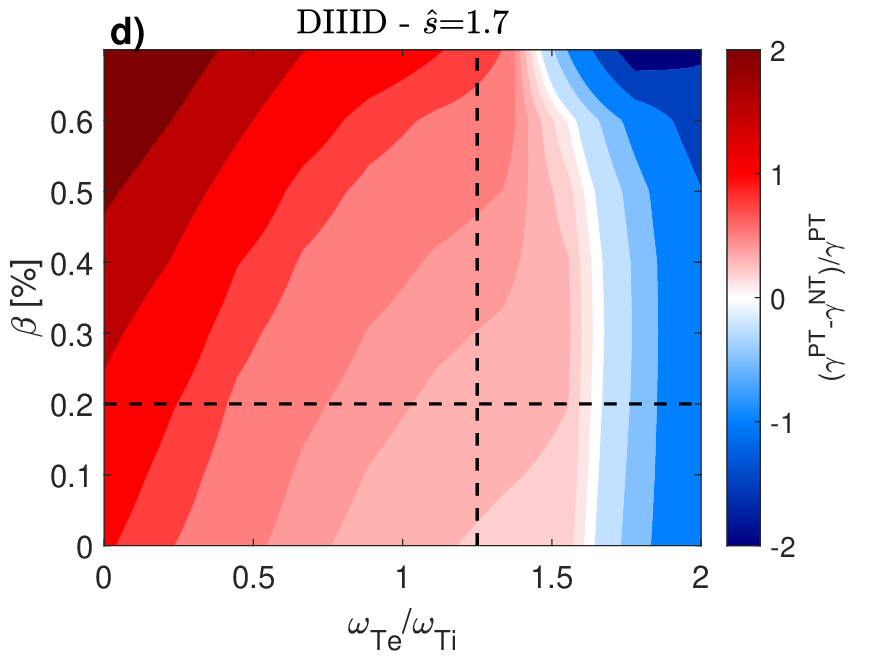}}
    \end{subfigure}
     \begin{subfigure}
    {\includegraphics[width=0.32\textwidth]{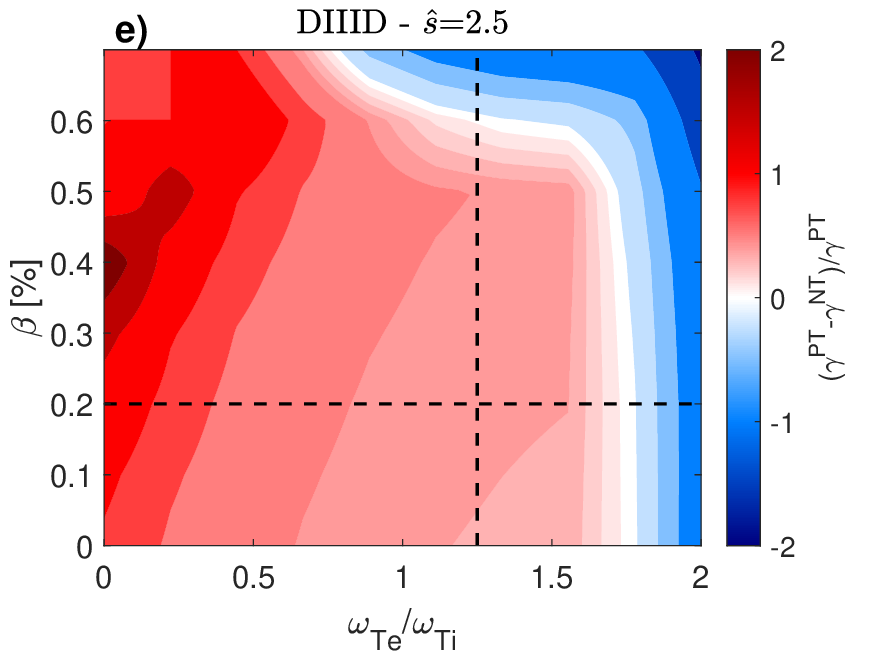}}
    \end{subfigure}
    \begin{subfigure}
    {\includegraphics[width=0.32\textwidth]{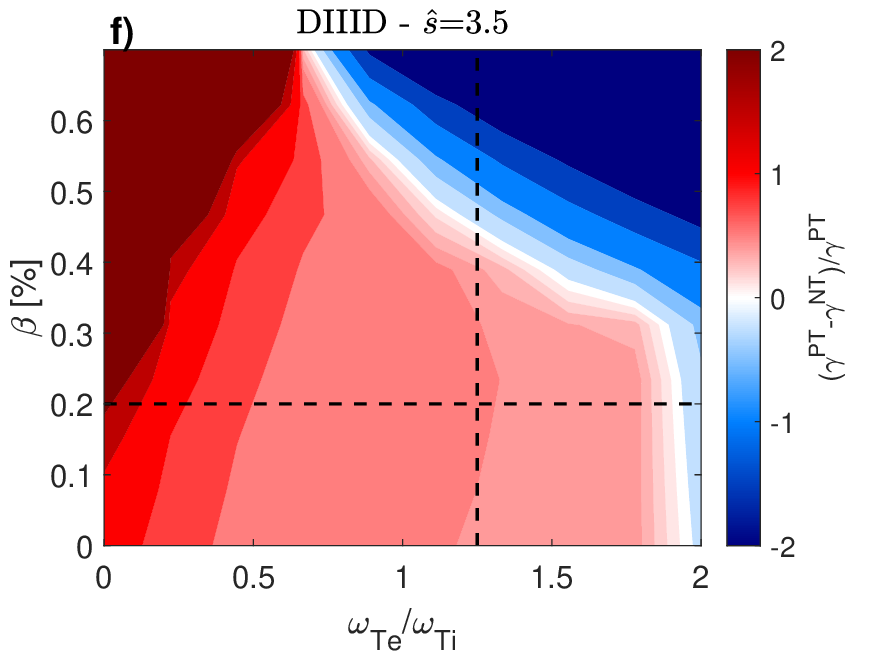}}
    \end{subfigure}
    \begin{subfigure}
    {\includegraphics[width=0.32\textwidth]{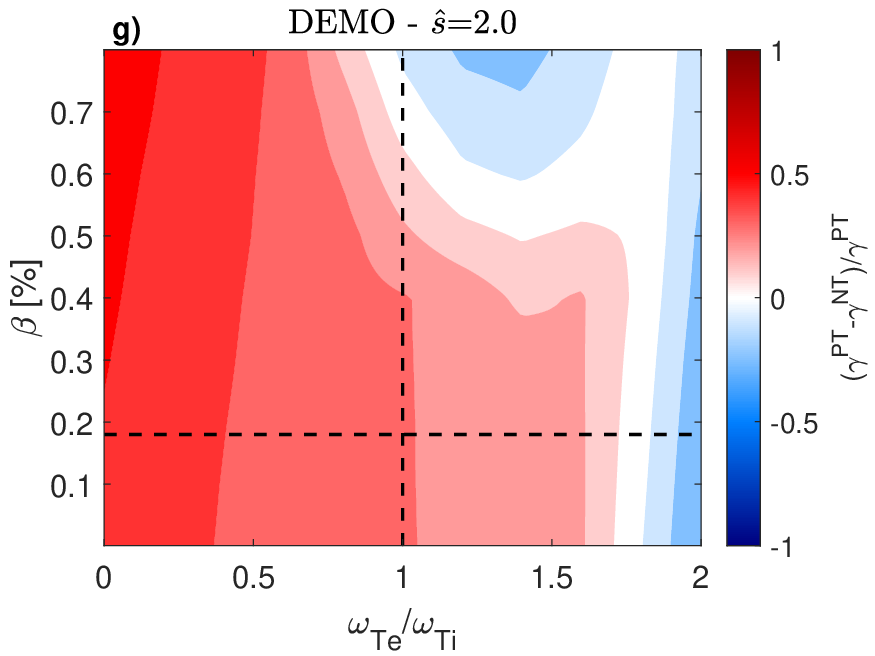}}
    \end{subfigure}
     \begin{subfigure}
    {\includegraphics[width=0.32\textwidth]{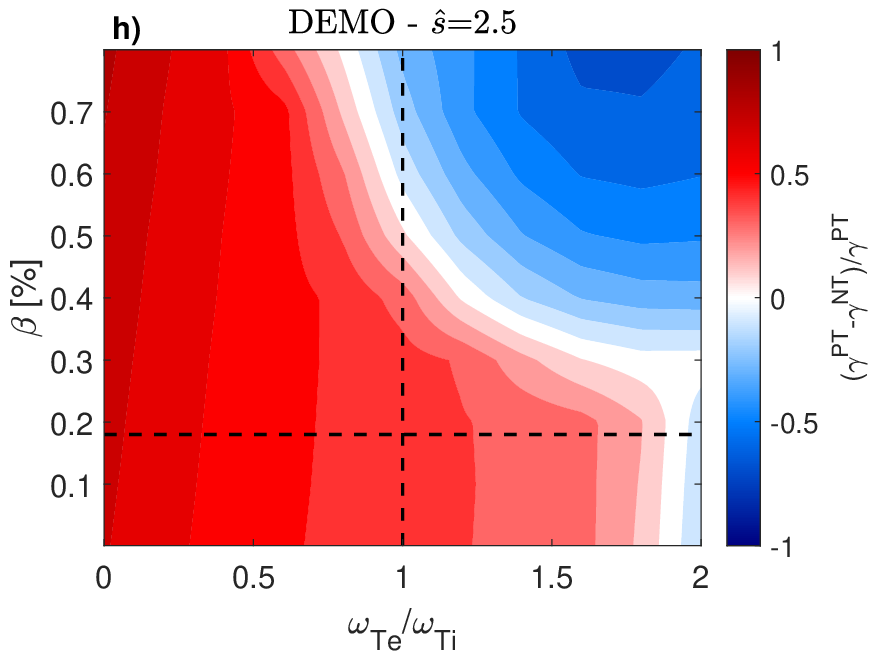}}
    \end{subfigure}
    \begin{subfigure}
    {\includegraphics[width=0.32\textwidth]{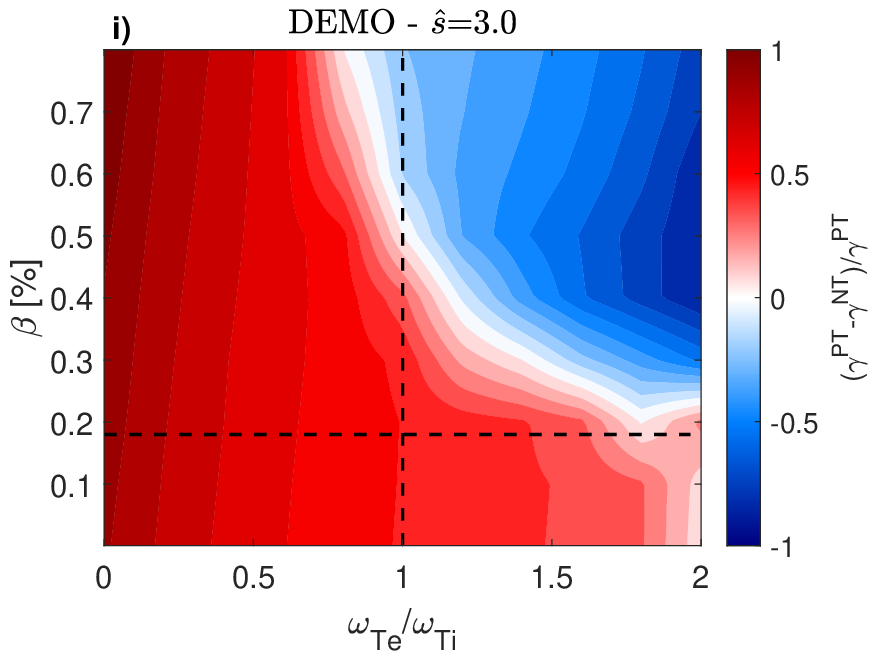}}
    \end{subfigure}
    \begin{subfigure}
    {\includegraphics[width=0.32\textwidth]{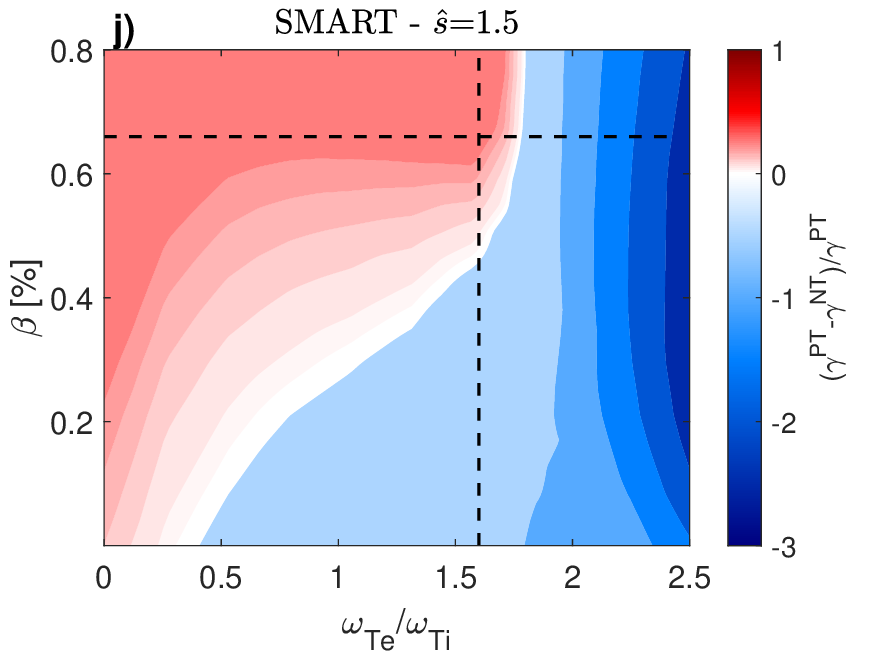}}
    \end{subfigure}
     \begin{subfigure}
    {\includegraphics[width=0.32\textwidth]{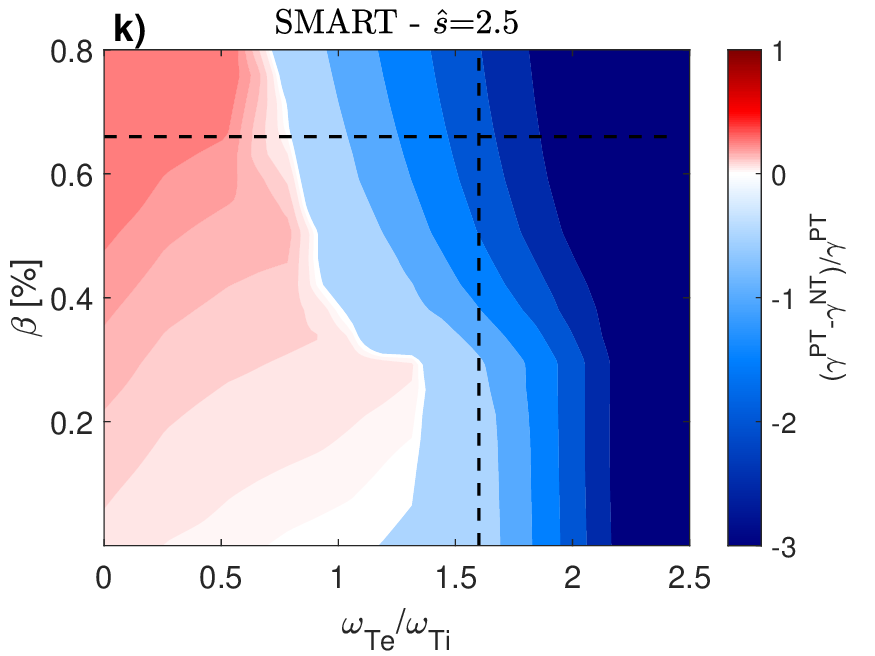}}
    \end{subfigure}
    \begin{subfigure}
    {\includegraphics[width=0.32\textwidth]{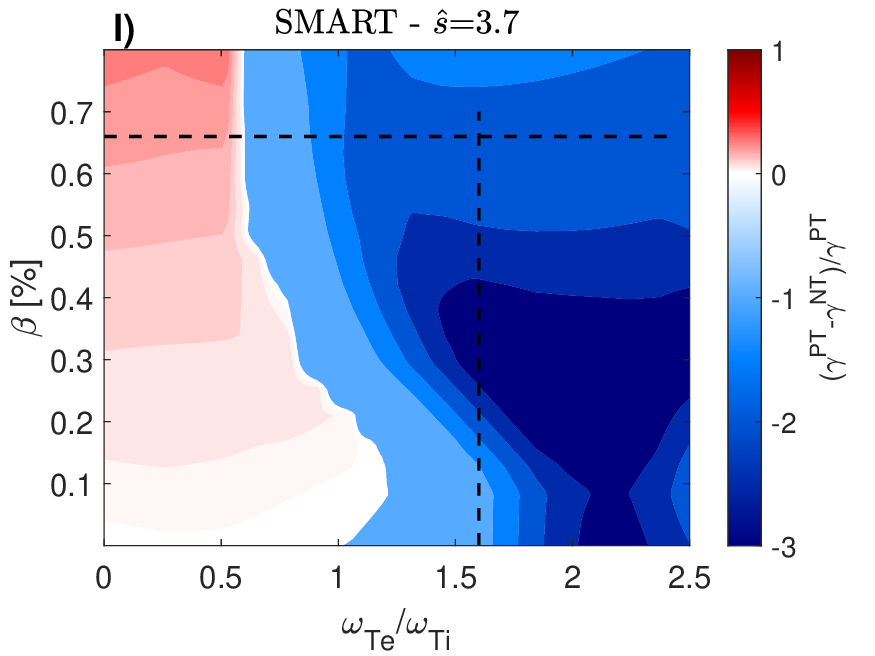}}
    \end{subfigure}
    \caption{Colormaps of the normalized difference of growth rates $(\gamma^{PT}-\gamma^{NT})/\gamma^{PT}$ between NT and PT scenarios as functions of $\beta$ and ratio of electron temperature gradient over ion temperature gradient $\omega_{Te}/\omega_{Ti}$. Different columns correspond to different values of magnetic shear $\hat{s}$, which increases from left to right. Different rows correspond to different machines, which have aspect ratios that decrease from top to bottom. The black dashed lines represent the nominal values of $\omega_{Te}/\omega_{Ti}$ and $\beta$. The nominal magnetic shear $\hat{s}$ is 2.5, 1.7, 2.0, 3.7 for each row respectively}
    \label{confr_gamma}
\end{figure*}

\begin{figure*}
    \centering
    \begin{subfigure}
    {\includegraphics[width=0.32\textwidth]{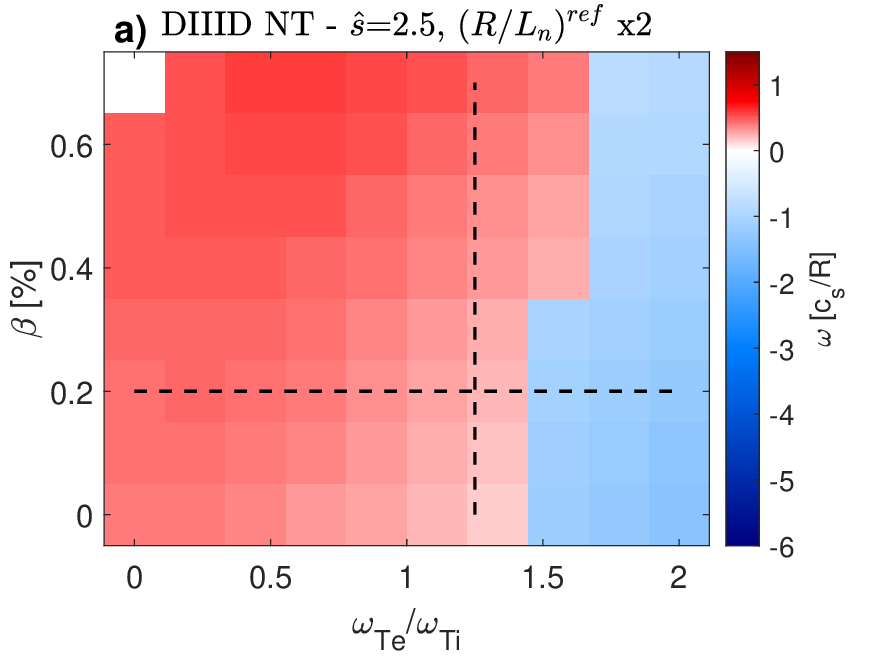}}
    \end{subfigure}
     \begin{subfigure}
    {\includegraphics[width=0.32\textwidth]{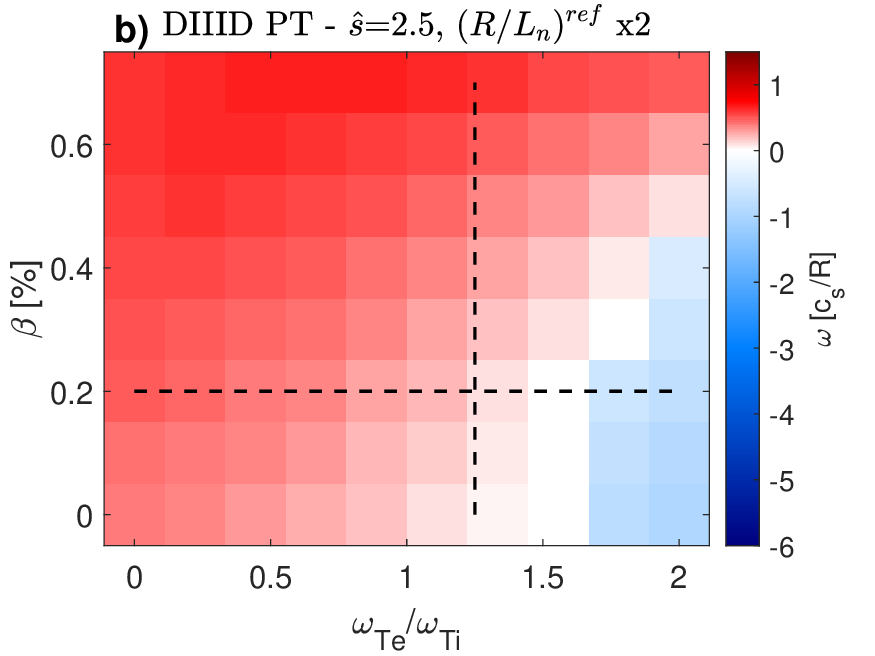}}
    \end{subfigure}
    \begin{subfigure}
    {\includegraphics[width=0.32\textwidth]{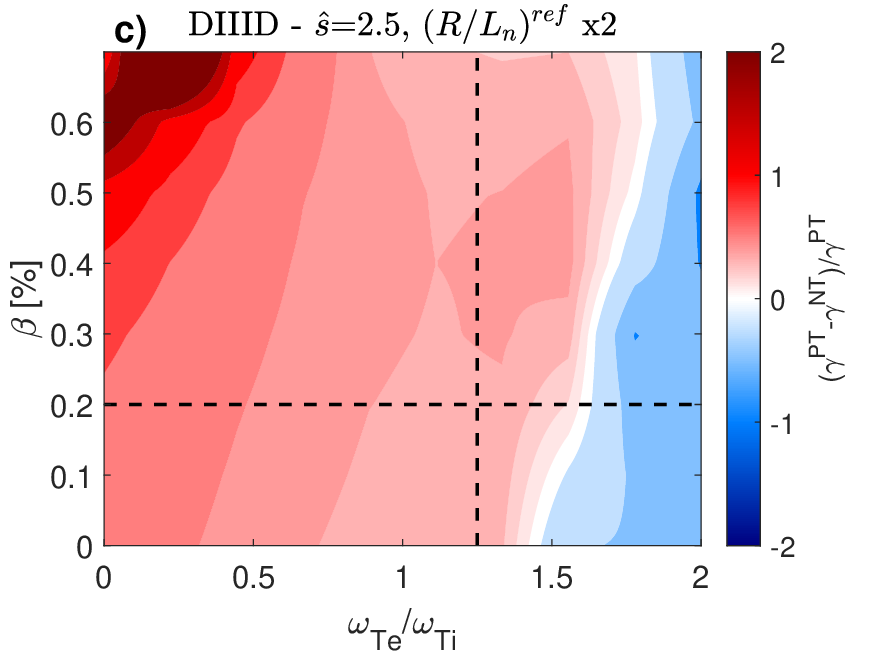}}
    \end{subfigure}
    \begin{subfigure}
    {\includegraphics[width=0.32\textwidth]{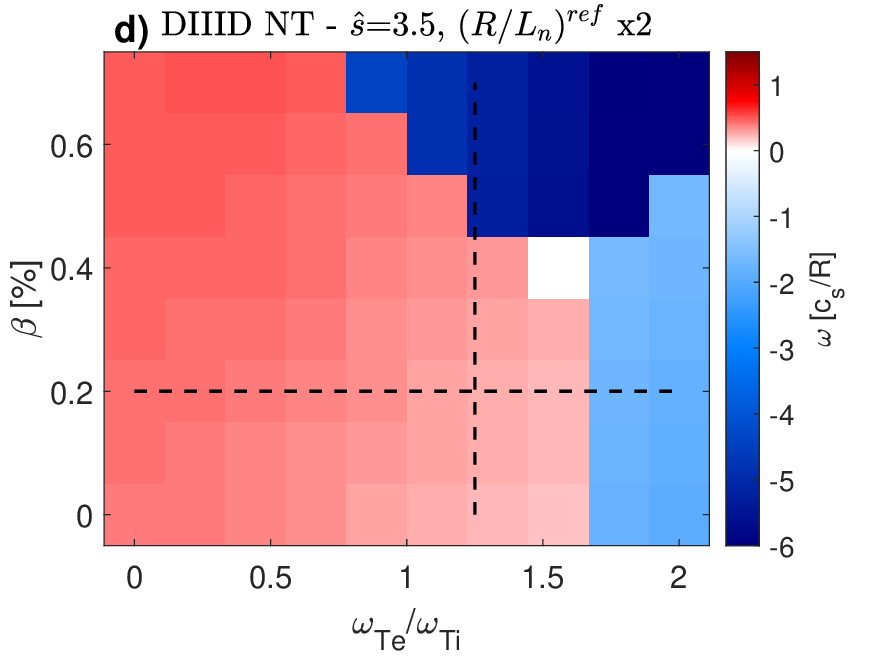}}
    \end{subfigure}
    \begin{subfigure}
    {\includegraphics[width=0.32\textwidth]{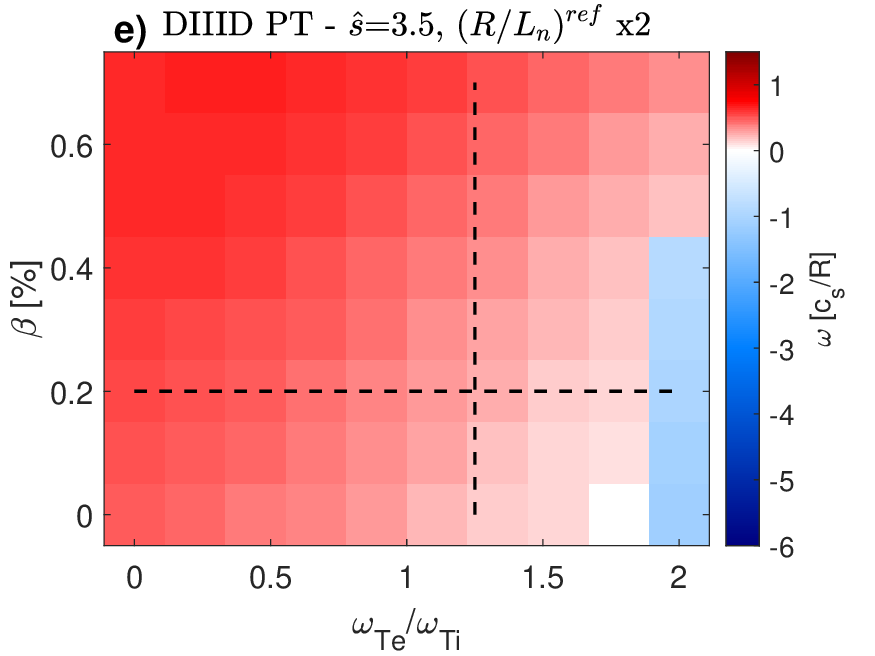}}
    \end{subfigure}
    \begin{subfigure}
    {\includegraphics[width=0.32\textwidth]{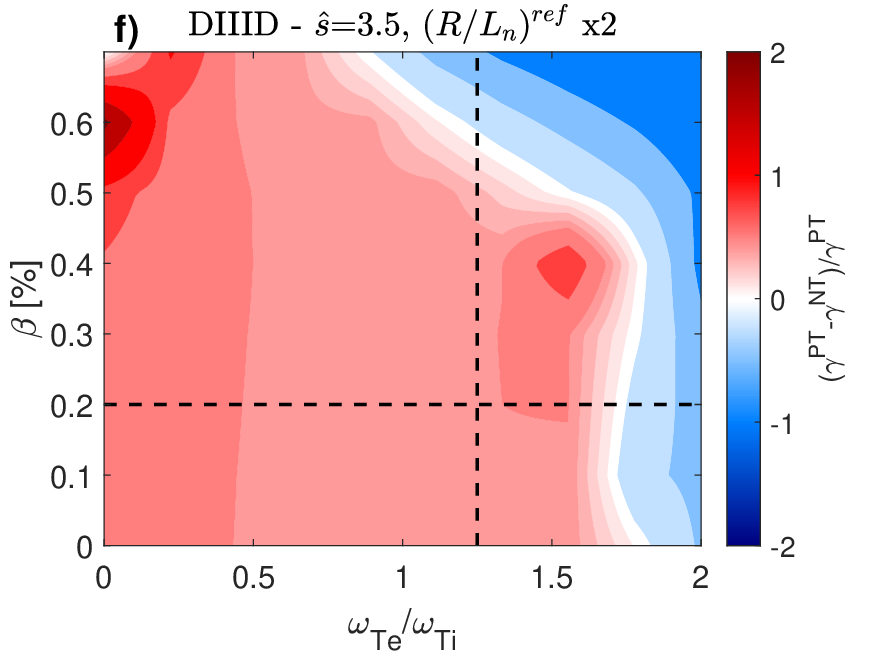}}
    \end{subfigure}
    \caption{Colormaps of the real frequency $\omega$ and the normalized difference of growth rates $(\gamma^{PT}-\gamma^{NT})/\gamma^{PT}$ between NT and PT scenarios as a function of $\beta$ and ratio of electron temperature gradient over ion temperature gradient $\omega_{Te}/\omega_{Ti}$, in the case of an artificially doubled logarithmic density gradient. Different columns correspond to different values of magnetic shear $\hat{s}$, which increases from left to right. The black dashed lines represent the nominal values of $\omega_{Te}/\omega_{Ti}$ and $\beta$.}
    \label{DIIID_dens}
\end{figure*}

To further investigate this behaviour and include the impact of the electron temperature gradient (another key parameter influencing MTM stability) we performed linear simulations where the values of $\beta$ and $\hat{s}$ have been changed alongside with the electron logarithmic temperature gradient $R/L_{Te}$ while keeping $R/L_{Ti}$ fixed. Figures \ref{TCV_omega}, \ref{DIIID_omega}, \ref{DEMO_omegaNT}, \ref{SMART_omega} and \ref{confr_gamma} show the results of these simulations for different scenarios with the mode $k_y\rho_i=0.2$, which is usually one of the modes that contributes the most in nonlinear simulations. In figures \ref{TCV_omega}, \ref{DIIID_omega}, \ref{DEMO_omegaNT} and \ref{SMART_omega} we plot colormaps of real frequency for NT and PT scenarios for TCV, DIII-D, DEMO and SMART respectively. Red color corresponds to positive values of $\omega$ and thus to ITG modes. Light shades of blue are identified as TEM and darker shades as MTMs (indicating their faster negative velocities). This mode separation has been confirmed by a more accurate analysis according to the rules of table \ref{modes_criteria}. Figures \ref{TCV_omega}, \ref{DIIID_omega}, \ref{DEMO_omegaNT} and \ref{SMART_omega} allow us to observe a consistent and striking difference between NT and PT. For NT scenarios, when magnetic shear is sufficiently large, we observe the opening and widening of an MTM-dominated region in the upper right corner of all the plots, i.e. at large $\beta$ and $\omega_{Te}/\omega_{Ti}$. Once more, we observe that this feature applies to all machines, regardless of the value of aspect ratio. In contrast, PT scenarios are much more resilient to MTMs. They either stay in an ITG-dominated regime at all the considered values of $\hat{s}$, $\beta$ and $\omega_{Te}/\omega_{Ti}$ (i.e. TCV and DIII-D) or transition to an MTM-dominated regime at larger values of $\hat{s}$, $\beta$ and $\omega_{Te}/\omega_{Ti}$ with respect to their NT counterparts (i.e. DEMO and SMART). This results in much smaller regions of parameter space where MTMs dominate in PT configurations. Therefore, these simulations show that NT is generally more unstable to MTM modes than PT and we observe that aspect ratio does not influence directly this picture. Nonetheless, we stress that, at nominal values of $\beta$, $\omega_{Te}/\omega_{Ti}$ and $\hat{s}$ only SMART is dominated by MTMs. 

However, we are not only interested in understanding when NT exhibits MTM turbulence. We also want to compare the growth rates of the fastest-growing mode between the NT and PT geometries. In figure \ref{confr_gamma} we show colormaps of the normalized difference between growth rates in PT and NT $(\gamma^{PT}-\gamma^{NT})/\gamma^{PT}$ for all the previous plots. In these plots, negative values (blue) mean that the NT is less stable than PT, while a positive value (red) indicates that NT is more stable. We observe a consistent picture across all the scenarios. The regions where ITG modes dominate are always red, meaning that  NT is always more stable than PT. This is consistent with our findings in \cite{Balestri_2024_2}, where we showed and explained why NT has a beneficial effect on ITG (i.e. faster magnetic drift velocities and stronger Finite Larmor Radius (FLR) effects at the outboard midplane). On the other hand, the regions of parameter space where MTMs dominate are those where we observe a fast degradation of NT stability and larger growth rates in NT than in PT, i.e. the blue regions. This is consistent with the observations made in the previous paragraphs, where we noticed stronger modes in NT when MTMs are the dominant type of instability. Finally, when TEMs dominate (i.e. regions at large $\omega_{Te}/\omega_{Ti}$, low $\beta$ and low $\hat{s}$ for NT scenarios), we observe the NT is more unstable than PT. How triangularity affects TEMs is not the subject of this work, however, we stress that this is consistent with \cite{Balestri_2024_2}, where we showed that NT can have either a stabilizing or destabilizing effect on TEMs depending on many other non-trivial factors.

To summarize, figure \ref{confr_gamma} shows that NT always has a stabilizing affect on ITG compared to PT. The effect on TEMs is not trivial and for the considered cases it is always destabilizing. Finally, when NT is dominated by MTMs, it has higher $\gamma$ than PT. This takes place when $\hat{s}\gtrsim2.5$, $\beta\gtrsim0.3\%$ and $\omega_{Te}/\omega_{Ti}>1$, regardless of aspect ratio. However, this should not be an issue for conventional aspect ratio tokamaks, as $\beta$ is usually much less than $0.3\%$ at these values of $\rho_{tor}$. On the other hand, spherical tokamaks, because of their larger $\beta$, could encounter MTMs and lose the beneficial effect of NT on confinement. In spite of this, it appears possible to avoid MTMs also in STs by keeping the magnetic shear sufficiently low, which may favour single-null operation.

Finally, we will assess the importance of the density gradient to understand if large density gradients can stabilize MTMs. We doubled the logarithmic density gradient of the DIII-D case and show the results in figure \ref{DIIID_dens} for $\hat{s}=2.5$ and $\hat{s}=3.5$. If we compare these plots with the corresponding ones showed in figures \ref{DIIID_omega} and \ref{confr_gamma}(e,f), we observe that MTMs are strongly reduced in the NT configuration. For the scenario with $\hat{s}=2.5$, the region where MTMs dominate completely disappears when $R/L_n$ is doubled. Moreover, the region where NT is less stable than PT is now limited to very large $R/L_{Te}$, i.e. the area where strong temperature-driven TEM dominate. Figure \ref{DIIID_dens}(d) shows similar behaviour for $\hat{s}=3.5$. It has a smaller MTM-dominated region than the case at nominal density gradient. This is also reflected in a smaller region where NT is more unstable than PT. We conclude that larger density gradients can weaken MTM modes and strengthen TEMs, thus creating another path to avoiding MTMs in NT.

\section{Multi-machine nonlinear simulations}\label{5}

Linear simulations suggest that NT geometry is more susceptible to MTMs, as they appear at lower values of $\beta$ and $\hat{s}$ than in PT. In addition, when NT develops MTMs, the growth rates of these modes are stronger than in PT configurations, making NT more linearly unstable than PT. In this section, we investigate whether this picture from linear simulations holds when nonlinear physics is introduced.

\begin{figure}
    \centering
    \includegraphics[width=\linewidth]{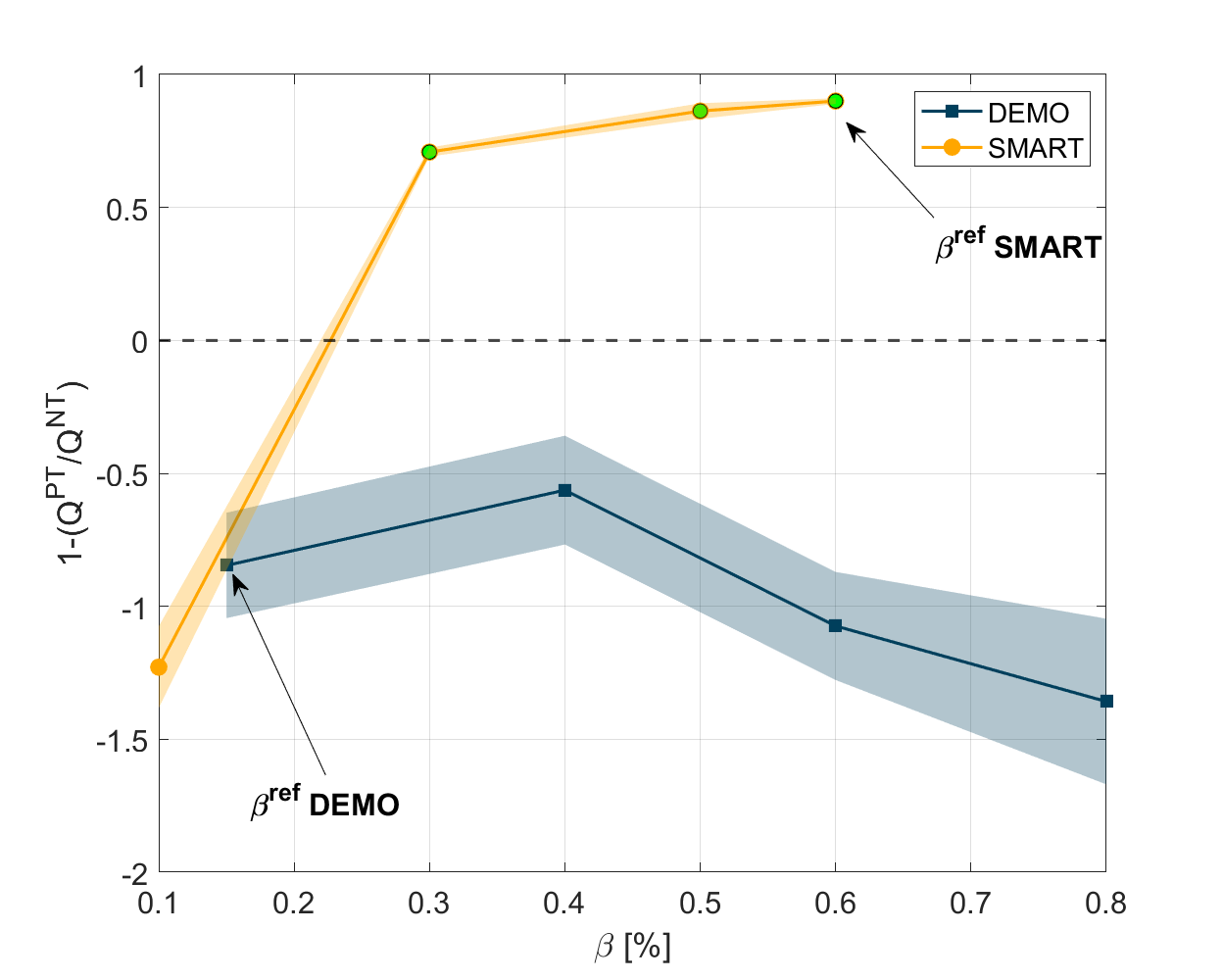}
    \caption{A comparison of the heat flux in PT ($Q^{PT})$ with that of the NT ($Q^{NT}$) as a function of $\beta$ for SMART (yellow circles) and DEMO (blue squares) with reference magnetic shear. The shaded areas represent standard deviations calculated from the time traces of the heat fluxes. Simulations where MTMs dominate are highlighted by green markers}
    \label{Q_vs_beta_DEMO_SMART}
\end{figure}

\begin{figure*}
    \centering
    \includegraphics[width=0.98\textwidth]{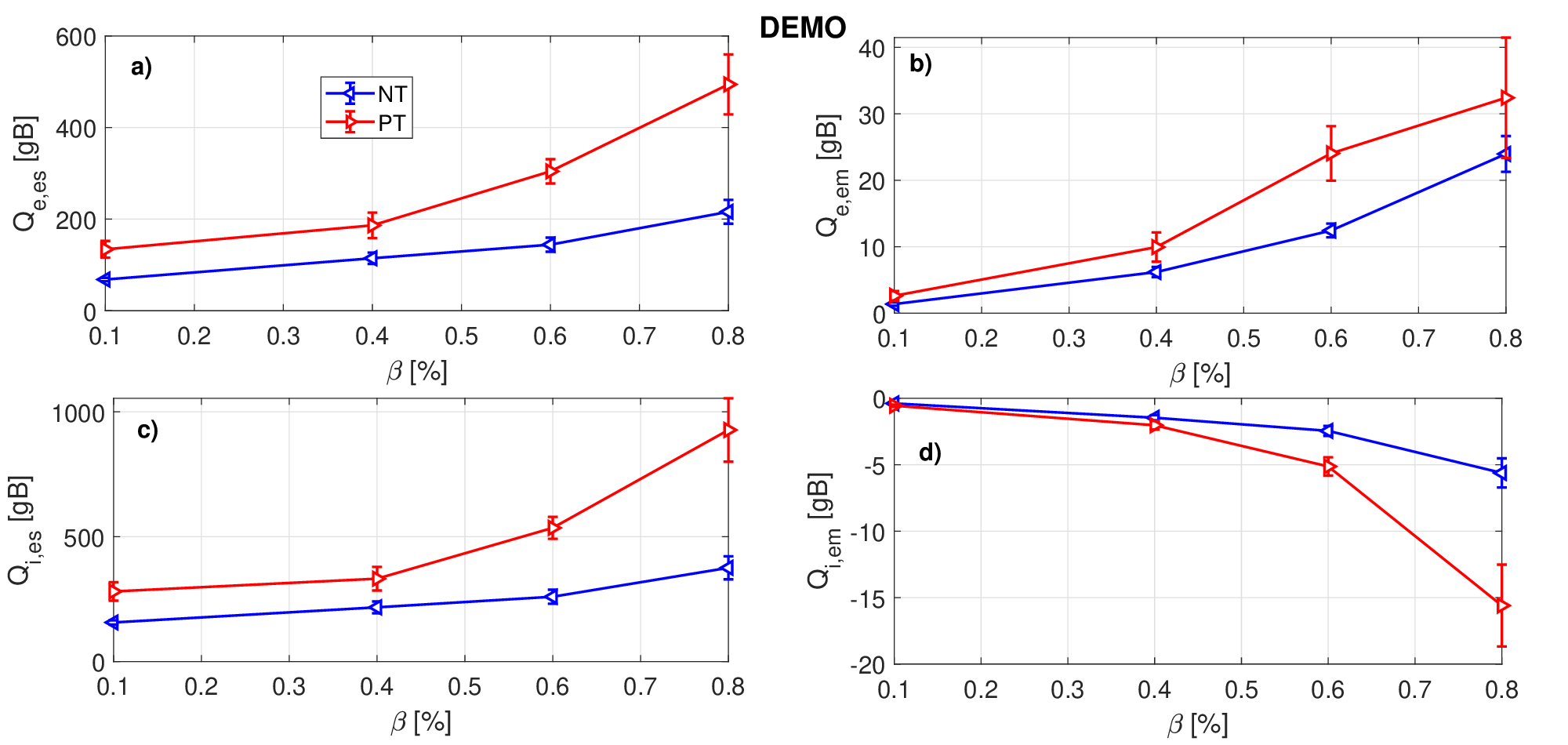}
    \caption{Components of the total heat flux in the NT (blue) and PT (red) DEMO geometries as functions of $\beta$.}
    \label{Q_vs_beta_DEMO}
\end{figure*}

\begin{figure*}
    \centering
    \includegraphics[width=0.94\textwidth]{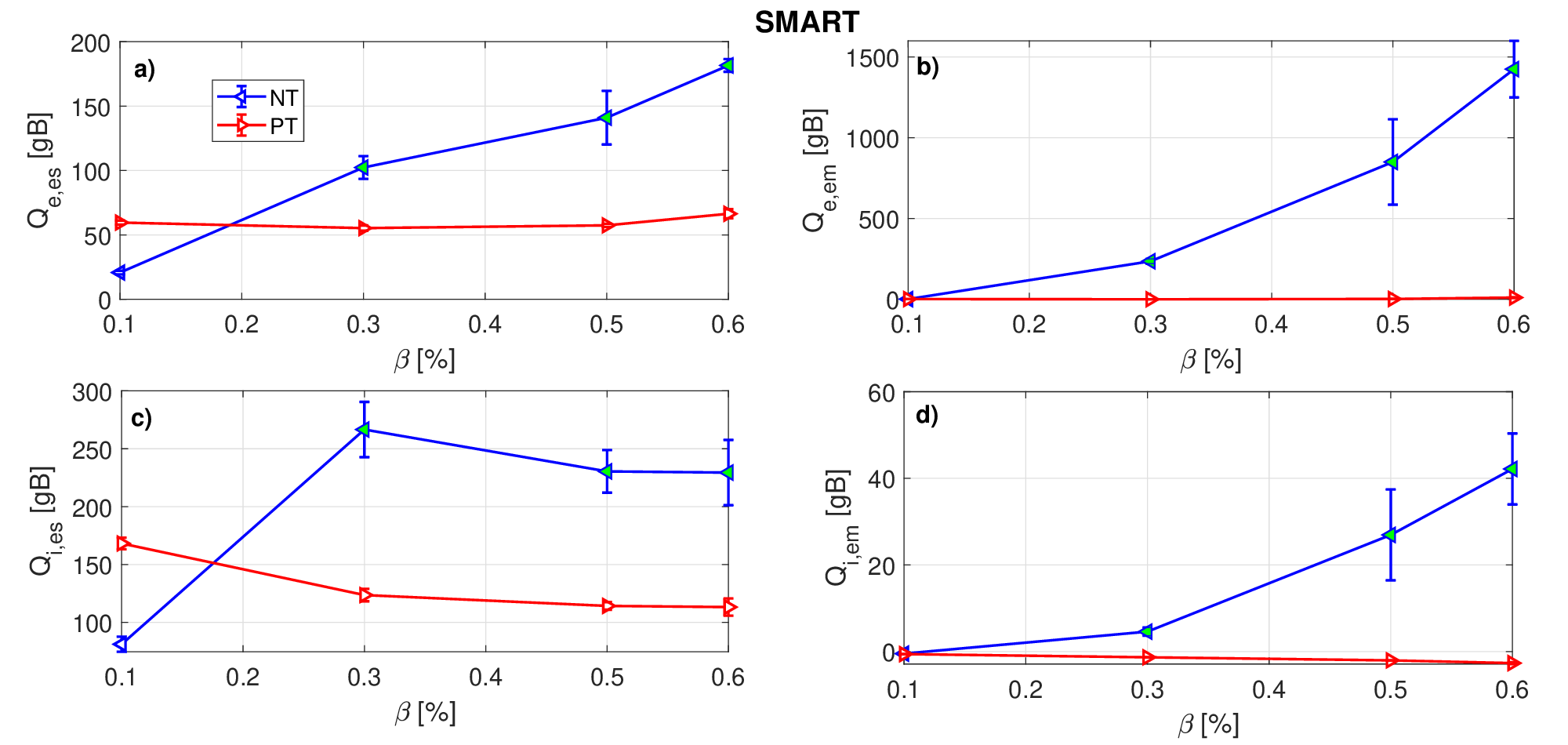}
    \caption{Components of the total heat flux in NT (blue) and PT (red) SMART geometries as functions of $\beta$. Simulations where MTMs dominate are highlighted by green markers.}
    \label{Q_vs_beta_SMART}
\end{figure*}

As mentioned before, nonlinear simulations involving strong MTM turbulence are very numerically challenging. These simulations require a very large box in the radial direction and very high resolutions to properly catch and resolve the evolution of the thin current layers. Moreover, including other species or other kinds of turbulence can trigger multi-mode turbulence interaction, making the simulations even heavier. Often simulations fail to converge to a quasi-stationary state. For these reasons, we performed nonlinear scans in $\beta$ only for the reference DEMO and SMART scenarios considered in the previous section. The nonlinear simulations for DEMO have been performed at reference parameters and only $\beta$ has been varied. The parameter $\alpha\equiv-q_0^2R_0d\beta/dr$ has been changed consistently. For SMART, all the nonlinear simulations were performed with a fixed ratio $\omega_{Te}/\omega_{Ti}=1$, to reduce the drive of MTMs and reach convergence more easily. All the simulations for SMART and DEMO were performed with $[n_{x},n_{y},n_z,n_{v_\parallel},n_\mu]=[384,64,48,40,10]$, $L_x\rho_i\sim250$ and $k_{y,min}\rho_i=0.03$. A convergence study in $n_x$ and $n_{y}$ was performed, showing that increasing the resolution in $x$ and $y$ affects the results by $10\%$. Moreover, all the nonlinear simulations were performed imposing a typical external $E\times B$ flow shear rate in GENE units, i.e. for SMART we used $\gamma_{ExB}=0.1\;c_s/R$ and $\gamma_{ExB}=0.05\;c_s/R$ for DEMO. These are reasonable values in experiment and weaken multi-mode interaction between MTM and ITG. Finally, we observed that turning off the Arakawa scheme \cite{ARAKAWA1966119} for the $z$ and $v_\parallel$ directions helped reaching convergence in simulations where MTMs dominate transport.

The results of the nonlinear $\beta$ scan are shown in figure \ref{Q_vs_beta_DEMO_SMART}. Here we display the normalized ratio of the total heat fluxes in PT over NT, i.e. $1-Q^{PT}/Q^{NT}$. A value larger than zero means that NT has worse energy confinement than PT, while negative values mean NT has better confinement than PT. These nonlinear simulations agree with the picture from linear simulations. According to the linear DEMO simulations (figures \ref{DEMO_omegaNT} and \ref{confr_gamma}(g)), we would expect NT to be more stable than PT at all values of $\beta$ (with the only exception of $\beta=0.8\%$, where NT and PT had very similar growth rates). This is consistent with what we see with nonlinear simulations, where NT has heat fluxes which are between 1.5 and 2 times lower than in PT. On the contrary, for SMART (with $\omega_{Te}/\omega_{Ti}=1$) figures \ref{SMART_omega} and \ref{confr_gamma}(l) indicate the NT scenario should be dominated by MTMs and more unstable than PT when $\beta\gtrsim0.3\%$. Once again, this is consistent with nonlinear simulations. NT has a heat flux which is 2 times lower than PT for $\beta=0.1\%$, but has heat fluxes that are 1.5 times larger than PT when $\beta\gtrsim0.3\%$

To understand which type of turbulence dominates the simulations, we can look at the various components of the heat fluxes, i.e. the electrostatic electron heat flux $Q^{es}_{e}$, electrostatic ion heat flux $Q^{es}_{i}$, electromagnetic electron heat flux $Q^{em}_{e}$ and electromagnetic ion heat flux $Q^{em}_{i}$. The results for DEMO are shown in figure \ref{Q_vs_beta_DEMO}. We see that the electrostatic components of the heat fluxes dominate over the electromagnetic ones for all values of $\beta$ for both NT and PT. Additionally, the ion channel is larger than the electron one. This implies that ITG is dominant. This appears to differ slightly from the linear simulations for the $k_y\rho_i=0.2$ mode shown in figure \ref{DEMO_omegaNT}, where we observed MTMs dominating in NT for $\beta\gtrsim0.6\%$. However, linear $k_y$ scans of NT DEMO with $\beta\gtrsim0.6\%$ (not shown here), show that modes with $k_y\rho_i>0.2$ are not MTMs, but rather ITG. This implies that, in this regime, MTMs are not strong enough linearly to dominate nonlinear transport and remain weak. The results from nonlinear simulations of SMART are entirely consistent with the linear simulations showed in the previous section (figures \ref{SMART_omega} and \ref{confr_gamma}(l)). Indeed, for $\omega_{Te}/\omega_{Ti}=1$, we would expect the NT scenario to be dominated by MTMs and more unstable than PT for $\beta\gtrsim0.3\%$. PT should be dominated by ITG for the full range of $\beta$ values. Indeed, these predictions are confirmed by the nonlinear simulations shown in figure \ref{Q_vs_beta_SMART}, where we observe that NT has larger heat fluxes than PT for $\beta\gtrsim0.3\%$. This behaviour is due to MTMs dominating transport, which can be seen from the steep increase in the electromagnetic component of the electron heat flux that occurs between $\beta=0.1\%$ and $\beta=0.3\%$. On the contrary, we see that the heat flux components in PT do not change much as $\beta$ is varied, indicating that PT is always dominated by ITG turbulence. Regarding the behaviour of NT SMART at $\beta=0.1\%$, we see that NT is more stable than PT and $Q_{e,em}\simeq0$, thus implying that ITG is dominant. This allows us to conclude that spherical tokamaks can also benefit from the reduction of turbulent transport by NT if the regime is dominated by ITG, as already observed in \cite{Balestri_2024_2}. We also note that, if converted to physical units (i.e. MW), we find the heating power that is necessary to sustain a scenario like NT SMART with $\beta\gtrsim0.4\%$ would be unrealistically high. Since we are performing gradient-driven simulations, this indicates that the gradients and kinetic profiles we are using would not be reached in a real experiment. Therefore, these simulations are telling us that, because of the strength of MTMs in NT, the actual performance of an NT scenario similar to the SMART one will be limited to a much lower $\beta$. On the contrary, for the DEMO scenario investigated here, NT is not affected by MTMs and large values of $\beta$ can be reached with a heating power that is lower than in the PT case.

\begin{figure}
    \centering
    \begin{subfigure}
    {\includegraphics[width=\linewidth]{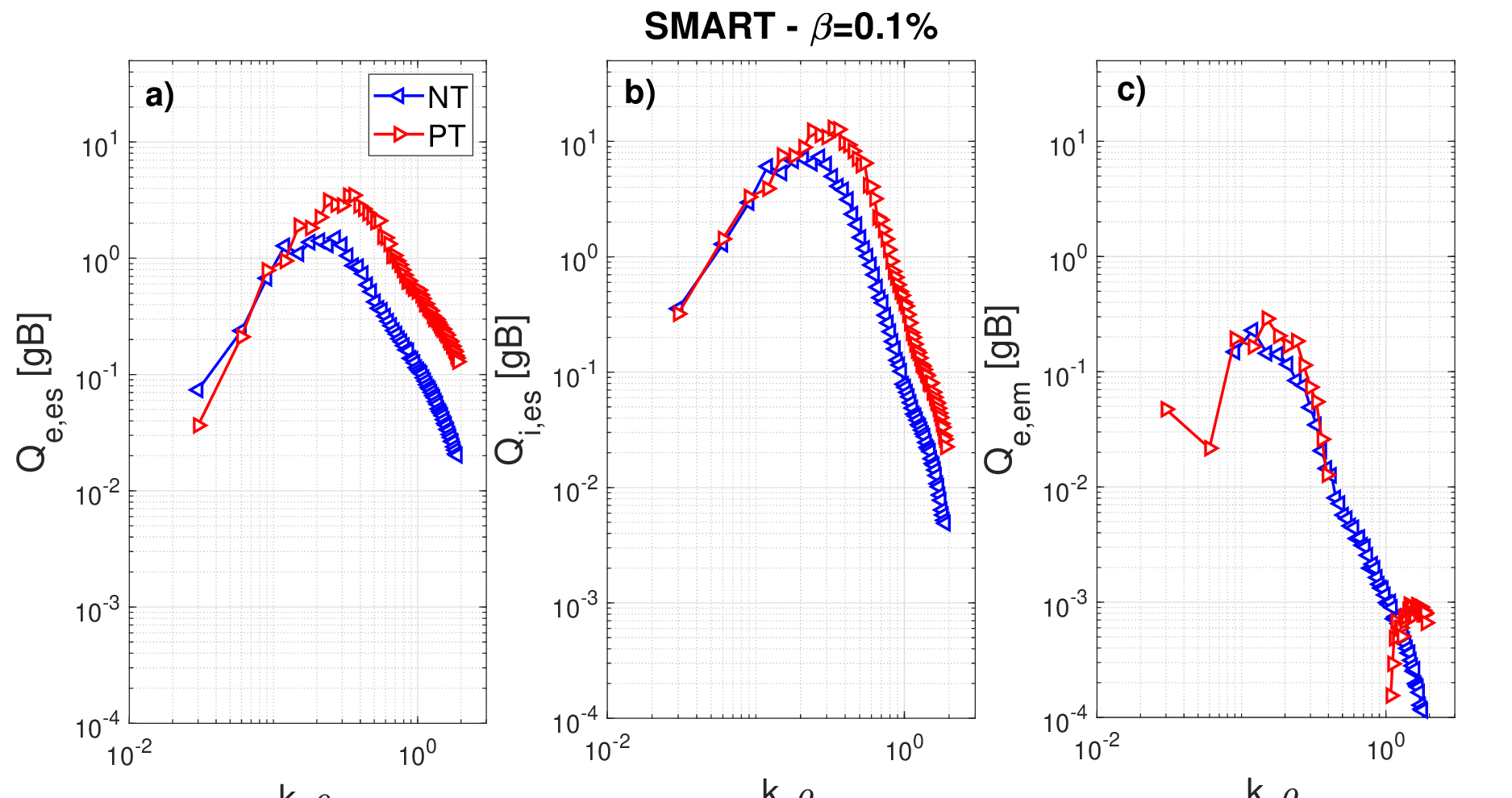}}
    \end{subfigure}
     \begin{subfigure}
    {\includegraphics[width=\linewidth]{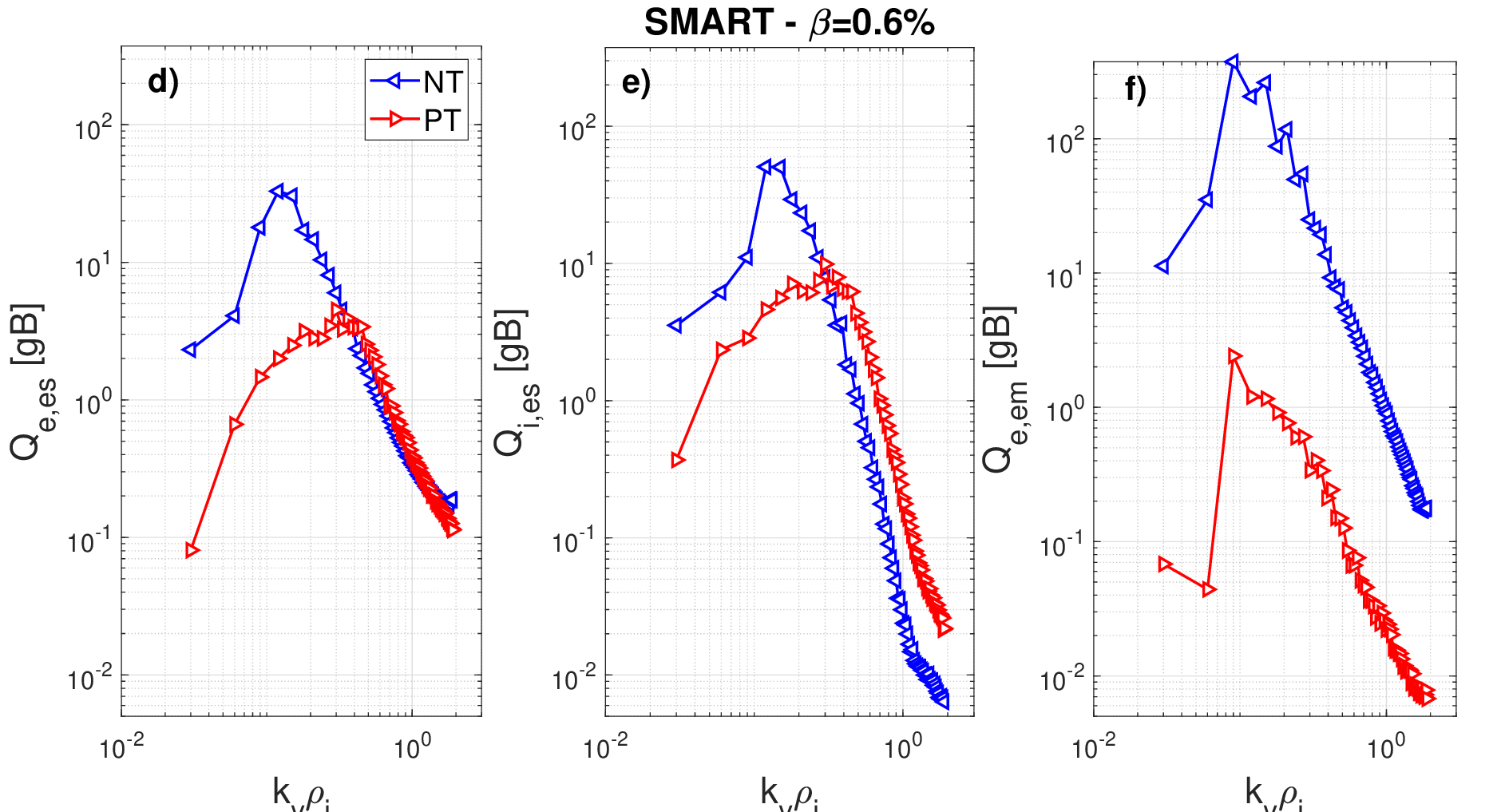}}
    \end{subfigure}
    \caption{Heat flux spectra of the components of heat flux $Q_{e,es}$, $Q_{i,es}$ and $Q_{e,em}$ for two different values of $\beta$ for the SMART scenarios. Results in blue are for NT and red for PT.}
    \label{Qspectra_SMART}
\end{figure}

\begin{figure}
    \centering
    \includegraphics[width=\linewidth]{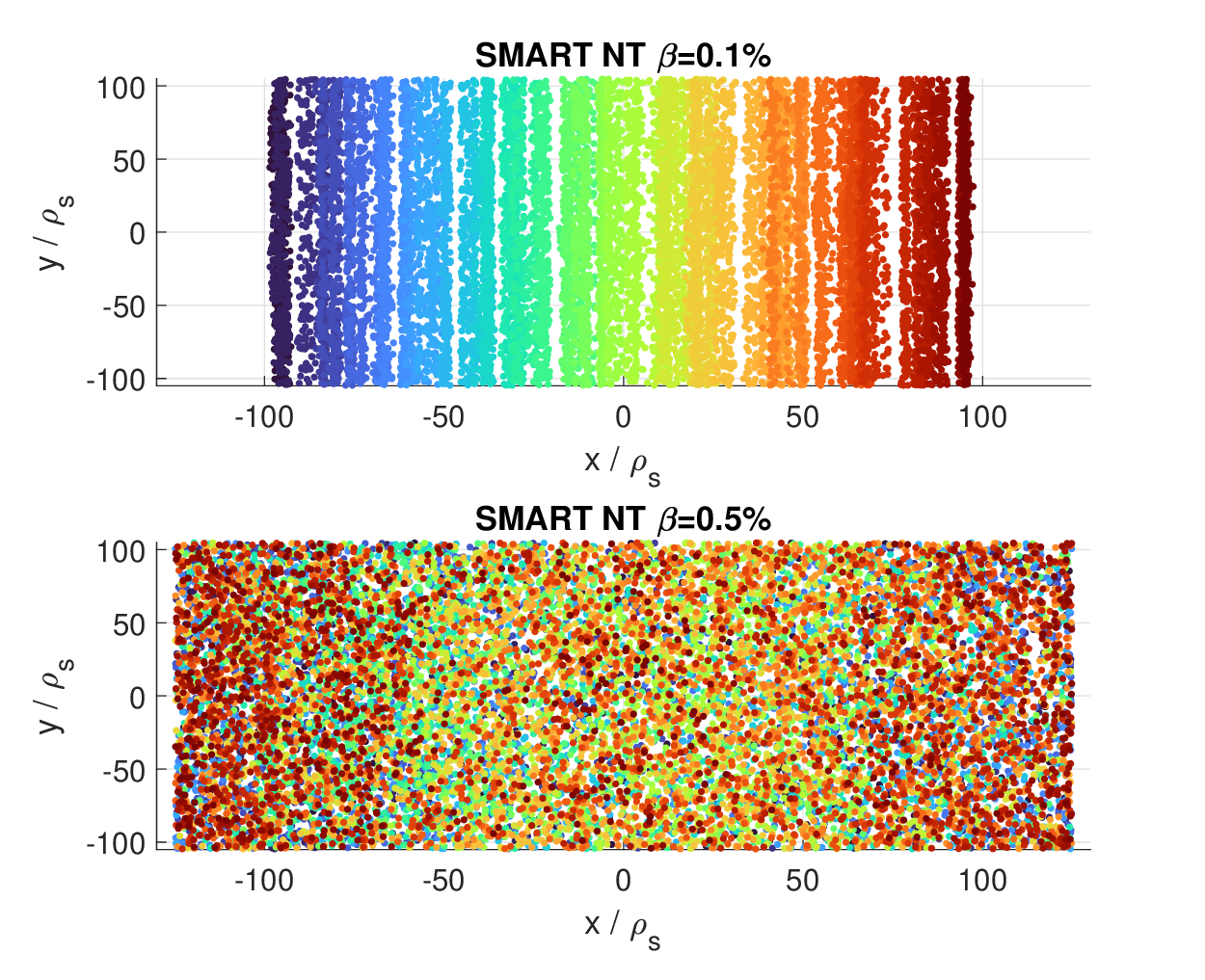}
    \caption{Poincar\'e section at the outboard midplane for NT SMART simulations with $\beta=0.1\%$ (top) and $\beta=0.5\%$ (bottom). Different colors correspond to different magnetic field lines (50 in total) seeded at $y=0$ and traced for 300 poloidal turns.}
    \label{SMART_poincare}
\end{figure}

With nonlinear simulations we can also study the heat flux spectra, i.e. the heat flux decomposed as a function of $k_y\rho_i$, in the two configurations at different values of $\beta$ and in different turbulent regimes. In figure \ref{Qspectra_SMART}, we show the heat flux spectra of $Q^{es}_{e}$, $Q^{es}_{i}$ and $Q^{em}_{e}$ for NT and PT SMART cases with $\beta=0.1\%$ and $\beta=0.6\%$. This comparison is useful because the turbulent regime changes from ITG to MTM when $\beta$ is increased. For $\beta=0.1\%$, we see that NT has a lower heat flux in each component than PT, while the spectra in both configurations peak around $k_y\rho_i=0.3$, a typical value for electrostatic turbulence. On the other hand, when $\beta=0.6\%$ MTMs dominate in NT and we see the spectra peak around $k_y\rho_i=0.1$. This was expected, given that MTMs tend to be stronger when their wavelength is larger \cite{2011_Doerk,Doerk_2012} (i.e. lower wavenumber).

Another way to look at transport due to MTMs is through magnetic field line stochasticity. As mentioned in section \ref{2}, MTMs can break magnetic field lines and form gyroradius-scale magnetic islands. Broken magnetic field lines are no longer forced to stay on a flux surface and can move radially, thus increasing electron transport. In nonlinear simulations, we can follow the magnetic field lines over many poloidal turns and see where they have moved  each time they cross the outboard midplane, i.e. $z=0$. We show this in figure \ref{SMART_poincare} in the form of Poincarè plots for the last timestep of the nonlinear SMART simulations with $\beta=0.1\%$ and $\beta=0.5\%$. We chose these two cases because electrostatic turbulence dominates in the former, while in the latter MTMs dominate. In figure \ref{SMART_poincare}, different colors represent different magnetic field lines when they cross the outboard midplane after each poloidal turn. We can see a striking difference between the two cases: when ITG dominates transport, we can clearly distinguish each flux surface, because magnetic field lines are confined to a flux surface and colors are well separated. A completely different picture emerges from the $\beta=0.5\%$ case, where we observe a mix of magnetic field lines and individual flux surfaces are no longer distinguishable. This is indeed the result of strong MTM transport, which breaks flux surfaces and leads to high radial transport, as observed in figure \ref{Q_vs_beta_SMART}.

\begin{figure}
    \centering
    \includegraphics[width=\linewidth]{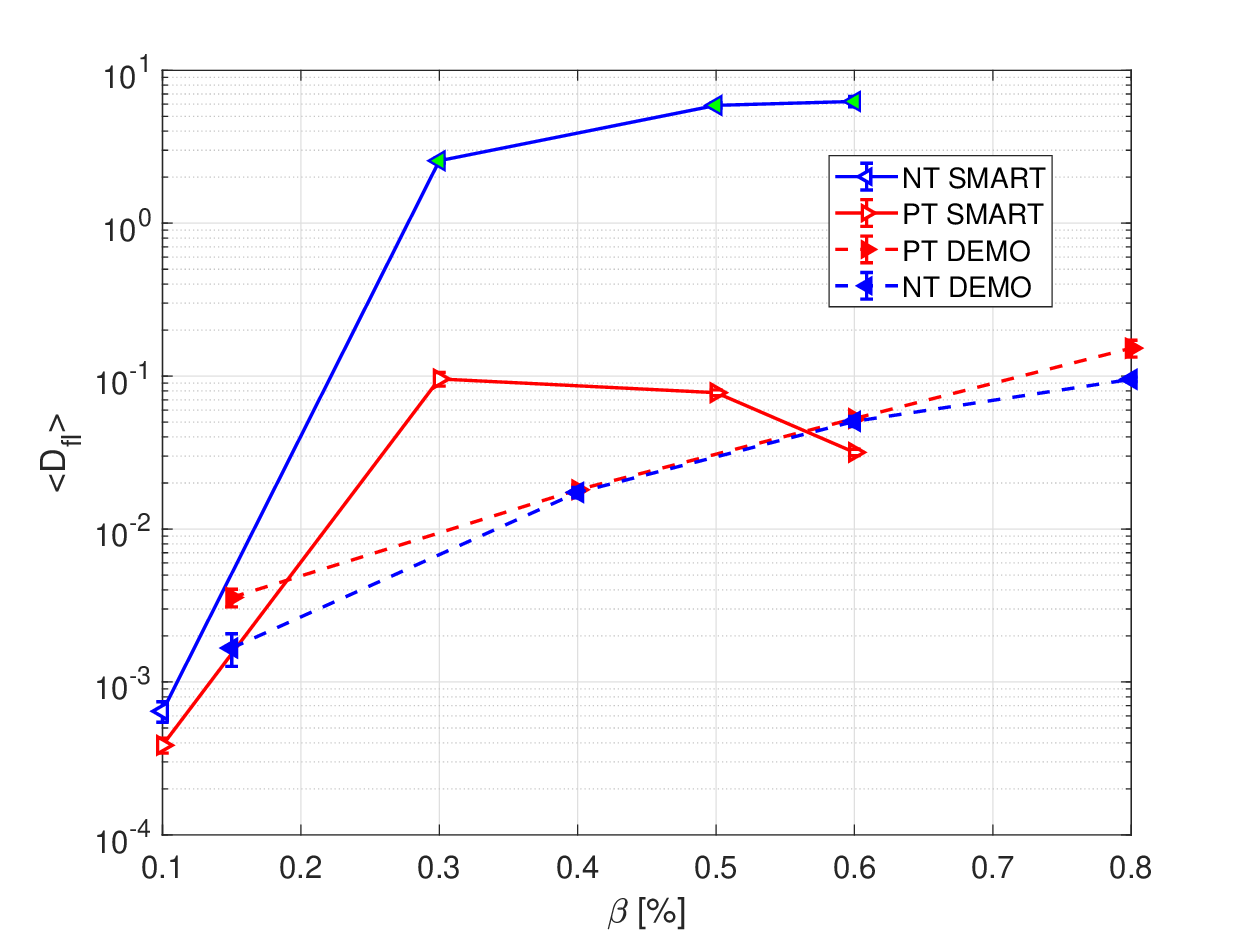}
    \caption{Average field line diffusivity as a function of $\beta$ from NT SMART (empty blue triangles), PT SMART (empty red triangles), NT DEMO (full blue triangles) and PT DEMO (full red triangles) nonlinear simulations. Simulations where MTMs dominate are highlighted by green markers.}
    \label{SMART_DEMO_diffusivity}
\end{figure}

We can better quantify magnetic field line diffusion by using the magnetic field lines diffusion coefficient $D_{fl}$ as defined in \cite{Pueschel_2013}
\begin{equation}
    D_{fl}(l,p)=\frac{\left[r(l,p)-r(l,0)\right]^2}{2\pi q_0R(p+1)},
\end{equation}
where $l$ identifies a magnetic field line, $p$ is the number of poloidal turns, $r(l,p)$ is the radial position at the outboard midplane of the magnetic field line $l$ after $p$ poloidal turns, $r(l,0)$ is initial position at the outboard midplane of magnetic field line $l$ and $q_0$ is the safety factor. One can then define an average diffusion coefficient $\langle D_{fl}\rangle$, where the average is performed over field lines and poloidal turns. In figure \ref{SMART_DEMO_diffusivity}, we show the average diffusion coefficient as a function of $\beta$ for PT and NT SMART and DEMO simulations. The magnetic diffusion coefficient is computed over 200 magnetic field lines for 300 poloidal turns. The diffusion coefficient is plotted in log scale.

From figure \ref{SMART_DEMO_diffusivity} we observe that magnetic diffusion correlates with the levels of electromagnetic transport. For DEMO simulations (full triangles), we see that $\langle D_{fl}\rangle$ is similar between NT and PT and stays well below unity for the whole range of $\beta$. This confirms that transport is essentially electrostatic and magnetic field lines are well confined on flux surfaces. On the other hand, for NT SMART simulations (empty blue triangles) we observe that the diffusion coefficient sharply increases for $\beta\gtrsim0.3$, reaching values that are  more than 2 orders of magnitude larger than in electrostatic simulations. This happens when MTMs dominate transport. By contrast, PT SMART simulations (empty red triangles) show levels of magnetic field line diffusion much lower than the NT counterpart, thus confirming that the PT scenario is dominated by electrostatic turbulence and MTMs are strongly subdominant.

\section{Physical picture of MTM}\label{6}

In the previous section, we observed that triangularity plays a crucial role in changing the stability of MTMs, with negative values of $\delta$ making these modes stronger compared to positive values of triangularity. Similarly to the work carried out in \cite{Balestri_2024_2} on ITG and TEM turbulence, we will investigate the role played by magnetic drifts and FLR effects in the destabilization of MTMs and how this relates to the effect of triangularity. As was done in \cite{Balestri_2024_2}, we consider the large aspect ratio limit $A\gg1$ because it simplifies th way geometry enters the gyrokinetic equation to only magnetic drifts and FLR effects.

\begin{figure*}
    \centering
    \includegraphics[width=\linewidth]{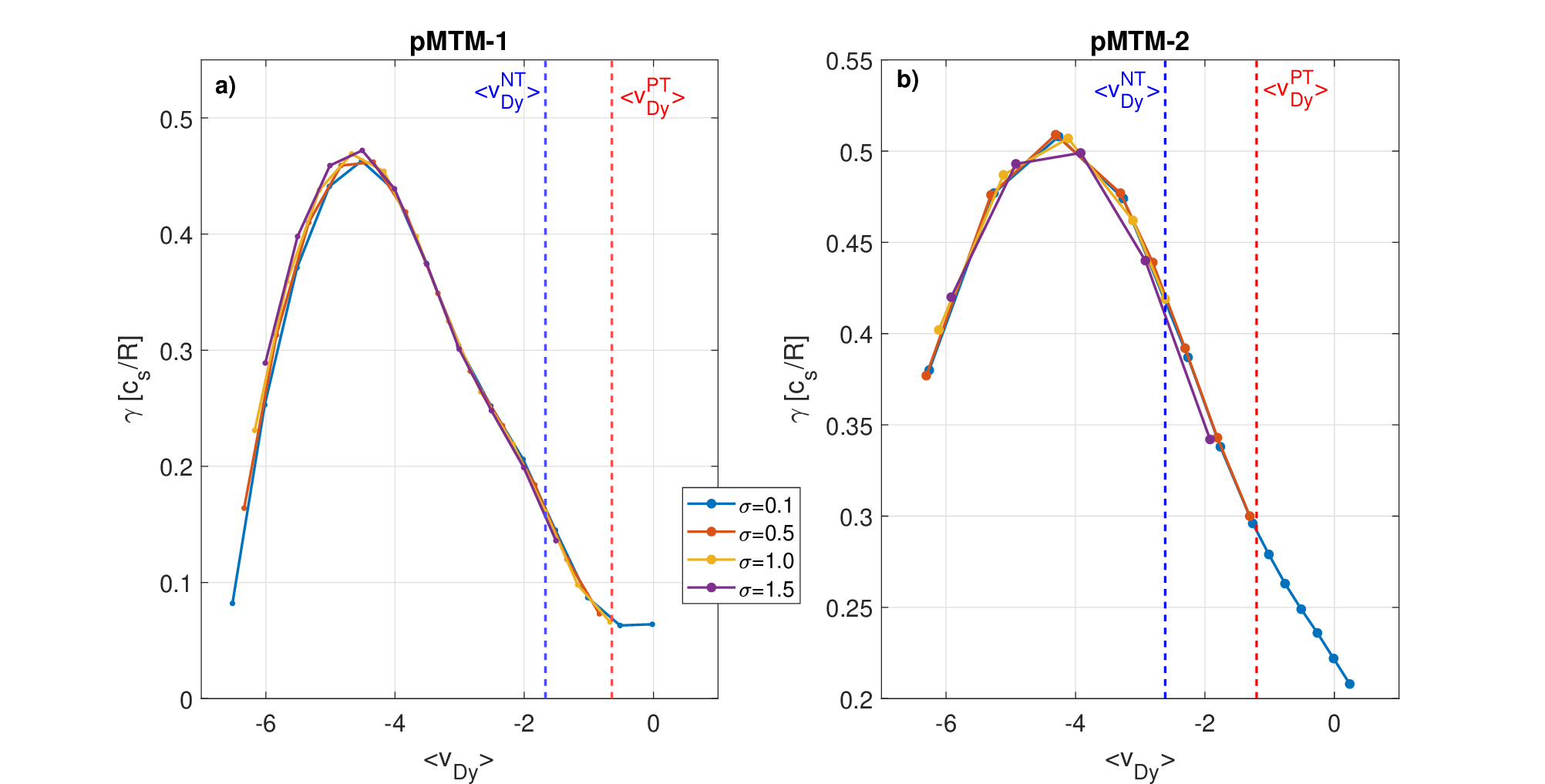}
    \caption{Linear growth rate $\gamma$ of the $[k_x\rho_i,k_y\rho_i]=[0,0.2]$ mode for different $v_{Dy}(z)$ curves as a function of the poloidally averaged binormal component of the magnetic drift velocity $\langle v_{Dy}\rangle$. The dashed blue and red lines represent the averaged magnetic drift velocity of the reference NT and PT scenarios respectively.}
    \label{curvature}
\end{figure*}

\begin{figure}
    \centering
    \includegraphics[width=\linewidth]{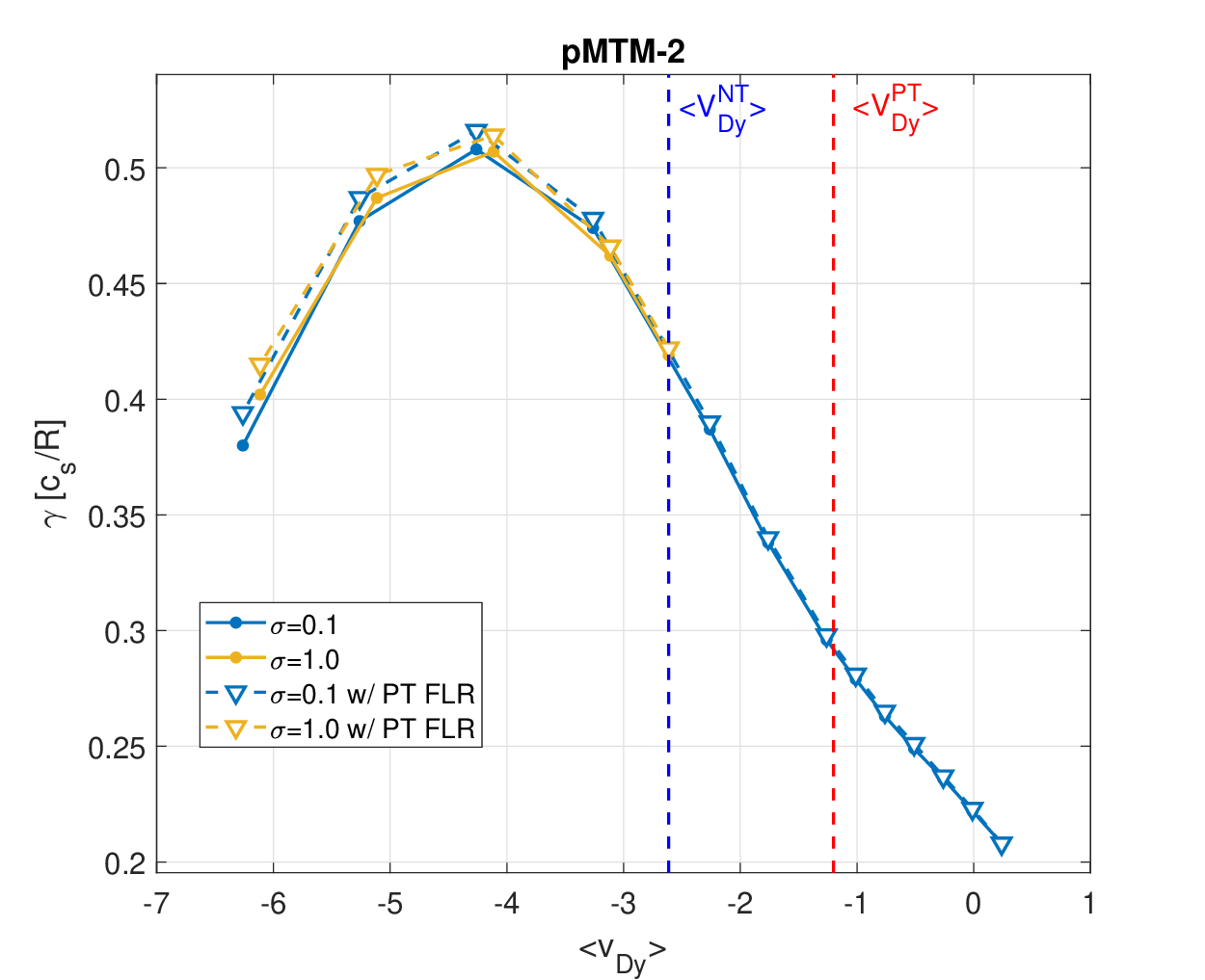}
    \caption{The linear growth rate $\gamma$ of the $[k_x\rho_i,k_y\rho_i]=[0,0.2]$ mode for different $v_{Dy}(z)$ curves as a function of the poloidally averaged binormal component of the magnetic drift velocity $\langle v_{Dy}\rangle$. Filled circles are NT simulations with self-consistent FLR effects. Empty triangles are NT simulations with FLR effects from the PT scenario. The dashed blue and red lines represent the averaged magnetic drift velocity of the reference NT and PT scenarios respectively.}
    \label{FLR}
\end{figure}

\begin{figure}
    \centering
    \includegraphics[width=\linewidth]{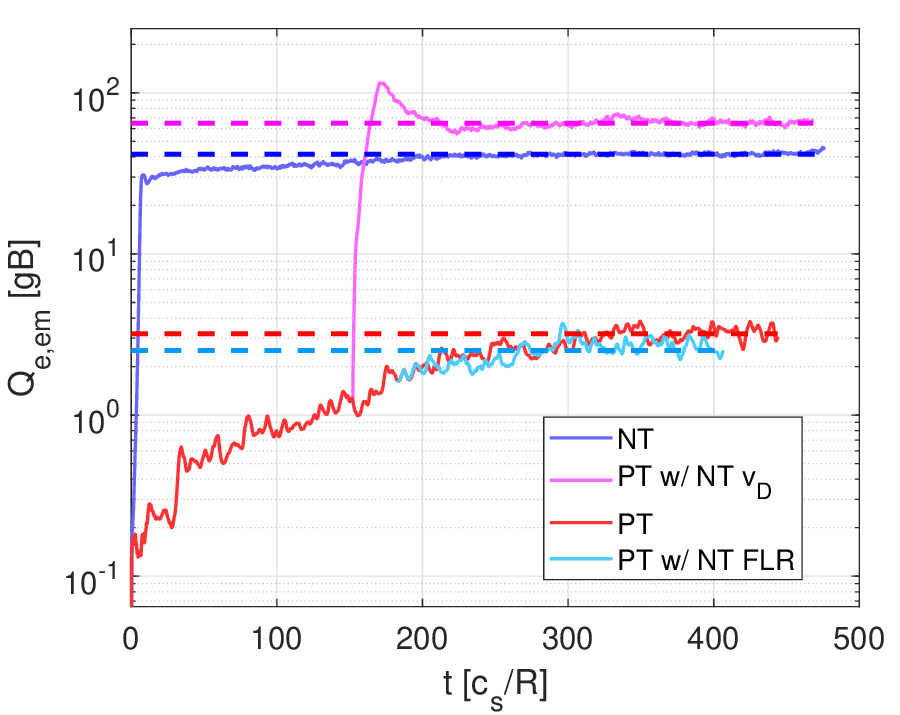}
    \caption{Time traces of the nonlinear electromagnetic component of the electron heat flux as a function of simulation time. The blue line is the self-consistent NT case, the red curve is the self-consistent PT case, the magenta line is the PT scenario with imposed magnetic drift profile from NT (starting at $t\sim150\;c_s/R$) and the light blue line is the PT case with imposed FLR effects from NT (starting at $t\sim180\;c_s/R$).}
    \label{HAM_NL}
\end{figure}

We considered two simplified scenarios where the only instability present are MTMs. The first scenario we used is taken from \cite{Hamed_2023}, where we added \textit{ad hoc} shaping and slightly increased magnetic shear and the electron temperature gradient to strengthen MTM turbulence. Details can be found in table \ref{input_pMTM}. This scenario will be called pMTM-1. The second scenario, called pMTM-2, is a modified version of the DIII-D scenario presented in the previous section, where we removed all the logarithmic gradients except for the electron temperature gradient.

\begin{table}[h]
    \centering
    \begin{tabular}{l|ccccc}
    \toprule
        &pMTM-1 & pMTM-2 \\ \hline
        $R/L_{Te}$& 12.0 & 18.0 \\ 
        $R/L_{Ti}$& 0.0 & 0.0 \\ 
        $R/L_{ne}$& 0.0 & 0.0\\ 
        $T_i/T_e$& 1.0 & 1.0\\ 
        $\beta [\%]$ & 1.55 & 0.8 \\
        $\nu_{C}$ & 0.0012 & 0.0079 \\
        \hline
        $A$& 50 & 50\\ 
        $q$& 1.4 & 1.99\\ 
        $\hat{s}$& 1.3 & 3.5\\
        $\kappa$ & 1.2 & 1.28 \\
        $|\delta|$ & 0.4 & 0.16\\
        $\zeta$ & 0.0 & -0.04 \\
        $s_\kappa$ & 0.1 & 0.04\\
        $|s_\delta|$& 0.4 & 0.34 \\
        $s_\zeta$ & 0.0 & -0.13 \\\toprule
    \end{tabular}
    \caption{\label{input_pMTM} Key physical parameters for the two pure MTM scenarios.}
\end{table}

\subsection{Linear simulations}

\subsubsection{Impact of magnetic drifts}

To study the impact of magnetic drifts on the stability of MTMs, we modified GENE to artificially change the profile of $v_{Dy}$ along the poloidal angle $z$ according to
\begin{equation}
    v_{Dy}(z)=\sigma v_{Dy}^{NT}(z) + v_{D0},
\end{equation}
where $v_{Dy}^{NT}$ is the original profile of the NT case, $\sigma$ is a scalar that modifies how much the profile varies along the poloidal angle and $v_{D0}$ is a scalar that shifts the whole drift profile vertically. More information can be found in \cite{Balestri_2024_2}, where we used the same procedure for ITG modes. We considered four values of $\sigma$ for each scenario and for each of them we changed the offset $v_{D0}$. For each profile, we performed a linear simulation for the mode $(k_x\rho_i,k_y\rho_i)=(0,0.2)$. In figure \ref{curvature}, we show the growth rate as a function of the poloidally averaged drift profile
\begin{equation}
    \langle v_{Dy}\rangle=\frac{\int_{-\pi}^{\pi}dz\,Jv_{Dy}}{\int_{-\pi}^{\pi}dz\,J}.
\end{equation}
For both scenarios we observe a consistent picture. The growth rate of the mode depends only on the averaged magnetic drift velocity and not on how (or if) the profile changes with poloidal angle. This can be observed by noticing that all the curves (which correspond to different values of $\sigma$) are identical. To some extent, this is not a surprising result because MTMs arise from the motion of passing electrons, which move very quickly along field lines and can sample the entire flux surface. This is different from ITG or TEM dynamics, as ions are much slower and trapped electrons cannot reach the inboard midplane. 

Another interesting observation is the non-trivial relation between growth rate and magnetic drift magnitude. The mode becomes more unstable with faster averaged magnetic drifts until a rollover point, where the trend flips and the mode becomes less unstable as the drift speed is increased. However, it must be noted that this rollover point is located at much larger values of drift than typical values for NT and PT geometries, as shown by the blue and red dashed lines in figure \ref{curvature}. Finally, we observe that NT has larger averaged drift velocity than PT, which can explain the much more unstable MTMs.

To conclude, we showed that MTMs are more unstable for faster poloidally averaged magnetic drift velocities $\langle v_{Dy}\rangle$. Because of this, NT geometries, which have larger values of $\langle v_{Dy}\rangle$ than PT, experience stronger MTMs.

\subsubsection{Impact of FLR effects}

To investigate the role FLR effects play in MTM turbulence, we built a new case with the pMTM-2 scenario. We imposed the FLR profile of the PT case in the NT case. Figure \ref{FLR} displays the results. Here, we plot the growth rates from the previous magnetic drift scan and we superimpose the results of the new NT simulations with the FLR effects profile from PT. The the solid circles show no difference from the empty triangles, suggesting that the difference between NT and PT geometries is not due to FLR effects.

\subsection{Nonlinear simulations}

From linear simulations we observed that MTMs are more unstable in NT geometry because of faster magnetic drifts. Here we perform nonlinear simulations to verify if this picture holds when nonlinear dynamics are retained.

As mentioned, nonlinear simulations involving strong MTM turbulence are very numerically challenging. These simulations require a very large box in the radial direction and very high resolutions to properly resolve the evolution of the thin current layers. Therefore, we performed nonlinear simulations only for the pMTM-1 scenario with adiabatic ions. We saw in the previous section that the ion heat flux is a marginal component of the total heat flux when MTMs strongly dominate transport. Thus, the assumption of adiabatic ions is reasonable if one wants to isolate the dynamics of MTMs. The extent of the simulation domain in the radial and binormal directions is $[L_x,L_y]=[250,251]\rho_i$, with a number of Fourier modes $[n_{k_x},n_{k_y}]=[384,56]$. The resolutions in the parallel direction and in velocity space are $[n_z,n_{v_\parallel},n_\mu]=[32,40,12]$ respectively. We recall that in this scenario the aspect ratio is $A=50$. This allows us to ignore all geometrical effects in the gyrokinetic equation except for FLR effects and magnetic drifts. Moreover, at very large aspect ratio, TEM turbulence cannot exist (due to low population of trapped particles) so we can be sure that only MTM can be the dominant instability. 

We performed four nonlinear simulations. One of the NT scenario, one of PT and two where we individually imposed the magnetic drift profiles and the FLR effects from NT geometry in PT. Figure \ref{HAM_NL} shows the time traces of the electromagnetic component of the heat flux for the four cases. We do not show the electrostatic component because it is much smaller. Comparing the self-consistent NT and PT scenarios, we observe that NT is much more unstable than PT (with a heat flux 10 times larger), in agreement with linear simulations. The other two simulations have been performed imposing magnetic drifts and FLR effects from NT, starting at a certain time of the self-consistent PT simulation. If we look at the magenta line in figure \ref{HAM_NL}, we see that, when the magnetic drifts from NT are imposed, turbulence is immediately destabilized dramatically. The heat flux settles around a value 1.5 times larger than the self-consistent NT case. This result is consistent with the linear simulations shown in the previous section and confirms that magnetic drift profile play a key role in influencing MTM dynamics. On the other hand, when the FLR effects from NT geometry are imposed in the PT simulations, we see that little changes, with the heat flux only slightly lower than self-consistent PT. This simulation verifies the linear simulations of figure \ref{FLR}, where very little effect was observed when FLR profiles were swapped between PT and NT geometries.

\section{Conclusions}\label{7}

In this work, we carried out a thorough analysis of the interplay between MTM turbulence and triangularity with linear and nonlinear flux tube GENE simulations of different scenarios, which all showed a coherent picture. We provide here a list of the most important takeaways.
\begin{itemize}
    \item NT geometry is more susceptible to MTMs than PT. This is due to faster poloidally averaged magnetic drift velocities in NT than PT.
    \item The key parameters in the destabilization of MTMs are large values of $\beta$, high $\hat{s}$, a relatively flat density gradient and a ratio of electron to ion temperature gradients ($\omega_{Te}/\omega_{Ti}$) larger than 1.
    \item All NT scenarios entered an MTM dominated regime when $\beta\gtrsim0.3\%$, $\hat{s}\gtrsim2.5$ and $\omega_{Te}/\omega_{Ti}>1$. For the same parameters, the PT scenarios remained dominated by electrostatic turbulence.
    \item When MTMs dominate, NT becomes much more unstable than PT, with heat fluxes that are several times larger than PT. Indeed, when MTMs dominate, magnetic stochasticity increases dramatically, leading to enormous electron electromagnetic heat flux. 
    \item Aspect ratio does not play a direct role in changing the threshold for MTMs destabilization in NT. However, the intrinsically high $\beta$ of the ST concept can make NT scenarios prone to MTMs. Nonetheless, the beneficial effect of NT on transport can be preserved in spherical tokamaks by lowering magnetic shear.
    \item NT conventional tokamaks typically operate away from the MTMs onset region and are not affected by them.  
\end{itemize}

To summarize further, NT plasmas are prone to developing MTM turbulence because geometry strongly increases the magnetic drift velocity of particles. Thus, in NT plasmas, MTM can be driven unstable more easily than in PT geometries. As a consequence, when the right conditions are met, i.e. sufficiently high $\beta$, large magnetic shear, a relatively flat density gradient and electron temperature gradient larger than the ion one, MTMs lead to heat fluxes much larger than in PT geometry. This increased level of heat transport is due to the stochastization of magnetic field lines. The heating power needed to reach a strongly MTM-dominated regime in NT is unrealistically high. However, since this mechanism blocks access to large values of $\beta$, one should be able to observe that the performance of a NT plasma becomes limited at certain temperature and density profiles.

The values of $\beta$ needed at the edge to significantly destabilize MTMs in NT geometry can only be achieved in STs and not in conventional tokamaks, for which the beneficial effect of NT on ITG and TEMs survives. Nonetheless, also in STs the dominance of MTM can be avoided by decreasing magnetic shear, which greatly stabilizes MTM turbulence and allows to recover the beneficial effect of NT on transport. We point out that this is a numerical study and experimental validation is fundamental. Hence, experiments on SMART are paramount.

Lastly, the observations made in this work may relate to the lack of H-mode in NT plasmas. H-mode density profiles are characterized by a flat region after the pedestal top (moving from edge to core). At this radial location, temperature gradients are still very large and magnetic shear is high. In addition, since the pedestal greatly increases temperature and density, large values of $\beta$ can be reached. These are the exact conditions where NT plasmas can be dominated by MTM transport. However, we found that the heat fluxes needed to sustain those kinetic profiles in an MTM dominated regime would be unrealistically high. Therefore, during the formation of the density pedestal, MTMs could be triggered in NT, clamping the electron temperature gradient and preventing the formation of the H-mode pedestal. This may play a role in preventing the formation of H-mode pedestals in NT. This will be the topic of future work.

\section*{Acknowledgments}
We acknowledge Diego José Cruz-Zabala, Manuel Garcia-Muñoz and Eleonora Viezzer for providing us the preliminary scenarios for SMART. We also acknowledge Stephan Brunner, Giovanni Di Giannatale, Alessandro Geraldini, Fabien Jeanquartier, Vincent Masson, Oleg Krutkin, Ralf Mackenbach, Haomin Sun and Arnas Volcokas for the constant help. We also greatly thank Felix Parra for the fruitful discussion and suggestions. This work has been carried out within the framework of the EUROfusion Consortium, via the Euratom Research and Training Programme (Grant Agreement No 101052200 — EUROfusion) and funded by the Swiss State Secretariat for Education, Research and Innovation (SERI). Views and opinions expressed are however those of the author(s) only and do not necessarily reflect those of the European Union, the European Commission, or SERI. Neither the European Union nor the European Commission nor SERI can be held responsible for them. This work was also carried out within the Theory, Simulation, Verification and Vailidation (TSVV) task 2. The authors acknowledge the CINECA award under the ISCRA initiative and the EUROfusion WP-AC, for the availability of high-performance computing resources and support. This work was also carried out (partially) using supercomputer resources provided under the EU-JA Broader Approach collaboration in the Computational Simulation Centre of International Fusion Energy Research Centre (IFERC-CSC). This work is also supported by a grant from the Swiss National Supercomputing Centre
(CSCS). 

% \appendix

% \textcolor{red}{\lipsum[1]}

\bibliography{refs}

\begin{thebibliography}{47}%
\makeatletter
\providecommand \@ifxundefined [1]{%
 \@ifx{#1\undefined}
}%
\providecommand \@ifnum [1]{%
 \ifnum #1\expandafter \@firstoftwo
 \else \expandafter \@secondoftwo
 \fi
}%
\providecommand \@ifx [1]{%
 \ifx #1\expandafter \@firstoftwo
 \else \expandafter \@secondoftwo
 \fi
}%
\providecommand \natexlab [1]{#1}%
\providecommand \enquote  [1]{``#1''}%
\providecommand \bibnamefont  [1]{#1}%
\providecommand \bibfnamefont [1]{#1}%
\providecommand \citenamefont [1]{#1}%
\providecommand \href@noop [0]{\@secondoftwo}%
\providecommand \href [0]{\begingroup \@sanitize@url \@href}%
\providecommand \@href[1]{\@@startlink{#1}\@@href}%
\providecommand \@@href[1]{\endgroup#1\@@endlink}%
\providecommand \@sanitize@url [0]{\catcode `\\12\catcode `\$12\catcode `\&12\catcode `\#12\catcode `\^12\catcode `\_12\catcode `\%12\relax}%
\providecommand \@@startlink[1]{}%
\providecommand \@@endlink[0]{}%
\providecommand \url  [0]{\begingroup\@sanitize@url \@url }%
\providecommand \@url [1]{\endgroup\@href {#1}{\urlprefix }}%
\providecommand \urlprefix  [0]{URL }%
\providecommand \Eprint [0]{\href }%
\providecommand \doibase [0]{http://dx.doi.org/}%
\providecommand \selectlanguage [0]{\@gobble}%
\providecommand \bibinfo  [0]{\@secondoftwo}%
\providecommand \bibfield  [0]{\@secondoftwo}%
\providecommand \translation [1]{[#1]}%
\providecommand \BibitemOpen [0]{}%
\providecommand \bibitemStop [0]{}%
\providecommand \bibitemNoStop [0]{.\EOS\space}%
\providecommand \EOS [0]{\spacefactor3000\relax}%
\providecommand \BibitemShut  [1]{\csname bibitem#1\endcsname}%
\let\auto@bib@innerbib\@empty
%</preamble>
\bibitem [{\citenamefont {Weisen}\ \emph {et~al.}(1997)\citenamefont {Weisen}, \citenamefont {Moret}, \citenamefont {Franke}, \citenamefont {Furno}, \citenamefont {Martin}, \citenamefont {Anton}, \citenamefont {Behn}, \citenamefont {Dutch}, \citenamefont {Duval}, \citenamefont {Hofmann}, \citenamefont {Joye}, \citenamefont {Nieswand}, \citenamefont {Pietrzyk},\ and\ \citenamefont {Toledo}}]{Weisen_1997}%
  \BibitemOpen
  \bibfield  {author} {\bibinfo {author} {\bibfnamefont {H.}~\bibnamefont {Weisen}}, \bibinfo {author} {\bibfnamefont {J.-M.}\ \bibnamefont {Moret}}, \bibinfo {author} {\bibfnamefont {S.}~\bibnamefont {Franke}}, \bibinfo {author} {\bibfnamefont {I.}~\bibnamefont {Furno}}, \bibinfo {author} {\bibfnamefont {Y.}~\bibnamefont {Martin}}, \bibinfo {author} {\bibfnamefont {M.}~\bibnamefont {Anton}}, \bibinfo {author} {\bibfnamefont {R.}~\bibnamefont {Behn}}, \bibinfo {author} {\bibfnamefont {M.}~\bibnamefont {Dutch}}, \bibinfo {author} {\bibfnamefont {B.}~\bibnamefont {Duval}}, \bibinfo {author} {\bibfnamefont {F.}~\bibnamefont {Hofmann}}, \bibinfo {author} {\bibfnamefont {B.}~\bibnamefont {Joye}}, \bibinfo {author} {\bibfnamefont {C.}~\bibnamefont {Nieswand}}, \bibinfo {author} {\bibfnamefont {Z.}~\bibnamefont {Pietrzyk}}, \ and\ \bibinfo {author} {\bibfnamefont {W.~V.}\ \bibnamefont {Toledo}},\ }\href {\doibase 10.1088/0029-5515/37/12/I07} {\bibfield  {journal} {\bibinfo  {journal} {Nuclear Fusion}\ }\textbf
  {\bibinfo {volume} {37}},\ \bibinfo {pages} {1741} (\bibinfo {year} {1997})}\BibitemShut {NoStop}%
\bibitem [{\citenamefont {Merle}\ \emph {et~al.}(2017)\citenamefont {Merle}, \citenamefont {Sauter},\ and\ \citenamefont {Medvedev}}]{Merle_2017}%
  \BibitemOpen
  \bibfield  {author} {\bibinfo {author} {\bibfnamefont {A.}~\bibnamefont {Merle}}, \bibinfo {author} {\bibfnamefont {O.}~\bibnamefont {Sauter}}, \ and\ \bibinfo {author} {\bibfnamefont {S.~Y.}\ \bibnamefont {Medvedev}},\ }\href {\doibase 10.1088/1361-6587/aa7ac0} {\bibfield  {journal} {\bibinfo  {journal} {Plasma Physics and Controlled Fusion}\ }\textbf {\bibinfo {volume} {59}},\ \bibinfo {pages} {104001} (\bibinfo {year} {2017})}\BibitemShut {NoStop}%
\bibitem [{\citenamefont {Nelson}\ \emph {et~al.}(2024)\citenamefont {Nelson}, \citenamefont {Schmitz}, \citenamefont {Cote}, \citenamefont {Parisi}, \citenamefont {Stewart}, \citenamefont {Paz-Soldan}, \citenamefont {Thome}, \citenamefont {Austin}, \citenamefont {Scotti}, \citenamefont {Barr}, \citenamefont {Hyatt}, \citenamefont {Leuthold}, \citenamefont {Marinoni}, \citenamefont {Neiser}, \citenamefont {Osborne}, \citenamefont {Richner}, \citenamefont {Welander}, \citenamefont {Wehner}, \citenamefont {Wilcox}, \citenamefont {Wilks}, \citenamefont {Yang},\ and\ \citenamefont {the DIII-D~Team}}]{Nelson_2024}%
  \BibitemOpen
  \bibfield  {author} {\bibinfo {author} {\bibfnamefont {A.~O.}\ \bibnamefont {Nelson}}, \bibinfo {author} {\bibfnamefont {L.}~\bibnamefont {Schmitz}}, \bibinfo {author} {\bibfnamefont {T.}~\bibnamefont {Cote}}, \bibinfo {author} {\bibfnamefont {J.~F.}\ \bibnamefont {Parisi}}, \bibinfo {author} {\bibfnamefont {S.}~\bibnamefont {Stewart}}, \bibinfo {author} {\bibfnamefont {C.}~\bibnamefont {Paz-Soldan}}, \bibinfo {author} {\bibfnamefont {K.~E.}\ \bibnamefont {Thome}}, \bibinfo {author} {\bibfnamefont {M.~E.}\ \bibnamefont {Austin}}, \bibinfo {author} {\bibfnamefont {F.}~\bibnamefont {Scotti}}, \bibinfo {author} {\bibfnamefont {J.~L.}\ \bibnamefont {Barr}}, \bibinfo {author} {\bibfnamefont {A.}~\bibnamefont {Hyatt}}, \bibinfo {author} {\bibfnamefont {N.}~\bibnamefont {Leuthold}}, \bibinfo {author} {\bibfnamefont {A.}~\bibnamefont {Marinoni}}, \bibinfo {author} {\bibfnamefont {T.}~\bibnamefont {Neiser}}, \bibinfo {author} {\bibfnamefont {T.}~\bibnamefont {Osborne}}, \bibinfo {author} {\bibfnamefont
  {N.}~\bibnamefont {Richner}}, \bibinfo {author} {\bibfnamefont {A.~S.}\ \bibnamefont {Welander}}, \bibinfo {author} {\bibfnamefont {W.~P.}\ \bibnamefont {Wehner}}, \bibinfo {author} {\bibfnamefont {R.}~\bibnamefont {Wilcox}}, \bibinfo {author} {\bibfnamefont {T.~M.}\ \bibnamefont {Wilks}}, \bibinfo {author} {\bibfnamefont {J.}~\bibnamefont {Yang}}, \ and\ \bibinfo {author} {\bibnamefont {the DIII-D~Team}},\ }\href {\doibase 10.1088/1361-6587/ad6a83} {\bibfield  {journal} {\bibinfo  {journal} {Plasma Physics and Controlled Fusion}\ }\textbf {\bibinfo {volume} {66}},\ \bibinfo {pages} {105014} (\bibinfo {year} {2024})}\BibitemShut {NoStop}%
\bibitem [{\citenamefont {Fontana}\ \emph {et~al.}(2017)\citenamefont {Fontana}, \citenamefont {Porte}, \citenamefont {Coda},\ and\ \citenamefont {and}}]{9}%
  \BibitemOpen
  \bibfield  {author} {\bibinfo {author} {\bibfnamefont {M.}~\bibnamefont {Fontana}}, \bibinfo {author} {\bibfnamefont {L.}~\bibnamefont {Porte}}, \bibinfo {author} {\bibfnamefont {S.}~\bibnamefont {Coda}}, \ and\ \bibinfo {author} {\bibfnamefont {O.~S.}\ \bibnamefont {and}},\ }\href {\doibase 10.1088/1741-4326/aa98f4} {\bibfield  {journal} {\bibinfo  {journal} {Nuclear Fusion}\ }\textbf {\bibinfo {volume} {58}},\ \bibinfo {pages} {024002} (\bibinfo {year} {2017})}\BibitemShut {NoStop}%
\bibitem [{\citenamefont {Marinoni}\ \emph {et~al.}(2019)\citenamefont {Marinoni}, \citenamefont {Austin}, \citenamefont {Hyatt}, \citenamefont {Walker}, \citenamefont {Candy}, \citenamefont {Chrystal}, \citenamefont {Lasnier}, \citenamefont {McKee}, \citenamefont {Odstrčil}, \citenamefont {Petty}, \citenamefont {Porkolab}, \citenamefont {Rost}, \citenamefont {Sauter}, \citenamefont {Smith}, \citenamefont {Staebler}, \citenamefont {Sung}, \citenamefont {Thome}, \citenamefont {Turnbull},\ and\ \citenamefont {Zeng}}]{7}%
  \BibitemOpen
  \bibfield  {author} {\bibinfo {author} {\bibfnamefont {A.}~\bibnamefont {Marinoni}}, \bibinfo {author} {\bibfnamefont {M.~E.}\ \bibnamefont {Austin}}, \bibinfo {author} {\bibfnamefont {A.~W.}\ \bibnamefont {Hyatt}}, \bibinfo {author} {\bibfnamefont {M.~L.}\ \bibnamefont {Walker}}, \bibinfo {author} {\bibfnamefont {J.}~\bibnamefont {Candy}}, \bibinfo {author} {\bibfnamefont {C.}~\bibnamefont {Chrystal}}, \bibinfo {author} {\bibfnamefont {C.~J.}\ \bibnamefont {Lasnier}}, \bibinfo {author} {\bibfnamefont {G.~R.}\ \bibnamefont {McKee}}, \bibinfo {author} {\bibfnamefont {T.}~\bibnamefont {Odstrčil}}, \bibinfo {author} {\bibfnamefont {C.~C.}\ \bibnamefont {Petty}}, \bibinfo {author} {\bibfnamefont {M.}~\bibnamefont {Porkolab}}, \bibinfo {author} {\bibfnamefont {J.~C.}\ \bibnamefont {Rost}}, \bibinfo {author} {\bibfnamefont {O.}~\bibnamefont {Sauter}}, \bibinfo {author} {\bibfnamefont {S.~P.}\ \bibnamefont {Smith}}, \bibinfo {author} {\bibfnamefont {G.~M.}\ \bibnamefont {Staebler}}, \bibinfo {author}
  {\bibfnamefont {C.}~\bibnamefont {Sung}}, \bibinfo {author} {\bibfnamefont {K.~E.}\ \bibnamefont {Thome}}, \bibinfo {author} {\bibfnamefont {A.~D.}\ \bibnamefont {Turnbull}}, \ and\ \bibinfo {author} {\bibfnamefont {L.}~\bibnamefont {Zeng}},\ }\href {\doibase 10.1063/1.5091802} {\bibfield  {journal} {\bibinfo  {journal} {Physics of Plasmas}\ }\textbf {\bibinfo {volume} {26}},\ \bibinfo {pages} {042515} (\bibinfo {year} {2019})},\ \Eprint {http://arxiv.org/abs/https://doi.org/10.1063/1.5091802} {https://doi.org/10.1063/1.5091802} \BibitemShut {NoStop}%
\bibitem [{\citenamefont {Austin}\ \emph {et~al.}(2019)\citenamefont {Austin}, \citenamefont {Marinoni}, \citenamefont {Walker}, \citenamefont {Brookman}, \citenamefont {deGrassie}, \citenamefont {Hyatt}, \citenamefont {McKee}, \citenamefont {Petty}, \citenamefont {Rhodes}, \citenamefont {Smith}, \citenamefont {Sung}, \citenamefont {Thome},\ and\ \citenamefont {Turnbull}}]{Austin_2019}%
  \BibitemOpen
  \bibfield  {author} {\bibinfo {author} {\bibfnamefont {M.~E.}\ \bibnamefont {Austin}}, \bibinfo {author} {\bibfnamefont {A.}~\bibnamefont {Marinoni}}, \bibinfo {author} {\bibfnamefont {M.~L.}\ \bibnamefont {Walker}}, \bibinfo {author} {\bibfnamefont {M.~W.}\ \bibnamefont {Brookman}}, \bibinfo {author} {\bibfnamefont {J.~S.}\ \bibnamefont {deGrassie}}, \bibinfo {author} {\bibfnamefont {A.~W.}\ \bibnamefont {Hyatt}}, \bibinfo {author} {\bibfnamefont {G.~R.}\ \bibnamefont {McKee}}, \bibinfo {author} {\bibfnamefont {C.~C.}\ \bibnamefont {Petty}}, \bibinfo {author} {\bibfnamefont {T.~L.}\ \bibnamefont {Rhodes}}, \bibinfo {author} {\bibfnamefont {S.~P.}\ \bibnamefont {Smith}}, \bibinfo {author} {\bibfnamefont {C.}~\bibnamefont {Sung}}, \bibinfo {author} {\bibfnamefont {K.~E.}\ \bibnamefont {Thome}}, \ and\ \bibinfo {author} {\bibfnamefont {A.~D.}\ \bibnamefont {Turnbull}},\ }\href {\doibase 10.1103/PhysRevLett.122.115001} {\bibfield  {journal} {\bibinfo  {journal} {Phys. Rev. Lett.}\ }\textbf {\bibinfo {volume}
  {122}},\ \bibinfo {pages} {115001} (\bibinfo {year} {2019})}\BibitemShut {NoStop}%
\bibitem [{\citenamefont {Paz-Soldan}\ and\ \citenamefont {the DIII-D~Team}(2021)}]{Paz-Soldan_2021}%
  \BibitemOpen
  \bibfield  {author} {\bibinfo {author} {\bibfnamefont {C.}~\bibnamefont {Paz-Soldan}}\ and\ \bibinfo {author} {\bibnamefont {the DIII-D~Team}},\ }\href {\doibase 10.1088/1361-6587/ac048b} {\bibfield  {journal} {\bibinfo  {journal} {Plasma Physics and Controlled Fusion}\ }\textbf {\bibinfo {volume} {63}},\ \bibinfo {pages} {083001} (\bibinfo {year} {2021})}\BibitemShut {NoStop}%
\bibitem [{\citenamefont {Coda}\ \emph {et~al.}(2021)\citenamefont {Coda}, \citenamefont {Merle}, \citenamefont {Sauter}, \citenamefont {Porte}, \citenamefont {Bagnato}, \citenamefont {Boedo}, \citenamefont {Bolzonella}, \citenamefont {Février}, \citenamefont {Labit}, \citenamefont {Marinoni}, \citenamefont {Pau}, \citenamefont {Pigatto}, \citenamefont {Sheikh}, \citenamefont {Tsui}, \citenamefont {Vallar}, \citenamefont {Vu},\ and\ \citenamefont {Team}}]{Coda_2022}%
  \BibitemOpen
  \bibfield  {author} {\bibinfo {author} {\bibfnamefont {S.}~\bibnamefont {Coda}}, \bibinfo {author} {\bibfnamefont {A.}~\bibnamefont {Merle}}, \bibinfo {author} {\bibfnamefont {O.}~\bibnamefont {Sauter}}, \bibinfo {author} {\bibfnamefont {L.}~\bibnamefont {Porte}}, \bibinfo {author} {\bibfnamefont {F.}~\bibnamefont {Bagnato}}, \bibinfo {author} {\bibfnamefont {J.}~\bibnamefont {Boedo}}, \bibinfo {author} {\bibfnamefont {T.}~\bibnamefont {Bolzonella}}, \bibinfo {author} {\bibfnamefont {O.}~\bibnamefont {Février}}, \bibinfo {author} {\bibfnamefont {B.}~\bibnamefont {Labit}}, \bibinfo {author} {\bibfnamefont {A.}~\bibnamefont {Marinoni}}, \bibinfo {author} {\bibfnamefont {A.}~\bibnamefont {Pau}}, \bibinfo {author} {\bibfnamefont {L.}~\bibnamefont {Pigatto}}, \bibinfo {author} {\bibfnamefont {U.}~\bibnamefont {Sheikh}}, \bibinfo {author} {\bibfnamefont {C.}~\bibnamefont {Tsui}}, \bibinfo {author} {\bibfnamefont {M.}~\bibnamefont {Vallar}}, \bibinfo {author} {\bibfnamefont {T.}~\bibnamefont {Vu}}, \ and\
  \bibinfo {author} {\bibfnamefont {T.~T.}\ \bibnamefont {Team}},\ }\href {\doibase 10.1088/1361-6587/ac3fec} {\bibfield  {journal} {\bibinfo  {journal} {Plasma Physics and Controlled Fusion}\ }\textbf {\bibinfo {volume} {64}},\ \bibinfo {pages} {014004} (\bibinfo {year} {2021})}\BibitemShut {NoStop}%
\bibitem [{\citenamefont {Happel}\ \emph {et~al.}(2022)\citenamefont {Happel}, \citenamefont {Pütterich}, \citenamefont {Told}, \citenamefont {Dunne}, \citenamefont {Fischer}, \citenamefont {Hobirk}, \citenamefont {McDermott}, \citenamefont {Plank},\ and\ \citenamefont {the}}]{Happel_2023}%
  \BibitemOpen
  \bibfield  {author} {\bibinfo {author} {\bibfnamefont {T.}~\bibnamefont {Happel}}, \bibinfo {author} {\bibfnamefont {T.}~\bibnamefont {Pütterich}}, \bibinfo {author} {\bibfnamefont {D.}~\bibnamefont {Told}}, \bibinfo {author} {\bibfnamefont {M.}~\bibnamefont {Dunne}}, \bibinfo {author} {\bibfnamefont {R.}~\bibnamefont {Fischer}}, \bibinfo {author} {\bibfnamefont {J.}~\bibnamefont {Hobirk}}, \bibinfo {author} {\bibfnamefont {R.}~\bibnamefont {McDermott}}, \bibinfo {author} {\bibfnamefont {U.}~\bibnamefont {Plank}}, \ and\ \bibinfo {author} {\bibfnamefont {A.~U.~T.}\ \bibnamefont {the}},\ }\href {\doibase 10.1088/1741-4326/ac8563} {\bibfield  {journal} {\bibinfo  {journal} {Nuclear Fusion}\ }\textbf {\bibinfo {volume} {63}},\ \bibinfo {pages} {016002} (\bibinfo {year} {2022})}\BibitemShut {NoStop}%
\bibitem [{\citenamefont {Nelson}\ \emph {et~al.}(2023)\citenamefont {Nelson}, \citenamefont {Schmitz}, \citenamefont {Paz-Soldan}, \citenamefont {Thome}, \citenamefont {Cote}, \citenamefont {Leuthold}, \citenamefont {Scotti}, \citenamefont {Austin}, \citenamefont {Hyatt},\ and\ \citenamefont {Osborne}}]{Nelson_PRL}%
  \BibitemOpen
  \bibfield  {author} {\bibinfo {author} {\bibfnamefont {A.~O.}\ \bibnamefont {Nelson}}, \bibinfo {author} {\bibfnamefont {L.}~\bibnamefont {Schmitz}}, \bibinfo {author} {\bibfnamefont {C.}~\bibnamefont {Paz-Soldan}}, \bibinfo {author} {\bibfnamefont {K.~E.}\ \bibnamefont {Thome}}, \bibinfo {author} {\bibfnamefont {T.~B.}\ \bibnamefont {Cote}}, \bibinfo {author} {\bibfnamefont {N.}~\bibnamefont {Leuthold}}, \bibinfo {author} {\bibfnamefont {F.}~\bibnamefont {Scotti}}, \bibinfo {author} {\bibfnamefont {M.~E.}\ \bibnamefont {Austin}}, \bibinfo {author} {\bibfnamefont {A.}~\bibnamefont {Hyatt}}, \ and\ \bibinfo {author} {\bibfnamefont {T.}~\bibnamefont {Osborne}},\ }\href {\doibase 10.1103/PhysRevLett.131.195101} {\bibfield  {journal} {\bibinfo  {journal} {Phys. Rev. Lett.}\ }\textbf {\bibinfo {volume} {131}},\ \bibinfo {pages} {195101} (\bibinfo {year} {2023})}\BibitemShut {NoStop}%
\bibitem [{\citenamefont {Balestri}\ \emph {et~al.}(2024{\natexlab{a}})\citenamefont {Balestri}, \citenamefont {Mantica}, \citenamefont {Mariani}, \citenamefont {Bagnato}, \citenamefont {Bolzonella}, \citenamefont {Ball}, \citenamefont {Coda}, \citenamefont {Dunne}, \citenamefont {Faitsch}, \citenamefont {Innocente}, \citenamefont {Muscente}, \citenamefont {Sauter}, \citenamefont {Vallar}, \citenamefont {Viezzer}, \citenamefont {the TCV~Team},\ and\ \citenamefont {the EUROfusion Tokamak Exploitation~Team}}]{Balestri_2024}%
  \BibitemOpen
  \bibfield  {author} {\bibinfo {author} {\bibfnamefont {A.}~\bibnamefont {Balestri}}, \bibinfo {author} {\bibfnamefont {P.}~\bibnamefont {Mantica}}, \bibinfo {author} {\bibfnamefont {A.}~\bibnamefont {Mariani}}, \bibinfo {author} {\bibfnamefont {F.}~\bibnamefont {Bagnato}}, \bibinfo {author} {\bibfnamefont {T.}~\bibnamefont {Bolzonella}}, \bibinfo {author} {\bibfnamefont {J.}~\bibnamefont {Ball}}, \bibinfo {author} {\bibfnamefont {S.}~\bibnamefont {Coda}}, \bibinfo {author} {\bibfnamefont {M.}~\bibnamefont {Dunne}}, \bibinfo {author} {\bibfnamefont {M.}~\bibnamefont {Faitsch}}, \bibinfo {author} {\bibfnamefont {P.}~\bibnamefont {Innocente}}, \bibinfo {author} {\bibfnamefont {P.}~\bibnamefont {Muscente}}, \bibinfo {author} {\bibfnamefont {O.}~\bibnamefont {Sauter}}, \bibinfo {author} {\bibfnamefont {M.}~\bibnamefont {Vallar}}, \bibinfo {author} {\bibfnamefont {E.}~\bibnamefont {Viezzer}}, \bibinfo {author} {\bibnamefont {the TCV~Team}}, \ and\ \bibinfo {author} {\bibnamefont {the EUROfusion Tokamak
  Exploitation~Team}},\ }\href {\doibase 10.1088/1361-6587/ad4674} {\bibfield  {journal} {\bibinfo  {journal} {Plasma Physics and Controlled Fusion}\ }\textbf {\bibinfo {volume} {66}},\ \bibinfo {pages} {065031} (\bibinfo {year} {2024}{\natexlab{a}})}\BibitemShut {NoStop}%
\bibitem [{\citenamefont {Aucone}\ \emph {et~al.}(2024)\citenamefont {Aucone}, \citenamefont {Mantica}, \citenamefont {Happel}, \citenamefont {Hobirk}, \citenamefont {Pütterich}, \citenamefont {Vanovac}, \citenamefont {Zimmermann}, \citenamefont {Bernert}, \citenamefont {Bolzonella}, \citenamefont {Cavedon}, \citenamefont {Dunne}, \citenamefont {Fischer}, \citenamefont {Innocente}, \citenamefont {Kappatou}, \citenamefont {McDermott}, \citenamefont {Mariani}, \citenamefont {Muscente}, \citenamefont {Plank}, \citenamefont {Sciortino}, \citenamefont {Tardini}, \citenamefont {WPTE~Team},\ and\ \citenamefont {Upgrade~Team}}]{Aucone_2024}%
  \BibitemOpen
  \bibfield  {author} {\bibinfo {author} {\bibfnamefont {L.}~\bibnamefont {Aucone}}, \bibinfo {author} {\bibfnamefont {P.}~\bibnamefont {Mantica}}, \bibinfo {author} {\bibfnamefont {T.}~\bibnamefont {Happel}}, \bibinfo {author} {\bibfnamefont {J.}~\bibnamefont {Hobirk}}, \bibinfo {author} {\bibfnamefont {T.}~\bibnamefont {Pütterich}}, \bibinfo {author} {\bibfnamefont {B.}~\bibnamefont {Vanovac}}, \bibinfo {author} {\bibfnamefont {C.~F.~B.}\ \bibnamefont {Zimmermann}}, \bibinfo {author} {\bibfnamefont {M.}~\bibnamefont {Bernert}}, \bibinfo {author} {\bibfnamefont {T.}~\bibnamefont {Bolzonella}}, \bibinfo {author} {\bibfnamefont {M.}~\bibnamefont {Cavedon}}, \bibinfo {author} {\bibfnamefont {M.}~\bibnamefont {Dunne}}, \bibinfo {author} {\bibfnamefont {R.}~\bibnamefont {Fischer}}, \bibinfo {author} {\bibfnamefont {P.}~\bibnamefont {Innocente}}, \bibinfo {author} {\bibfnamefont {A.}~\bibnamefont {Kappatou}}, \bibinfo {author} {\bibfnamefont {R.~M.}\ \bibnamefont {McDermott}}, \bibinfo {author} {\bibfnamefont
  {A.}~\bibnamefont {Mariani}}, \bibinfo {author} {\bibfnamefont {P.}~\bibnamefont {Muscente}}, \bibinfo {author} {\bibfnamefont {U.}~\bibnamefont {Plank}}, \bibinfo {author} {\bibfnamefont {F.}~\bibnamefont {Sciortino}}, \bibinfo {author} {\bibfnamefont {G.}~\bibnamefont {Tardini}}, \bibinfo {author} {\bibfnamefont {t.~E.}\ \bibnamefont {WPTE~Team}}, \ and\ \bibinfo {author} {\bibfnamefont {t.~A.}\ \bibnamefont {Upgrade~Team}},\ }\href {\doibase 10.1088/1361-6587/ad4d1c} {\bibfield  {journal} {\bibinfo  {journal} {Plasma Physics and Controlled Fusion}\ }\textbf {\bibinfo {volume} {66}},\ \bibinfo {pages} {075013} (\bibinfo {year} {2024})}\BibitemShut {NoStop}%
\bibitem [{\citenamefont {Mariani}\ \emph {et~al.}(2024)\citenamefont {Mariani}, \citenamefont {Aucone}, \citenamefont {Balestri}, \citenamefont {Mantica}, \citenamefont {Merlo}, \citenamefont {Ambrosino}, \citenamefont {Bagnato}, \citenamefont {Balbinot}, \citenamefont {Ball}, \citenamefont {Bolzonella}, \citenamefont {Brioschi}, \citenamefont {Casiraghi}, \citenamefont {Castaldo}, \citenamefont {Coda}, \citenamefont {Frassinetti}, \citenamefont {Fusco}, \citenamefont {Happel}, \citenamefont {Hobirk}, \citenamefont {Innocente}, \citenamefont {McDermott}, \citenamefont {Muscente}, \citenamefont {Pütterich}, \citenamefont {Sauter}, \citenamefont {Sciortino}, \citenamefont {Vallar}, \citenamefont {Vanovac}, \citenamefont {Vianello}, \citenamefont {Vlad}, \citenamefont {Zimmermann}, \citenamefont {the EUROfusion Tokamak Exploitation~team}, \citenamefont {the TCV~Team},\ and\ \citenamefont {the ASDEX Upgrade~Team}}]{Mariani_2024}%
  \BibitemOpen
  \bibfield  {author} {\bibinfo {author} {\bibfnamefont {A.}~\bibnamefont {Mariani}}, \bibinfo {author} {\bibfnamefont {L.}~\bibnamefont {Aucone}}, \bibinfo {author} {\bibfnamefont {A.}~\bibnamefont {Balestri}}, \bibinfo {author} {\bibfnamefont {P.}~\bibnamefont {Mantica}}, \bibinfo {author} {\bibfnamefont {G.}~\bibnamefont {Merlo}}, \bibinfo {author} {\bibfnamefont {R.}~\bibnamefont {Ambrosino}}, \bibinfo {author} {\bibfnamefont {F.}~\bibnamefont {Bagnato}}, \bibinfo {author} {\bibfnamefont {L.}~\bibnamefont {Balbinot}}, \bibinfo {author} {\bibfnamefont {J.}~\bibnamefont {Ball}}, \bibinfo {author} {\bibfnamefont {T.}~\bibnamefont {Bolzonella}}, \bibinfo {author} {\bibfnamefont {D.}~\bibnamefont {Brioschi}}, \bibinfo {author} {\bibfnamefont {I.}~\bibnamefont {Casiraghi}}, \bibinfo {author} {\bibfnamefont {A.}~\bibnamefont {Castaldo}}, \bibinfo {author} {\bibfnamefont {S.}~\bibnamefont {Coda}}, \bibinfo {author} {\bibfnamefont {L.}~\bibnamefont {Frassinetti}}, \bibinfo {author} {\bibfnamefont {V.}~\bibnamefont
  {Fusco}}, \bibinfo {author} {\bibfnamefont {T.}~\bibnamefont {Happel}}, \bibinfo {author} {\bibfnamefont {J.}~\bibnamefont {Hobirk}}, \bibinfo {author} {\bibfnamefont {P.}~\bibnamefont {Innocente}}, \bibinfo {author} {\bibfnamefont {R.}~\bibnamefont {McDermott}}, \bibinfo {author} {\bibfnamefont {P.}~\bibnamefont {Muscente}}, \bibinfo {author} {\bibfnamefont {T.}~\bibnamefont {Pütterich}}, \bibinfo {author} {\bibfnamefont {O.}~\bibnamefont {Sauter}}, \bibinfo {author} {\bibfnamefont {F.}~\bibnamefont {Sciortino}}, \bibinfo {author} {\bibfnamefont {M.}~\bibnamefont {Vallar}}, \bibinfo {author} {\bibfnamefont {B.}~\bibnamefont {Vanovac}}, \bibinfo {author} {\bibfnamefont {N.}~\bibnamefont {Vianello}}, \bibinfo {author} {\bibfnamefont {G.}~\bibnamefont {Vlad}}, \bibinfo {author} {\bibfnamefont {C.}~\bibnamefont {Zimmermann}}, \bibinfo {author} {\bibnamefont {the EUROfusion Tokamak Exploitation~team}}, \bibinfo {author} {\bibnamefont {the TCV~Team}}, \ and\ \bibinfo {author} {\bibnamefont {the ASDEX
  Upgrade~Team}},\ }\href {\doibase 10.1088/1741-4326/ad6ea0} {\bibfield  {journal} {\bibinfo  {journal} {Nuclear Fusion}\ }\textbf {\bibinfo {volume} {64}},\ \bibinfo {pages} {106024} (\bibinfo {year} {2024})}\BibitemShut {NoStop}%
\bibitem [{\citenamefont {Thome}\ \emph {et~al.}(2024)\citenamefont {Thome}, \citenamefont {Austin}, \citenamefont {Hyatt}, \citenamefont {Marinoni}, \citenamefont {Nelson}, \citenamefont {Paz-Soldan}, \citenamefont {Scotti}, \citenamefont {Boyes}, \citenamefont {Casali}, \citenamefont {Chrystal}, \citenamefont {Ding}, \citenamefont {Du}, \citenamefont {Eldon}, \citenamefont {Ernst}, \citenamefont {Hong}, \citenamefont {McKee}, \citenamefont {Mordijck}, \citenamefont {Sauter}, \citenamefont {Schmitz}, \citenamefont {Barr}, \citenamefont {Burke}, \citenamefont {Coda}, \citenamefont {Cote}, \citenamefont {Fenstermacher}, \citenamefont {Garofalo}, \citenamefont {Khabanov}, \citenamefont {Kramer}, \citenamefont {Lasnier}, \citenamefont {Logan}, \citenamefont {Lunia}, \citenamefont {McLean}, \citenamefont {Okabayashi}, \citenamefont {Shiraki}, \citenamefont {Stewart}, \citenamefont {Takemura}, \citenamefont {Truong}, \citenamefont {Osborne}, \citenamefont {Zeeland}, \citenamefont {Victor}, \citenamefont {Wang},
  \citenamefont {Watkins}, \citenamefont {Wehner}, \citenamefont {Welander}, \citenamefont {Wilks}, \citenamefont {Yang}, \citenamefont {Yu}, \citenamefont {Zeng},\ and\ \citenamefont {the DIII-D~Team}}]{Thome_2024}%
  \BibitemOpen
  \bibfield  {author} {\bibinfo {author} {\bibfnamefont {K.~E.}\ \bibnamefont {Thome}}, \bibinfo {author} {\bibfnamefont {M.~E.}\ \bibnamefont {Austin}}, \bibinfo {author} {\bibfnamefont {A.}~\bibnamefont {Hyatt}}, \bibinfo {author} {\bibfnamefont {A.}~\bibnamefont {Marinoni}}, \bibinfo {author} {\bibfnamefont {A.~O.}\ \bibnamefont {Nelson}}, \bibinfo {author} {\bibfnamefont {C.}~\bibnamefont {Paz-Soldan}}, \bibinfo {author} {\bibfnamefont {F.}~\bibnamefont {Scotti}}, \bibinfo {author} {\bibfnamefont {W.}~\bibnamefont {Boyes}}, \bibinfo {author} {\bibfnamefont {L.}~\bibnamefont {Casali}}, \bibinfo {author} {\bibfnamefont {C.}~\bibnamefont {Chrystal}}, \bibinfo {author} {\bibfnamefont {S.}~\bibnamefont {Ding}}, \bibinfo {author} {\bibfnamefont {X.~D.}\ \bibnamefont {Du}}, \bibinfo {author} {\bibfnamefont {D.}~\bibnamefont {Eldon}}, \bibinfo {author} {\bibfnamefont {D.}~\bibnamefont {Ernst}}, \bibinfo {author} {\bibfnamefont {R.}~\bibnamefont {Hong}}, \bibinfo {author} {\bibfnamefont {G.~R.}\ \bibnamefont
  {McKee}}, \bibinfo {author} {\bibfnamefont {S.}~\bibnamefont {Mordijck}}, \bibinfo {author} {\bibfnamefont {O.}~\bibnamefont {Sauter}}, \bibinfo {author} {\bibfnamefont {L.}~\bibnamefont {Schmitz}}, \bibinfo {author} {\bibfnamefont {J.~L.}\ \bibnamefont {Barr}}, \bibinfo {author} {\bibfnamefont {M.~G.}\ \bibnamefont {Burke}}, \bibinfo {author} {\bibfnamefont {S.}~\bibnamefont {Coda}}, \bibinfo {author} {\bibfnamefont {T.~B.}\ \bibnamefont {Cote}}, \bibinfo {author} {\bibfnamefont {M.~E.}\ \bibnamefont {Fenstermacher}}, \bibinfo {author} {\bibfnamefont {A.}~\bibnamefont {Garofalo}}, \bibinfo {author} {\bibfnamefont {F.~O.}\ \bibnamefont {Khabanov}}, \bibinfo {author} {\bibfnamefont {G.~J.}\ \bibnamefont {Kramer}}, \bibinfo {author} {\bibfnamefont {C.~J.}\ \bibnamefont {Lasnier}}, \bibinfo {author} {\bibfnamefont {N.~C.}\ \bibnamefont {Logan}}, \bibinfo {author} {\bibfnamefont {P.}~\bibnamefont {Lunia}}, \bibinfo {author} {\bibfnamefont {A.~G.}\ \bibnamefont {McLean}}, \bibinfo {author} {\bibfnamefont
  {M.}~\bibnamefont {Okabayashi}}, \bibinfo {author} {\bibfnamefont {D.}~\bibnamefont {Shiraki}}, \bibinfo {author} {\bibfnamefont {S.}~\bibnamefont {Stewart}}, \bibinfo {author} {\bibfnamefont {Y.}~\bibnamefont {Takemura}}, \bibinfo {author} {\bibfnamefont {D.~D.}\ \bibnamefont {Truong}}, \bibinfo {author} {\bibfnamefont {T.}~\bibnamefont {Osborne}}, \bibinfo {author} {\bibfnamefont {M.~A.~V.}\ \bibnamefont {Zeeland}}, \bibinfo {author} {\bibfnamefont {B.~S.}\ \bibnamefont {Victor}}, \bibinfo {author} {\bibfnamefont {H.~Q.}\ \bibnamefont {Wang}}, \bibinfo {author} {\bibfnamefont {J.~G.}\ \bibnamefont {Watkins}}, \bibinfo {author} {\bibfnamefont {W.~P.}\ \bibnamefont {Wehner}}, \bibinfo {author} {\bibfnamefont {A.~S.}\ \bibnamefont {Welander}}, \bibinfo {author} {\bibfnamefont {T.~M.}\ \bibnamefont {Wilks}}, \bibinfo {author} {\bibfnamefont {J.}~\bibnamefont {Yang}}, \bibinfo {author} {\bibfnamefont {G.}~\bibnamefont {Yu}}, \bibinfo {author} {\bibfnamefont {L.}~\bibnamefont {Zeng}}, \ and\ \bibinfo {author}
  {\bibnamefont {the DIII-D~Team}},\ }\href {\doibase 10.1088/1361-6587/ad6f40} {\bibfield  {journal} {\bibinfo  {journal} {Plasma Physics and Controlled Fusion}\ }\textbf {\bibinfo {volume} {66}},\ \bibinfo {pages} {105018} (\bibinfo {year} {2024})}\BibitemShut {NoStop}%
\bibitem [{\citenamefont {Paz-Soldan}\ \emph {et~al.}(2024)\citenamefont {Paz-Soldan}, \citenamefont {Chrystal}, \citenamefont {Lunia}, \citenamefont {Nelson}, \citenamefont {Thome}, \citenamefont {Austin}, \citenamefont {Cote}, \citenamefont {Hyatt}, \citenamefont {Leuthold}, \citenamefont {Marinoni}, \citenamefont {Osborne}, \citenamefont {Pharr}, \citenamefont {Sauter}, \citenamefont {Scotti}, \citenamefont {Wilks},\ and\ \citenamefont {Wilson}}]{Paz-Soldan_2024}%
  \BibitemOpen
  \bibfield  {author} {\bibinfo {author} {\bibfnamefont {C.}~\bibnamefont {Paz-Soldan}}, \bibinfo {author} {\bibfnamefont {C.}~\bibnamefont {Chrystal}}, \bibinfo {author} {\bibfnamefont {P.}~\bibnamefont {Lunia}}, \bibinfo {author} {\bibfnamefont {A.}~\bibnamefont {Nelson}}, \bibinfo {author} {\bibfnamefont {K.}~\bibnamefont {Thome}}, \bibinfo {author} {\bibfnamefont {M.}~\bibnamefont {Austin}}, \bibinfo {author} {\bibfnamefont {T.}~\bibnamefont {Cote}}, \bibinfo {author} {\bibfnamefont {A.}~\bibnamefont {Hyatt}}, \bibinfo {author} {\bibfnamefont {N.}~\bibnamefont {Leuthold}}, \bibinfo {author} {\bibfnamefont {A.}~\bibnamefont {Marinoni}}, \bibinfo {author} {\bibfnamefont {T.}~\bibnamefont {Osborne}}, \bibinfo {author} {\bibfnamefont {M.}~\bibnamefont {Pharr}}, \bibinfo {author} {\bibfnamefont {O.}~\bibnamefont {Sauter}}, \bibinfo {author} {\bibfnamefont {F.}~\bibnamefont {Scotti}}, \bibinfo {author} {\bibfnamefont {T.}~\bibnamefont {Wilks}}, \ and\ \bibinfo {author} {\bibfnamefont {H.}~\bibnamefont
  {Wilson}},\ }\href {\doibase 10.1088/1741-4326/ad69a4} {\bibfield  {journal} {\bibinfo  {journal} {Nuclear Fusion}\ }\textbf {\bibinfo {volume} {64}},\ \bibinfo {pages} {094002} (\bibinfo {year} {2024})}\BibitemShut {NoStop}%
\bibitem [{\citenamefont {{The MANTA Collaboration}}\ \emph {et~al.}(2024)\citenamefont {{The MANTA Collaboration}}, \citenamefont {Rutherford}, \citenamefont {Wilson}, \citenamefont {Saltzman}, \citenamefont {Arnold}, \citenamefont {Ball}, \citenamefont {Benjamin}, \citenamefont {Bielajew}, \citenamefont {de~Boucaud}, \citenamefont {Calvo-Carrera}, \citenamefont {Chandra}, \citenamefont {Choudhury}, \citenamefont {Cummings}, \citenamefont {Corsaro}, \citenamefont {DaSilva}, \citenamefont {Diab}, \citenamefont {Devitre}, \citenamefont {Ferry}, \citenamefont {Frank}, \citenamefont {Hansen}, \citenamefont {Jerkins}, \citenamefont {Johnson}, \citenamefont {Lunia}, \citenamefont {van~de Lindt}, \citenamefont {Mackie}, \citenamefont {Maris}, \citenamefont {Mandell}, \citenamefont {Miller}, \citenamefont {Mouratidis}, \citenamefont {Nelson}, \citenamefont {Pharr}, \citenamefont {Peterson}, \citenamefont {Rodriguez-Fernandez}, \citenamefont {Segantin}, \citenamefont {Tobin}, \citenamefont {Velberg}, \citenamefont
  {Wang}, \citenamefont {Wigram}, \citenamefont {Witham}, \citenamefont {Paz-Soldan},\ and\ \citenamefont {Whyte}}]{MANTA_2024}%
  \BibitemOpen
  \bibfield  {author} {\bibinfo {author} {\bibnamefont {{The MANTA Collaboration}}}, \bibinfo {author} {\bibfnamefont {G.}~\bibnamefont {Rutherford}}, \bibinfo {author} {\bibfnamefont {H.~S.}\ \bibnamefont {Wilson}}, \bibinfo {author} {\bibfnamefont {A.}~\bibnamefont {Saltzman}}, \bibinfo {author} {\bibfnamefont {D.}~\bibnamefont {Arnold}}, \bibinfo {author} {\bibfnamefont {J.~L.}\ \bibnamefont {Ball}}, \bibinfo {author} {\bibfnamefont {S.}~\bibnamefont {Benjamin}}, \bibinfo {author} {\bibfnamefont {R.}~\bibnamefont {Bielajew}}, \bibinfo {author} {\bibfnamefont {N.}~\bibnamefont {de~Boucaud}}, \bibinfo {author} {\bibfnamefont {M.}~\bibnamefont {Calvo-Carrera}}, \bibinfo {author} {\bibfnamefont {R.}~\bibnamefont {Chandra}}, \bibinfo {author} {\bibfnamefont {H.}~\bibnamefont {Choudhury}}, \bibinfo {author} {\bibfnamefont {C.}~\bibnamefont {Cummings}}, \bibinfo {author} {\bibfnamefont {L.}~\bibnamefont {Corsaro}}, \bibinfo {author} {\bibfnamefont {N.}~\bibnamefont {DaSilva}}, \bibinfo {author} {\bibfnamefont
  {R.}~\bibnamefont {Diab}}, \bibinfo {author} {\bibfnamefont {A.~R.}\ \bibnamefont {Devitre}}, \bibinfo {author} {\bibfnamefont {S.}~\bibnamefont {Ferry}}, \bibinfo {author} {\bibfnamefont {S.~J.}\ \bibnamefont {Frank}}, \bibinfo {author} {\bibfnamefont {C.~J.}\ \bibnamefont {Hansen}}, \bibinfo {author} {\bibfnamefont {J.}~\bibnamefont {Jerkins}}, \bibinfo {author} {\bibfnamefont {J.~D.}\ \bibnamefont {Johnson}}, \bibinfo {author} {\bibfnamefont {P.}~\bibnamefont {Lunia}}, \bibinfo {author} {\bibfnamefont {J.}~\bibnamefont {van~de Lindt}}, \bibinfo {author} {\bibfnamefont {S.}~\bibnamefont {Mackie}}, \bibinfo {author} {\bibfnamefont {A.~D.}\ \bibnamefont {Maris}}, \bibinfo {author} {\bibfnamefont {N.~R.}\ \bibnamefont {Mandell}}, \bibinfo {author} {\bibfnamefont {M.~A.}\ \bibnamefont {Miller}}, \bibinfo {author} {\bibfnamefont {T.}~\bibnamefont {Mouratidis}}, \bibinfo {author} {\bibfnamefont {A.~O.}\ \bibnamefont {Nelson}}, \bibinfo {author} {\bibfnamefont {M.}~\bibnamefont {Pharr}}, \bibinfo {author}
  {\bibfnamefont {E.~E.}\ \bibnamefont {Peterson}}, \bibinfo {author} {\bibfnamefont {P.}~\bibnamefont {Rodriguez-Fernandez}}, \bibinfo {author} {\bibfnamefont {S.}~\bibnamefont {Segantin}}, \bibinfo {author} {\bibfnamefont {M.}~\bibnamefont {Tobin}}, \bibinfo {author} {\bibfnamefont {A.}~\bibnamefont {Velberg}}, \bibinfo {author} {\bibfnamefont {A.~M.}\ \bibnamefont {Wang}}, \bibinfo {author} {\bibfnamefont {M.}~\bibnamefont {Wigram}}, \bibinfo {author} {\bibfnamefont {J.}~\bibnamefont {Witham}}, \bibinfo {author} {\bibfnamefont {C.}~\bibnamefont {Paz-Soldan}}, \ and\ \bibinfo {author} {\bibfnamefont {D.~G.}\ \bibnamefont {Whyte}},\ }\href {\doibase 10.1088/1361-6587/ad6708} {\bibfield  {journal} {\bibinfo  {journal} {Plasma Physics and Controlled Fusion}\ }\textbf {\bibinfo {volume} {66}},\ \bibinfo {pages} {105006} (\bibinfo {year} {2024})}\BibitemShut {NoStop}%
\bibitem [{\citenamefont {Wilson}\ \emph {et~al.}(2024)\citenamefont {Wilson}, \citenamefont {Nelson}, \citenamefont {McClenaghan}, \citenamefont {Rodriguez-Fernandez}, \citenamefont {Parisi},\ and\ \citenamefont {Paz-Soldan}}]{Wilson_2025}%
  \BibitemOpen
  \bibfield  {author} {\bibinfo {author} {\bibfnamefont {H.~S.}\ \bibnamefont {Wilson}}, \bibinfo {author} {\bibfnamefont {A.~O.}\ \bibnamefont {Nelson}}, \bibinfo {author} {\bibfnamefont {J.}~\bibnamefont {McClenaghan}}, \bibinfo {author} {\bibfnamefont {P.}~\bibnamefont {Rodriguez-Fernandez}}, \bibinfo {author} {\bibfnamefont {J.}~\bibnamefont {Parisi}}, \ and\ \bibinfo {author} {\bibfnamefont {C.}~\bibnamefont {Paz-Soldan}},\ }\href {\doibase 10.1088/1361-6587/ad9be5} {\bibfield  {journal} {\bibinfo  {journal} {Plasma Physics and Controlled Fusion}\ }\textbf {\bibinfo {volume} {67}},\ \bibinfo {pages} {015026} (\bibinfo {year} {2024})}\BibitemShut {NoStop}%
\bibitem [{\citenamefont {Marinoni}\ \emph {et~al.}(2009)\citenamefont {Marinoni}, \citenamefont {Brunner}, \citenamefont {Camenen}, \citenamefont {Coda}, \citenamefont {Graves}, \citenamefont {Lapillonne}, \citenamefont {Pochelon}, \citenamefont {Sauter},\ and\ \citenamefont {Villard}}]{Marinoni_2009}%
  \BibitemOpen
  \bibfield  {author} {\bibinfo {author} {\bibfnamefont {A.}~\bibnamefont {Marinoni}}, \bibinfo {author} {\bibfnamefont {S.}~\bibnamefont {Brunner}}, \bibinfo {author} {\bibfnamefont {Y.}~\bibnamefont {Camenen}}, \bibinfo {author} {\bibfnamefont {S.}~\bibnamefont {Coda}}, \bibinfo {author} {\bibfnamefont {J.~P.}\ \bibnamefont {Graves}}, \bibinfo {author} {\bibfnamefont {X.}~\bibnamefont {Lapillonne}}, \bibinfo {author} {\bibfnamefont {A.}~\bibnamefont {Pochelon}}, \bibinfo {author} {\bibfnamefont {O.}~\bibnamefont {Sauter}}, \ and\ \bibinfo {author} {\bibfnamefont {L.}~\bibnamefont {Villard}},\ }\href {\doibase 10.1088/0741-3335/51/5/055016} {\bibfield  {journal} {\bibinfo  {journal} {Plasma Physics and Controlled Fusion}\ }\textbf {\bibinfo {volume} {51}},\ \bibinfo {pages} {055016} (\bibinfo {year} {2009})}\BibitemShut {NoStop}%
\bibitem [{\citenamefont {Merlo}\ \emph {et~al.}(2019)\citenamefont {Merlo}, \citenamefont {Fontana}, \citenamefont {Coda}, \citenamefont {Hatch}, \citenamefont {Janhunen}, \citenamefont {Porte},\ and\ \citenamefont {Jenko}}]{5}%
  \BibitemOpen
  \bibfield  {author} {\bibinfo {author} {\bibfnamefont {G.}~\bibnamefont {Merlo}}, \bibinfo {author} {\bibfnamefont {M.}~\bibnamefont {Fontana}}, \bibinfo {author} {\bibfnamefont {S.}~\bibnamefont {Coda}}, \bibinfo {author} {\bibfnamefont {D.}~\bibnamefont {Hatch}}, \bibinfo {author} {\bibfnamefont {S.}~\bibnamefont {Janhunen}}, \bibinfo {author} {\bibfnamefont {L.}~\bibnamefont {Porte}}, \ and\ \bibinfo {author} {\bibfnamefont {F.}~\bibnamefont {Jenko}},\ }\href {\doibase 10.1063/1.5115390} {\bibfield  {journal} {\bibinfo  {journal} {Physics of Plasmas}\ }\textbf {\bibinfo {volume} {26}},\ \bibinfo {pages} {102302} (\bibinfo {year} {2019})},\ \Eprint {http://arxiv.org/abs/https://doi.org/10.1063/1.5115390} {https://doi.org/10.1063/1.5115390} \BibitemShut {NoStop}%
\bibitem [{\citenamefont {Balestri}\ \emph {et~al.}(2024{\natexlab{b}})\citenamefont {Balestri}, \citenamefont {Ball}, \citenamefont {Coda}, \citenamefont {Cruz-Zabala}, \citenamefont {Garcia-Munoz},\ and\ \citenamefont {Viezzer}}]{Balestri_2024_2}%
  \BibitemOpen
  \bibfield  {author} {\bibinfo {author} {\bibfnamefont {A.}~\bibnamefont {Balestri}}, \bibinfo {author} {\bibfnamefont {J.}~\bibnamefont {Ball}}, \bibinfo {author} {\bibfnamefont {S.}~\bibnamefont {Coda}}, \bibinfo {author} {\bibfnamefont {D.~J.}\ \bibnamefont {Cruz-Zabala}}, \bibinfo {author} {\bibfnamefont {M.}~\bibnamefont {Garcia-Munoz}}, \ and\ \bibinfo {author} {\bibfnamefont {E.}~\bibnamefont {Viezzer}},\ }\href {\doibase 10.1088/1361-6587/ad4d1d} {\bibfield  {journal} {\bibinfo  {journal} {Plasma Physics and Controlled Fusion}\ }\textbf {\bibinfo {volume} {66}},\ \bibinfo {pages} {075012} (\bibinfo {year} {2024}{\natexlab{b}})}\BibitemShut {NoStop}%
\bibitem [{\citenamefont {Giannatale}\ \emph {et~al.}(2024)\citenamefont {Giannatale}, \citenamefont {Bottino}, \citenamefont {Brunner}, \citenamefont {Murugappan},\ and\ \citenamefont {Villard}}]{Di_Giannatale_2024}%
  \BibitemOpen
  \bibfield  {author} {\bibinfo {author} {\bibfnamefont {G.~D.}\ \bibnamefont {Giannatale}}, \bibinfo {author} {\bibfnamefont {A.}~\bibnamefont {Bottino}}, \bibinfo {author} {\bibfnamefont {S.}~\bibnamefont {Brunner}}, \bibinfo {author} {\bibfnamefont {M.}~\bibnamefont {Murugappan}}, \ and\ \bibinfo {author} {\bibfnamefont {L.}~\bibnamefont {Villard}},\ }\href {\doibase 10.1088/1361-6587/ad5df9} {\bibfield  {journal} {\bibinfo  {journal} {Plasma Physics and Controlled Fusion}\ }\textbf {\bibinfo {volume} {66}},\ \bibinfo {pages} {095003} (\bibinfo {year} {2024})}\BibitemShut {NoStop}%
\bibitem [{\citenamefont {Hoffmann}\ \emph {et~al.}(2024)\citenamefont {Hoffmann}, \citenamefont {Balestri},\ and\ \citenamefont {Ricci}}]{Hoffmann_2025}%
  \BibitemOpen
  \bibfield  {author} {\bibinfo {author} {\bibfnamefont {A.~C.~D.}\ \bibnamefont {Hoffmann}}, \bibinfo {author} {\bibfnamefont {A.}~\bibnamefont {Balestri}}, \ and\ \bibinfo {author} {\bibfnamefont {P.}~\bibnamefont {Ricci}},\ }\href {\doibase 10.1088/1361-6587/ad9e6f} {\bibfield  {journal} {\bibinfo  {journal} {Plasma Physics and Controlled Fusion}\ }\textbf {\bibinfo {volume} {67}},\ \bibinfo {pages} {015031} (\bibinfo {year} {2024})}\BibitemShut {NoStop}%
\bibitem [{\citenamefont {Merlo}\ and\ \citenamefont {Jenko}(2023)}]{merlo_jenko_2023}%
  \BibitemOpen
  \bibfield  {author} {\bibinfo {author} {\bibfnamefont {G.}~\bibnamefont {Merlo}}\ and\ \bibinfo {author} {\bibfnamefont {F.}~\bibnamefont {Jenko}},\ }\href {\doibase 10.1017/S0022377822001076} {\bibfield  {journal} {\bibinfo  {journal} {Journal of Plasma Physics}\ }\textbf {\bibinfo {volume} {89}},\ \bibinfo {pages} {905890104} (\bibinfo {year} {2023})}\BibitemShut {NoStop}%
\bibitem [{\citenamefont {Merlo}\ \emph {et~al.}(2023)\citenamefont {Merlo}, \citenamefont {Dicorato}, \citenamefont {Allen}, \citenamefont {Dannert}, \citenamefont {Germaschewski},\ and\ \citenamefont {Jenko}}]{Merlo_ITG}%
  \BibitemOpen
  \bibfield  {author} {\bibinfo {author} {\bibfnamefont {G.}~\bibnamefont {Merlo}}, \bibinfo {author} {\bibfnamefont {M.}~\bibnamefont {Dicorato}}, \bibinfo {author} {\bibfnamefont {B.}~\bibnamefont {Allen}}, \bibinfo {author} {\bibfnamefont {T.}~\bibnamefont {Dannert}}, \bibinfo {author} {\bibfnamefont {K.}~\bibnamefont {Germaschewski}}, \ and\ \bibinfo {author} {\bibfnamefont {F.}~\bibnamefont {Jenko}},\ }\href {\doibase 10.1063/5.0167292} {\bibfield  {journal} {\bibinfo  {journal} {Physics of Plasmas}\ }\textbf {\bibinfo {volume} {30}},\ \bibinfo {pages} {102302} (\bibinfo {year} {2023})},\ \Eprint {http://arxiv.org/abs/https://pubs.aip.org/aip/pop/article-pdf/doi/10.1063/5.0167292/18156680/102302\_1\_5.0167292.pdf} {https://pubs.aip.org/aip/pop/article-pdf/doi/10.1063/5.0167292/18156680/102302\_1\_5.0167292.pdf} \BibitemShut {NoStop}%
\bibitem [{\citenamefont {Pueschel}\ \emph {et~al.}(2025)\citenamefont {Pueschel}, \citenamefont {Mulholland}, \citenamefont {Sauter}, \citenamefont {Coosemans}, \citenamefont {Ball}, \citenamefont {Hamed}, \citenamefont {Vallar},\ and\ \citenamefont {the TCV~Team}}]{Pueschel_2025}%
  \BibitemOpen
  \bibfield  {author} {\bibinfo {author} {\bibfnamefont {M.~J.}\ \bibnamefont {Pueschel}}, \bibinfo {author} {\bibfnamefont {P.}~\bibnamefont {Mulholland}}, \bibinfo {author} {\bibfnamefont {O.}~\bibnamefont {Sauter}}, \bibinfo {author} {\bibfnamefont {R.}~\bibnamefont {Coosemans}}, \bibinfo {author} {\bibfnamefont {J.}~\bibnamefont {Ball}}, \bibinfo {author} {\bibfnamefont {M.}~\bibnamefont {Hamed}}, \bibinfo {author} {\bibfnamefont {M.}~\bibnamefont {Vallar}}, \ and\ \bibinfo {author} {\bibnamefont {the TCV~Team}},\ }\href {\doibase 10.1063/5.0255557} {\bibfield  {journal} {\bibinfo  {journal} {Physics of Plasmas}\ }\textbf {\bibinfo {volume} {32}},\ \bibinfo {pages} {052302} (\bibinfo {year} {2025})},\ \Eprint {http://arxiv.org/abs/https://pubs.aip.org/aip/pop/article-pdf/doi/10.1063/5.0255557/20520339/052302\_1\_5.0255557.pdf} {https://pubs.aip.org/aip/pop/article-pdf/doi/10.1063/5.0255557/20520339/052302\_1\_5.0255557.pdf} \BibitemShut {NoStop}%
\bibitem [{\citenamefont {Guttenfelder}\ \emph {et~al.}(2011)\citenamefont {Guttenfelder}, \citenamefont {Candy}, \citenamefont {Kaye}, \citenamefont {Nevins}, \citenamefont {Wang}, \citenamefont {Bell}, \citenamefont {Hammett}, \citenamefont {LeBlanc}, \citenamefont {Mikkelsen},\ and\ \citenamefont {Yuh}}]{Guttenfelder_2011}%
  \BibitemOpen
  \bibfield  {author} {\bibinfo {author} {\bibfnamefont {W.}~\bibnamefont {Guttenfelder}}, \bibinfo {author} {\bibfnamefont {J.}~\bibnamefont {Candy}}, \bibinfo {author} {\bibfnamefont {S.~M.}\ \bibnamefont {Kaye}}, \bibinfo {author} {\bibfnamefont {W.~M.}\ \bibnamefont {Nevins}}, \bibinfo {author} {\bibfnamefont {E.}~\bibnamefont {Wang}}, \bibinfo {author} {\bibfnamefont {R.~E.}\ \bibnamefont {Bell}}, \bibinfo {author} {\bibfnamefont {G.~W.}\ \bibnamefont {Hammett}}, \bibinfo {author} {\bibfnamefont {B.~P.}\ \bibnamefont {LeBlanc}}, \bibinfo {author} {\bibfnamefont {D.~R.}\ \bibnamefont {Mikkelsen}}, \ and\ \bibinfo {author} {\bibfnamefont {H.}~\bibnamefont {Yuh}},\ }\href {\doibase 10.1103/PhysRevLett.106.155004} {\bibfield  {journal} {\bibinfo  {journal} {Phys. Rev. Lett.}\ }\textbf {\bibinfo {volume} {106}},\ \bibinfo {pages} {155004} (\bibinfo {year} {2011})}\BibitemShut {NoStop}%
\bibitem [{\citenamefont {Doerk}\ \emph {et~al.}(2011)\citenamefont {Doerk}, \citenamefont {Jenko}, \citenamefont {Pueschel},\ and\ \citenamefont {Hatch}}]{2011_Doerk}%
  \BibitemOpen
  \bibfield  {author} {\bibinfo {author} {\bibfnamefont {H.}~\bibnamefont {Doerk}}, \bibinfo {author} {\bibfnamefont {F.}~\bibnamefont {Jenko}}, \bibinfo {author} {\bibfnamefont {M.~J.}\ \bibnamefont {Pueschel}}, \ and\ \bibinfo {author} {\bibfnamefont {D.~R.}\ \bibnamefont {Hatch}},\ }\href {\doibase 10.1103/PhysRevLett.106.155003} {\bibfield  {journal} {\bibinfo  {journal} {Phys. Rev. Lett.}\ }\textbf {\bibinfo {volume} {106}},\ \bibinfo {pages} {155003} (\bibinfo {year} {2011})}\BibitemShut {NoStop}%
\bibitem [{\citenamefont {Doerk}\ \emph {et~al.}(2012)\citenamefont {Doerk}, \citenamefont {Jenko}, \citenamefont {Görler}, \citenamefont {Told}, \citenamefont {Pueschel},\ and\ \citenamefont {Hatch}}]{Doerk_2012}%
  \BibitemOpen
  \bibfield  {author} {\bibinfo {author} {\bibfnamefont {H.}~\bibnamefont {Doerk}}, \bibinfo {author} {\bibfnamefont {F.}~\bibnamefont {Jenko}}, \bibinfo {author} {\bibfnamefont {T.}~\bibnamefont {Görler}}, \bibinfo {author} {\bibfnamefont {D.}~\bibnamefont {Told}}, \bibinfo {author} {\bibfnamefont {M.~J.}\ \bibnamefont {Pueschel}}, \ and\ \bibinfo {author} {\bibfnamefont {D.~R.}\ \bibnamefont {Hatch}},\ }\href {\doibase 10.1063/1.3694663} {\bibfield  {journal} {\bibinfo  {journal} {Physics of Plasmas}\ }\textbf {\bibinfo {volume} {19}},\ \bibinfo {pages} {055907} (\bibinfo {year} {2012})},\ \Eprint {http://arxiv.org/abs/https://pubs.aip.org/aip/pop/article-pdf/doi/10.1063/1.3694663/13988983/055907\_1\_online.pdf} {https://pubs.aip.org/aip/pop/article-pdf/doi/10.1063/1.3694663/13988983/055907\_1\_online.pdf} \BibitemShut {NoStop}%
\bibitem [{\citenamefont {Maeyama}\ \emph {et~al.}(2017)\citenamefont {Maeyama}, \citenamefont {Watanabe},\ and\ \citenamefont {Ishizawa}}]{Maeyama_2017}%
  \BibitemOpen
  \bibfield  {author} {\bibinfo {author} {\bibfnamefont {S.}~\bibnamefont {Maeyama}}, \bibinfo {author} {\bibfnamefont {T.-H.}\ \bibnamefont {Watanabe}}, \ and\ \bibinfo {author} {\bibfnamefont {A.}~\bibnamefont {Ishizawa}},\ }\href {\doibase 10.1103/PhysRevLett.119.195002} {\bibfield  {journal} {\bibinfo  {journal} {Phys. Rev. Lett.}\ }\textbf {\bibinfo {volume} {119}},\ \bibinfo {pages} {195002} (\bibinfo {year} {2017})}\BibitemShut {NoStop}%
\bibitem [{\citenamefont {Hamed}\ \emph {et~al.}(2023)\citenamefont {Hamed}, \citenamefont {Pueschel}, \citenamefont {Citrin}, \citenamefont {Muraglia}, \citenamefont {Garbet},\ and\ \citenamefont {Camenen}}]{Hamed_2023}%
  \BibitemOpen
  \bibfield  {author} {\bibinfo {author} {\bibfnamefont {M.}~\bibnamefont {Hamed}}, \bibinfo {author} {\bibfnamefont {M.~J.}\ \bibnamefont {Pueschel}}, \bibinfo {author} {\bibfnamefont {J.}~\bibnamefont {Citrin}}, \bibinfo {author} {\bibfnamefont {M.}~\bibnamefont {Muraglia}}, \bibinfo {author} {\bibfnamefont {X.}~\bibnamefont {Garbet}}, \ and\ \bibinfo {author} {\bibfnamefont {Y.}~\bibnamefont {Camenen}},\ }\href {\doibase 10.1063/5.0104879} {\bibfield  {journal} {\bibinfo  {journal} {Physics of Plasmas}\ }\textbf {\bibinfo {volume} {30}},\ \bibinfo {pages} {042303} (\bibinfo {year} {2023})},\ \Eprint {http://arxiv.org/abs/https://pubs.aip.org/aip/pop/article-pdf/doi/10.1063/5.0104879/16831647/042303\_1\_5.0104879.pdf} {https://pubs.aip.org/aip/pop/article-pdf/doi/10.1063/5.0104879/16831647/042303\_1\_5.0104879.pdf} \BibitemShut {NoStop}%
\bibitem [{\citenamefont {Hazeltine}\ \emph {et~al.}(1975)\citenamefont {Hazeltine}, \citenamefont {Dobrott},\ and\ \citenamefont {Wang}}]{Hazeltine_1975}%
  \BibitemOpen
  \bibfield  {author} {\bibinfo {author} {\bibfnamefont {R.~D.}\ \bibnamefont {Hazeltine}}, \bibinfo {author} {\bibfnamefont {D.}~\bibnamefont {Dobrott}}, \ and\ \bibinfo {author} {\bibfnamefont {T.~S.}\ \bibnamefont {Wang}},\ }\href {\doibase 10.1063/1.861097} {\bibfield  {journal} {\bibinfo  {journal} {The Physics of Fluids}\ }\textbf {\bibinfo {volume} {18}},\ \bibinfo {pages} {1778} (\bibinfo {year} {1975})},\ \Eprint {http://arxiv.org/abs/https://pubs.aip.org/aip/pfl/article-pdf/18/12/1778/12677837/1778\_1\_online.pdf} {https://pubs.aip.org/aip/pfl/article-pdf/18/12/1778/12677837/1778\_1\_online.pdf} \BibitemShut {NoStop}%
\bibitem [{\citenamefont {Gladd}\ \emph {et~al.}(1980)\citenamefont {Gladd}, \citenamefont {Drake}, \citenamefont {Chang},\ and\ \citenamefont {Liu}}]{1980_Gladd}%
  \BibitemOpen
  \bibfield  {author} {\bibinfo {author} {\bibfnamefont {N.~T.}\ \bibnamefont {Gladd}}, \bibinfo {author} {\bibfnamefont {J.~F.}\ \bibnamefont {Drake}}, \bibinfo {author} {\bibfnamefont {C.~L.}\ \bibnamefont {Chang}}, \ and\ \bibinfo {author} {\bibfnamefont {C.~S.}\ \bibnamefont {Liu}},\ }\href {\doibase 10.1063/1.863119} {\bibfield  {journal} {\bibinfo  {journal} {The Physics of Fluids}\ }\textbf {\bibinfo {volume} {23}},\ \bibinfo {pages} {1182} (\bibinfo {year} {1980})},\ \Eprint {http://arxiv.org/abs/https://pubs.aip.org/aip/pfl/article-pdf/23/6/1182/12463057/1182\_1\_online.pdf} {https://pubs.aip.org/aip/pfl/article-pdf/23/6/1182/12463057/1182\_1\_online.pdf} \BibitemShut {NoStop}%
\bibitem [{\citenamefont {Drake}\ and\ \citenamefont {Lee}(1977)}]{Drake_1977}%
  \BibitemOpen
  \bibfield  {author} {\bibinfo {author} {\bibfnamefont {J.~F.}\ \bibnamefont {Drake}}\ and\ \bibinfo {author} {\bibfnamefont {Y.~C.}\ \bibnamefont {Lee}},\ }\href {\doibase 10.1063/1.862017} {\bibfield  {journal} {\bibinfo  {journal} {The Physics of Fluids}\ }\textbf {\bibinfo {volume} {20}},\ \bibinfo {pages} {1341} (\bibinfo {year} {1977})},\ \Eprint {http://arxiv.org/abs/https://pubs.aip.org/aip/pfl/article-pdf/20/8/1341/12612734/1341\_1\_online.pdf} {https://pubs.aip.org/aip/pfl/article-pdf/20/8/1341/12612734/1341\_1\_online.pdf} \BibitemShut {NoStop}%
\bibitem [{\citenamefont {Catto}\ and\ \citenamefont {Rosenbluth}(1981)}]{catto1981trapped}%
  \BibitemOpen
  \bibfield  {author} {\bibinfo {author} {\bibfnamefont {P.~J.}\ \bibnamefont {Catto}}\ and\ \bibinfo {author} {\bibfnamefont {M.}~\bibnamefont {Rosenbluth}},\ }\href@noop {} {\bibfield  {journal} {\bibinfo  {journal} {The Physics of Fluids}\ }\textbf {\bibinfo {volume} {24}},\ \bibinfo {pages} {243} (\bibinfo {year} {1981})}\BibitemShut {NoStop}%
\bibitem [{\citenamefont {Garbet}\ \emph {et~al.}(1990)\citenamefont {Garbet}, \citenamefont {Mourgues},\ and\ \citenamefont {Samain}}]{1990_Garbet}%
  \BibitemOpen
  \bibfield  {author} {\bibinfo {author} {\bibfnamefont {X.}~\bibnamefont {Garbet}}, \bibinfo {author} {\bibfnamefont {F.}~\bibnamefont {Mourgues}}, \ and\ \bibinfo {author} {\bibfnamefont {A.}~\bibnamefont {Samain}},\ }\href {\doibase 10.1088/0741-3335/32/2/004} {\bibfield  {journal} {\bibinfo  {journal} {Plasma Physics and Controlled Fusion}\ }\textbf {\bibinfo {volume} {32}},\ \bibinfo {pages} {131} (\bibinfo {year} {1990})}\BibitemShut {NoStop}%
\bibitem [{\citenamefont {Applegate}\ \emph {et~al.}(2007)\citenamefont {Applegate}, \citenamefont {Roach}, \citenamefont {Connor}, \citenamefont {Cowley}, \citenamefont {Dorland}, \citenamefont {Hastie},\ and\ \citenamefont {Joiner}}]{Applegate_2007}%
  \BibitemOpen
  \bibfield  {author} {\bibinfo {author} {\bibfnamefont {D.~J.}\ \bibnamefont {Applegate}}, \bibinfo {author} {\bibfnamefont {C.~M.}\ \bibnamefont {Roach}}, \bibinfo {author} {\bibfnamefont {J.~W.}\ \bibnamefont {Connor}}, \bibinfo {author} {\bibfnamefont {S.~C.}\ \bibnamefont {Cowley}}, \bibinfo {author} {\bibfnamefont {W.}~\bibnamefont {Dorland}}, \bibinfo {author} {\bibfnamefont {R.~J.}\ \bibnamefont {Hastie}}, \ and\ \bibinfo {author} {\bibfnamefont {N.}~\bibnamefont {Joiner}},\ }\href {\doibase 10.1088/0741-3335/49/8/001} {\bibfield  {journal} {\bibinfo  {journal} {Plasma Physics and Controlled Fusion}\ }\textbf {\bibinfo {volume} {49}},\ \bibinfo {pages} {1113} (\bibinfo {year} {2007})}\BibitemShut {NoStop}%
\bibitem [{\citenamefont {Hamed}\ \emph {et~al.}(2019)\citenamefont {Hamed}, \citenamefont {Muraglia}, \citenamefont {Camenen}, \citenamefont {Garbet},\ and\ \citenamefont {Agullo}}]{2019_Hamed}%
  \BibitemOpen
  \bibfield  {author} {\bibinfo {author} {\bibfnamefont {M.}~\bibnamefont {Hamed}}, \bibinfo {author} {\bibfnamefont {M.}~\bibnamefont {Muraglia}}, \bibinfo {author} {\bibfnamefont {Y.}~\bibnamefont {Camenen}}, \bibinfo {author} {\bibfnamefont {X.}~\bibnamefont {Garbet}}, \ and\ \bibinfo {author} {\bibfnamefont {O.}~\bibnamefont {Agullo}},\ }\href {\doibase 10.1063/1.5111701} {\bibfield  {journal} {\bibinfo  {journal} {Physics of Plasmas}\ }\textbf {\bibinfo {volume} {26}},\ \bibinfo {pages} {092506} (\bibinfo {year} {2019})},\ \Eprint {http://arxiv.org/abs/https://pubs.aip.org/aip/pop/article-pdf/doi/10.1063/1.5111701/14850651/092506\_1\_online.pdf} {https://pubs.aip.org/aip/pop/article-pdf/doi/10.1063/1.5111701/14850651/092506\_1\_online.pdf} \BibitemShut {NoStop}%
\bibitem [{\citenamefont {Miller}\ \emph {et~al.}(1998)\citenamefont {Miller}, \citenamefont {Chu}, \citenamefont {Greene}, \citenamefont {Lin-Liu},\ and\ \citenamefont {Waltz}}]{miller}%
  \BibitemOpen
  \bibfield  {author} {\bibinfo {author} {\bibfnamefont {R.~L.}\ \bibnamefont {Miller}}, \bibinfo {author} {\bibfnamefont {M.~S.}\ \bibnamefont {Chu}}, \bibinfo {author} {\bibfnamefont {J.~M.}\ \bibnamefont {Greene}}, \bibinfo {author} {\bibfnamefont {Y.~R.}\ \bibnamefont {Lin-Liu}}, \ and\ \bibinfo {author} {\bibfnamefont {R.~E.}\ \bibnamefont {Waltz}},\ }\href {\doibase 10.1063/1.872666} {\bibfield  {journal} {\bibinfo  {journal} {Physics of Plasmas}\ }\textbf {\bibinfo {volume} {5}},\ \bibinfo {pages} {973} (\bibinfo {year} {1998})},\ \Eprint {http://arxiv.org/abs/https://doi.org/10.1063/1.872666} {https://doi.org/10.1063/1.872666} \BibitemShut {NoStop}%
\bibitem [{\citenamefont {Candy}(2009)}]{Candy_2009}%
  \BibitemOpen
  \bibfield  {author} {\bibinfo {author} {\bibfnamefont {J.}~\bibnamefont {Candy}},\ }\href {\doibase 10.1088/0741-3335/51/10/105009} {\bibfield  {journal} {\bibinfo  {journal} {Plasma Physics and Controlled Fusion}\ }\textbf {\bibinfo {volume} {51}},\ \bibinfo {pages} {105009} (\bibinfo {year} {2009})}\BibitemShut {NoStop}%
\bibitem [{\citenamefont {Parisi}\ \emph {et~al.}(2024)\citenamefont {Parisi}, \citenamefont {Guttenfelder}, \citenamefont {Nelson}, \citenamefont {Gaur}, \citenamefont {Kleiner}, \citenamefont {Lampert}, \citenamefont {Avdeeva}, \citenamefont {Berkery}, \citenamefont {Clauser}, \citenamefont {Curie}, \citenamefont {Diallo}, \citenamefont {Dorland}, \citenamefont {Kaye}, \citenamefont {McClenaghan},\ and\ \citenamefont {Parra}}]{Parisi_2024}%
  \BibitemOpen
  \bibfield  {author} {\bibinfo {author} {\bibfnamefont {J.}~\bibnamefont {Parisi}}, \bibinfo {author} {\bibfnamefont {W.}~\bibnamefont {Guttenfelder}}, \bibinfo {author} {\bibfnamefont {A.}~\bibnamefont {Nelson}}, \bibinfo {author} {\bibfnamefont {R.}~\bibnamefont {Gaur}}, \bibinfo {author} {\bibfnamefont {A.}~\bibnamefont {Kleiner}}, \bibinfo {author} {\bibfnamefont {M.}~\bibnamefont {Lampert}}, \bibinfo {author} {\bibfnamefont {G.}~\bibnamefont {Avdeeva}}, \bibinfo {author} {\bibfnamefont {J.}~\bibnamefont {Berkery}}, \bibinfo {author} {\bibfnamefont {C.}~\bibnamefont {Clauser}}, \bibinfo {author} {\bibfnamefont {M.}~\bibnamefont {Curie}}, \bibinfo {author} {\bibfnamefont {A.}~\bibnamefont {Diallo}}, \bibinfo {author} {\bibfnamefont {W.}~\bibnamefont {Dorland}}, \bibinfo {author} {\bibfnamefont {S.}~\bibnamefont {Kaye}}, \bibinfo {author} {\bibfnamefont {J.}~\bibnamefont {McClenaghan}}, \ and\ \bibinfo {author} {\bibfnamefont {F.}~\bibnamefont {Parra}},\ }\href {\doibase 10.1088/1741-4326/ad39fb}
  {\bibfield  {journal} {\bibinfo  {journal} {Nuclear Fusion}\ }\textbf {\bibinfo {volume} {64}},\ \bibinfo {pages} {054002} (\bibinfo {year} {2024})}\BibitemShut {NoStop}%
\bibitem [{\citenamefont {Doyle}\ \emph {et~al.}(2021{\natexlab{a}})\citenamefont {Doyle}, \citenamefont {Mancini}, \citenamefont {Agredano-Torres}, \citenamefont {Garcia-Sanchez}, \citenamefont {Segado-Fernandez}, \citenamefont {Ayllon-Guerola}, \citenamefont {Garcia-Munoz}, \citenamefont {Viezzer}, \citenamefont {Garcia-Lopez}, \citenamefont {Hwang},\ and\ \citenamefont {Chung}}]{SMART1}%
  \BibitemOpen
  \bibfield  {author} {\bibinfo {author} {\bibfnamefont {S.~J.}\ \bibnamefont {Doyle}}, \bibinfo {author} {\bibfnamefont {A.}~\bibnamefont {Mancini}}, \bibinfo {author} {\bibfnamefont {M.}~\bibnamefont {Agredano-Torres}}, \bibinfo {author} {\bibfnamefont {J.~L.}\ \bibnamefont {Garcia-Sanchez}}, \bibinfo {author} {\bibfnamefont {J.}~\bibnamefont {Segado-Fernandez}}, \bibinfo {author} {\bibfnamefont {J.}~\bibnamefont {Ayllon-Guerola}}, \bibinfo {author} {\bibfnamefont {M.}~\bibnamefont {Garcia-Munoz}}, \bibinfo {author} {\bibfnamefont {E.}~\bibnamefont {Viezzer}}, \bibinfo {author} {\bibfnamefont {J.}~\bibnamefont {Garcia-Lopez}}, \bibinfo {author} {\bibfnamefont {Y.~S.}\ \bibnamefont {Hwang}}, \ and\ \bibinfo {author} {\bibfnamefont {K.~J.}\ \bibnamefont {Chung}},\ }\href {\doibase 10.1088/2516-1067/ac2a0e} {\bibfield  {journal} {\bibinfo  {journal} {Plasma Research Express}\ }\textbf {\bibinfo {volume} {3}},\ \bibinfo {pages} {044001} (\bibinfo {year} {2021}{\natexlab{a}})}\BibitemShut {NoStop}%
\bibitem [{\citenamefont {Doyle}\ \emph {et~al.}(2021{\natexlab{b}})\citenamefont {Doyle}, \citenamefont {Lopez-Aires}, \citenamefont {Mancini}, \citenamefont {Agredano-Torres}, \citenamefont {Garcia-Sanchez}, \citenamefont {Segado-Fernandez}, \citenamefont {Ayllon-Guerola}, \citenamefont {Garcia-Muñoz}, \citenamefont {Viezzer}, \citenamefont {Soria-Hoyo}, \citenamefont {Garcia-Lopez}, \citenamefont {Cunningham}, \citenamefont {Buxton}, \citenamefont {Gryaznevich}, \citenamefont {Hwang},\ and\ \citenamefont {Chung}}]{SMART2}%
  \BibitemOpen
  \bibfield  {author} {\bibinfo {author} {\bibfnamefont {S.}~\bibnamefont {Doyle}}, \bibinfo {author} {\bibfnamefont {D.}~\bibnamefont {Lopez-Aires}}, \bibinfo {author} {\bibfnamefont {A.}~\bibnamefont {Mancini}}, \bibinfo {author} {\bibfnamefont {M.}~\bibnamefont {Agredano-Torres}}, \bibinfo {author} {\bibfnamefont {J.}~\bibnamefont {Garcia-Sanchez}}, \bibinfo {author} {\bibfnamefont {J.}~\bibnamefont {Segado-Fernandez}}, \bibinfo {author} {\bibfnamefont {J.}~\bibnamefont {Ayllon-Guerola}}, \bibinfo {author} {\bibfnamefont {M.}~\bibnamefont {Garcia-Muñoz}}, \bibinfo {author} {\bibfnamefont {E.}~\bibnamefont {Viezzer}}, \bibinfo {author} {\bibfnamefont {C.}~\bibnamefont {Soria-Hoyo}}, \bibinfo {author} {\bibfnamefont {J.}~\bibnamefont {Garcia-Lopez}}, \bibinfo {author} {\bibfnamefont {G.}~\bibnamefont {Cunningham}}, \bibinfo {author} {\bibfnamefont {P.}~\bibnamefont {Buxton}}, \bibinfo {author} {\bibfnamefont {M.}~\bibnamefont {Gryaznevich}}, \bibinfo {author} {\bibfnamefont {Y.}~\bibnamefont {Hwang}}, \
  and\ \bibinfo {author} {\bibfnamefont {K.}~\bibnamefont {Chung}},\ }\href {\doibase https://doi.org/10.1016/j.fusengdes.2021.112706} {\bibfield  {journal} {\bibinfo  {journal} {Fusion Engineering and Design}\ }\textbf {\bibinfo {volume} {171}},\ \bibinfo {pages} {112706} (\bibinfo {year} {2021}{\natexlab{b}})}\BibitemShut {NoStop}%
\bibitem [{\citenamefont {Cruz-Zabala}\ \emph {et~al.}(2024)\citenamefont {Cruz-Zabala}, \citenamefont {Podestà}, \citenamefont {Poli}, \citenamefont {Kaye}, \citenamefont {Garcia-Munoz}, \citenamefont {Viezzer},\ and\ \citenamefont {Berkery}}]{Cruz-Zabala_2024}%
  \BibitemOpen
  \bibfield  {author} {\bibinfo {author} {\bibfnamefont {D.}~\bibnamefont {Cruz-Zabala}}, \bibinfo {author} {\bibfnamefont {M.}~\bibnamefont {Podestà}}, \bibinfo {author} {\bibfnamefont {F.}~\bibnamefont {Poli}}, \bibinfo {author} {\bibfnamefont {S.}~\bibnamefont {Kaye}}, \bibinfo {author} {\bibfnamefont {M.}~\bibnamefont {Garcia-Munoz}}, \bibinfo {author} {\bibfnamefont {E.}~\bibnamefont {Viezzer}}, \ and\ \bibinfo {author} {\bibfnamefont {J.}~\bibnamefont {Berkery}},\ }\href {\doibase 10.1088/1741-4326/ad8a70} {\bibfield  {journal} {\bibinfo  {journal} {Nuclear Fusion}\ }\textbf {\bibinfo {volume} {64}},\ \bibinfo {pages} {126071} (\bibinfo {year} {2024})}\BibitemShut {NoStop}%
\bibitem [{\citenamefont {Pereverzev}\ and\ \citenamefont {Yushmanov}(2002)}]{ASTRA}%
  \BibitemOpen
  \bibfield  {author} {\bibinfo {author} {\bibfnamefont {G.~V.}\ \bibnamefont {Pereverzev}}\ and\ \bibinfo {author} {\bibfnamefont {P.~N.}\ \bibnamefont {Yushmanov}},\ }\href@noop {} {\emph {\bibinfo {title} {{ASTRA Automated System for TRansport Analysis}}}} (\bibinfo {year} {2002})\BibitemShut {NoStop}%
\bibitem [{\citenamefont {Staebler}\ \emph {et~al.}(2007)\citenamefont {Staebler}, \citenamefont {Kinsey},\ and\ \citenamefont {Waltz}}]{TGLF}%
  \BibitemOpen
  \bibfield  {author} {\bibinfo {author} {\bibfnamefont {G.~M.}\ \bibnamefont {Staebler}}, \bibinfo {author} {\bibfnamefont {J.~E.}\ \bibnamefont {Kinsey}}, \ and\ \bibinfo {author} {\bibfnamefont {R.~E.}\ \bibnamefont {Waltz}},\ }\href {\doibase 10.1063/1.2436852} {\bibfield  {journal} {\bibinfo  {journal} {Physics of Plasmas}\ }\textbf {\bibinfo {volume} {14}},\ \bibinfo {pages} {055909} (\bibinfo {year} {2007})},\ \Eprint {http://arxiv.org/abs/https://doi.org/10.1063/1.2436852} {https://doi.org/10.1063/1.2436852} \BibitemShut {NoStop}%
\bibitem [{\citenamefont {Arakawa}(1966)}]{ARAKAWA1966119}%
  \BibitemOpen
  \bibfield  {author} {\bibinfo {author} {\bibfnamefont {A.}~\bibnamefont {Arakawa}},\ }\href {\doibase https://doi.org/10.1016/0021-9991(66)90015-5} {\bibfield  {journal} {\bibinfo  {journal} {Journal of Computational Physics}\ }\textbf {\bibinfo {volume} {1}},\ \bibinfo {pages} {119} (\bibinfo {year} {1966})}\BibitemShut {NoStop}%
\bibitem [{\citenamefont {Pueschel}\ \emph {et~al.}(2013)\citenamefont {Pueschel}, \citenamefont {Hatch}, \citenamefont {Görler}, \citenamefont {Nevins}, \citenamefont {Jenko}, \citenamefont {Terry},\ and\ \citenamefont {Told}}]{Pueschel_2013}%
  \BibitemOpen
  \bibfield  {author} {\bibinfo {author} {\bibfnamefont {M.~J.}\ \bibnamefont {Pueschel}}, \bibinfo {author} {\bibfnamefont {D.~R.}\ \bibnamefont {Hatch}}, \bibinfo {author} {\bibfnamefont {T.}~\bibnamefont {Görler}}, \bibinfo {author} {\bibfnamefont {W.~M.}\ \bibnamefont {Nevins}}, \bibinfo {author} {\bibfnamefont {F.}~\bibnamefont {Jenko}}, \bibinfo {author} {\bibfnamefont {P.~W.}\ \bibnamefont {Terry}}, \ and\ \bibinfo {author} {\bibfnamefont {D.}~\bibnamefont {Told}},\ }\href {\doibase 10.1063/1.4823717} {\bibfield  {journal} {\bibinfo  {journal} {Physics of Plasmas}\ }\textbf {\bibinfo {volume} {20}},\ \bibinfo {pages} {102301} (\bibinfo {year} {2013})},\ \Eprint {http://arxiv.org/abs/https://pubs.aip.org/aip/pop/article-pdf/doi/10.1063/1.4823717/15683690/102301\_1\_online.pdf} {https://pubs.aip.org/aip/pop/article-pdf/doi/10.1063/1.4823717/15683690/102301\_1\_online.pdf} \BibitemShut {NoStop}%
\end{thebibliography}%
% \bibliography{apssamp}% Produces the bibliography via BibTeX.

\end{document}